\documentclass[12pt]{iopart}
\usepackage{makeidx}
\usepackage{eurosym}
\usepackage{amssymb}
\usepackage{graphicx}
\begin{document}

\review[Atomic Physics Techniques for Nuclear Physics with Radioactive Beams]{Precision Atomic Physics Techniques for Nuclear Physics with Radioactive Beams}

\author{Klaus Blaum$\dag$, Jens Dilling$\ddag$ and Wilfried N\"ortersh\"auser%
\S }

\begin{abstract}
Atomic physics techniques for the determination of ground-state properties of radioactive isotopes are very sensitive and provide accurate masses, binding energies, $Q$-values, charge radii, spins, and electromagnetic moments. Many fields in nuclear physics benefit from these highly accurate numbers. They give insight into details of the nuclear structure for a better understanding of the underlying effective interactions, provide important input for studies of fundamental symmetries in physics, and help to understand the nucleosynthesis processes that are responsible for the observed chemical abundances in the Universe. Penning-trap and and storage-ring mass spectrometry as well as laser spectroscopy of radioactive nuclei have now been used for a long time but significant progress has been achieved in these fields within the last decade. The basic principles of laser spectroscopic investigations, Penning-trap and storage-ring mass measurements of short-lived nuclei are summarized and selected physics results are discussed.
\end{abstract}

\address{$\dag$ Max-Planck-Institut f\"ur Kernphysik, Saupfercheckweg 1, 69117 Heidelberg, Germany} \address{$\ddag$ TRIUMF, 4004 Wesbrook Mall, Vancouver, British Columbia, V6T 2A3 Canada \\ Department of Physics and Astronomy, University of British Columbia, Vancouver, British Columbia, V6T 1Z1 Canada} \address{$\S$ Technische Universit\"at Darmstadt, Institut f\"ur Kernphysik, Schlossgartenstr. 9, 64289 Darmstadt Germany \\ Johannes Gutenberg-Universit\"at Mainz, Institut f\"ur Kernchemie, Fritz-Stra\ss{}mann-Weg 2, 55128 Mainz, Germany \\ Helmholtzzentrum f\"ur Schwerionenforschung GSI, Planckstr. 1, 64291 Darmstadt, Germany}
\ead{\mailto{klaus.blaum@mpi-hd.mpg.de}, \mailto{jdilling@triumf.ca},
\mailto{wnoertershaeuser@ikp.tu-darmstadt.de}}

%Uncomment for PACS numbers title message
%\pacs{00.00, 20.00, 42.10}
% Keywords required only for MST, PB, PMB, PM, JOA, JOB?
%\vspace{2pc}
%\noindent{\it Keywords}: Article preparation, IOP journals
% Uncomment for Submitted to journal title message

%\submitto{\JPA}
% Comment out if separate title page not required
%\maketitle

\section{Introduction}

Masses, charge radii and nuclear moments of nuclei can be determined with
high accuracy using atomic physics techniques. Excellent progress has been made in recent years to adapt techniques to measure these properties to the requirements of radioactive beams \cite{Lunn2003,Blau2006,Neugart2006,Cheal2010}. The sensitivity and precision are at a level where detailed insight can be gained beyond bulk-properties. For example, the level of precision for mass measurements that allowed the discovery of magic numbers, which eventually led to the development of the Nuclear Shell Model, was on the order of $\delta m \approx 500$~keV -- 1~MeV. This was sufficient to identify extra binding and abnormalities in the general trend for example of the separation energies, which are on the order of $5 - 7$~MeV. Nowadays, the reachable precision is, even for radioactive isotopes, between (or often below) $\delta m \approx 1 - 10$~keV. Similar precision improvements have been achieved for laser spectroscopy, which is employed for radii and momentum determinations \cite{Cheal2010,Neugart1981,Otten1989,Billowes1995,Kluge2003}. Moreover, a key improvement is the level of sensitivity that was demonstrated, which is currently on the single ion level.\\
These sophisticated tools are powerful means to study the complex nuclear interaction with all its effective forces and interactions, and the resulting nuclear structure. Moreover, they provide sensitive tests of fundamental symmetries in physics, e.g., probing the nature of neutrinos, and help to understand the production paths in the nucleosynthesis and therewith the chemical abundances in the Universe. An overview of the current state of atomic physics techniques which are applied to radioactive isotopes is given in the following. This is showcased on selected physics examples.

\section{Fundamental Principles}
\subsection{Techniques for Mass Determination}
\subsubsection{Penning-Trap Mass Spectrometry}
The Penning trap technique for mass determinations is one of the most powerful tools to reach the highest precision, and moreover, allows one to maintain control over precision and accuracy \cite{Blau2006}. Penning traps were first conceived as storage devices for precision experiments on isolated single electrons. In fact, the Noble Prize for Physics was awarded in 1989 for 'the development of the ion trap technique' in part to H.G.~Dehmelt and W.~Paul. H.G.~Dehmelt invented and used the Penning trap system to successfully determine the $g$-factor of the electron \cite{Dehmelt1990}. W.~Paul invented the so-called Paul trap \cite{Paul1990}, which plays an important role for mass spectrometry as devices to prepare the ions prior to the injection in the mass measurement Penning trap system.
Nowadays Penning traps are used for mass measurements of stable and unstable atomic species, as well as molecules and antiparticles. For atomic mass measurements of stable species precisions of $\delta m/m = 10^{-11}$ \cite{Pritchard2004} have been reached. Penning-trap mass measurement systems are currently in use (or planned) at practically all radioactive beam facilities around the world (see table \ref{tab:PT_world}).

\begin{table}
\caption{Table of all currently operational, under commissioning, or planned Penning-trap mass measurement facilities at existing or planned radioactive beam facilities around the world. \vspace{0.5cm}}
\label{tab:PT_world}
\lineup
\begin{tabular}{|l|l|l|l|l|}
\hline\hline
Continent & Country & Facility & Penning Trap & Status \\ \hline\hline
North America & USA & Argonne & CPT & operational \\ \hline
& USA & MSU/NSCL & LEBIT & operational \\ \hline
& USA & Texas A\&M & TAMU-TRAP & planning  \\ \hline
& Canada & TRIUMF & TITAN & operational \\ \hline
Europe & Finland & Jyv\"{a}skyl\"{a} & JYFLTRAP & operational \\ \hline
& France & GANIL & DESIR-TRAP & planning \\ \hline
& Germany & GSI & SHIPTRAP & operational \\ \hline
& Germany & GSI & HITRAP & commissioning \\ \hline
& Germany & GSI/FAIR & MATS & planning \\ \hline
& Germany & Mainz University & TRIGA-TRAP & operational \\ \hline
& Germany & Munich University & MLL-TRAP & commissioning \\ \hline
& Switzerland & ISOLDE & ISOLTRAP & operational \\ \hline
Asia & China & IMP Lanzhou & Lanzhou-TRAP & planning \\ \hline
& China & CIAE, BRIF & BRIF-TRAP & planning \\ \hline
& China & CIAE, CARIF & CARIF-TRAP & planning \\ \hline
& India & VECC & VECC-TRAP & planning \\ \hline
& Japan & RIKEN & RIKEN-TRAP & planning \\ \hline
& South Korea & RISP & RISP-TRAP & planning \\ \hline\hline
\end{tabular}
\end{table}

%\begin{figure}[tbh]
%\begin{center}
%\includegraphics[width=15cm]{PT_world_table_small.eps}
%\end{center}
%\caption{Table of all currently operational, under commissioning, or planned Penning-trap mass measurement facilities at existing or planned radioactive beam facilities around the world.}
%\label{PT_world}
%\end{figure}

\begin{figure}[tbh]
\begin{center}
\includegraphics[width=7cm]{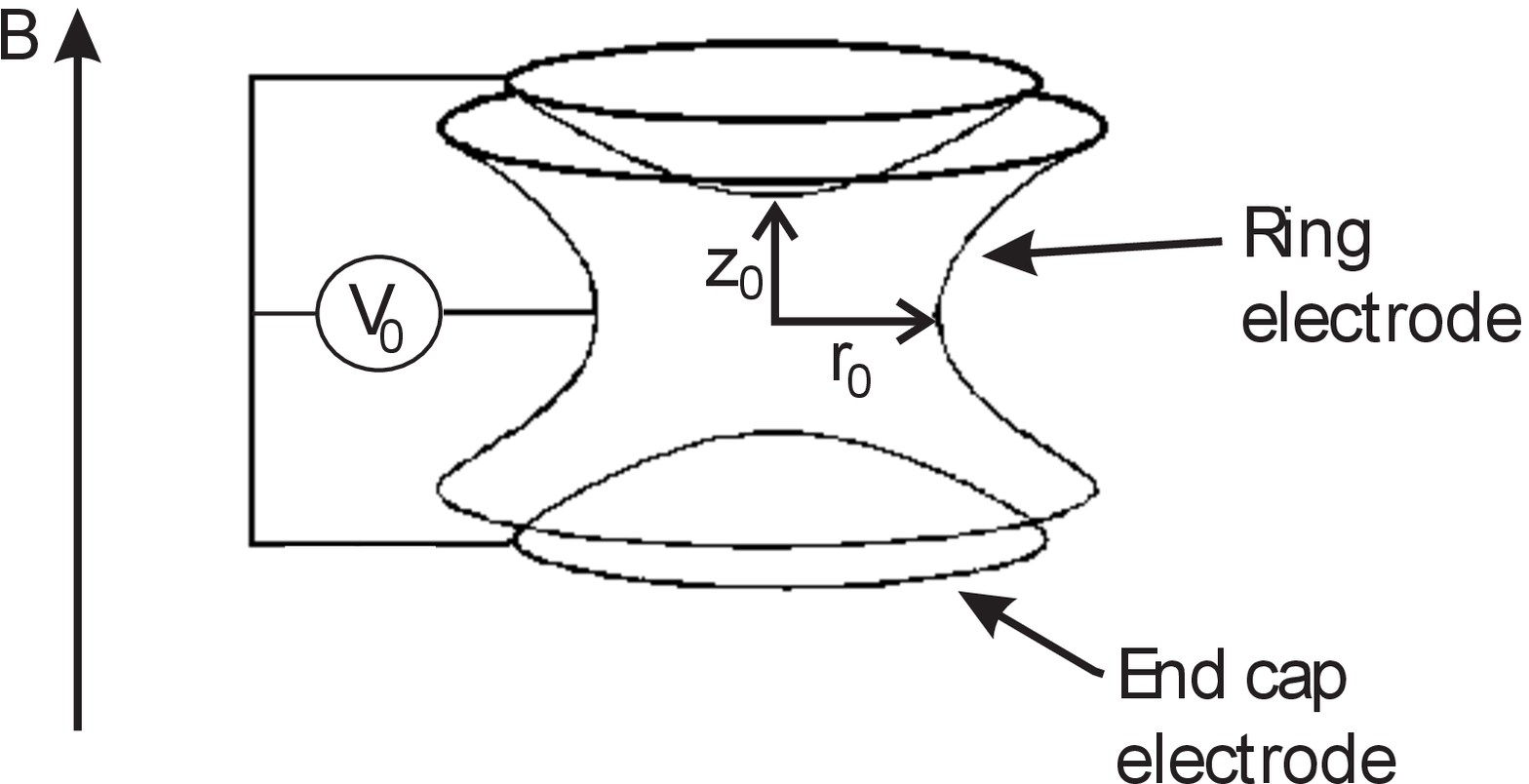}
\includegraphics[width=8cm]{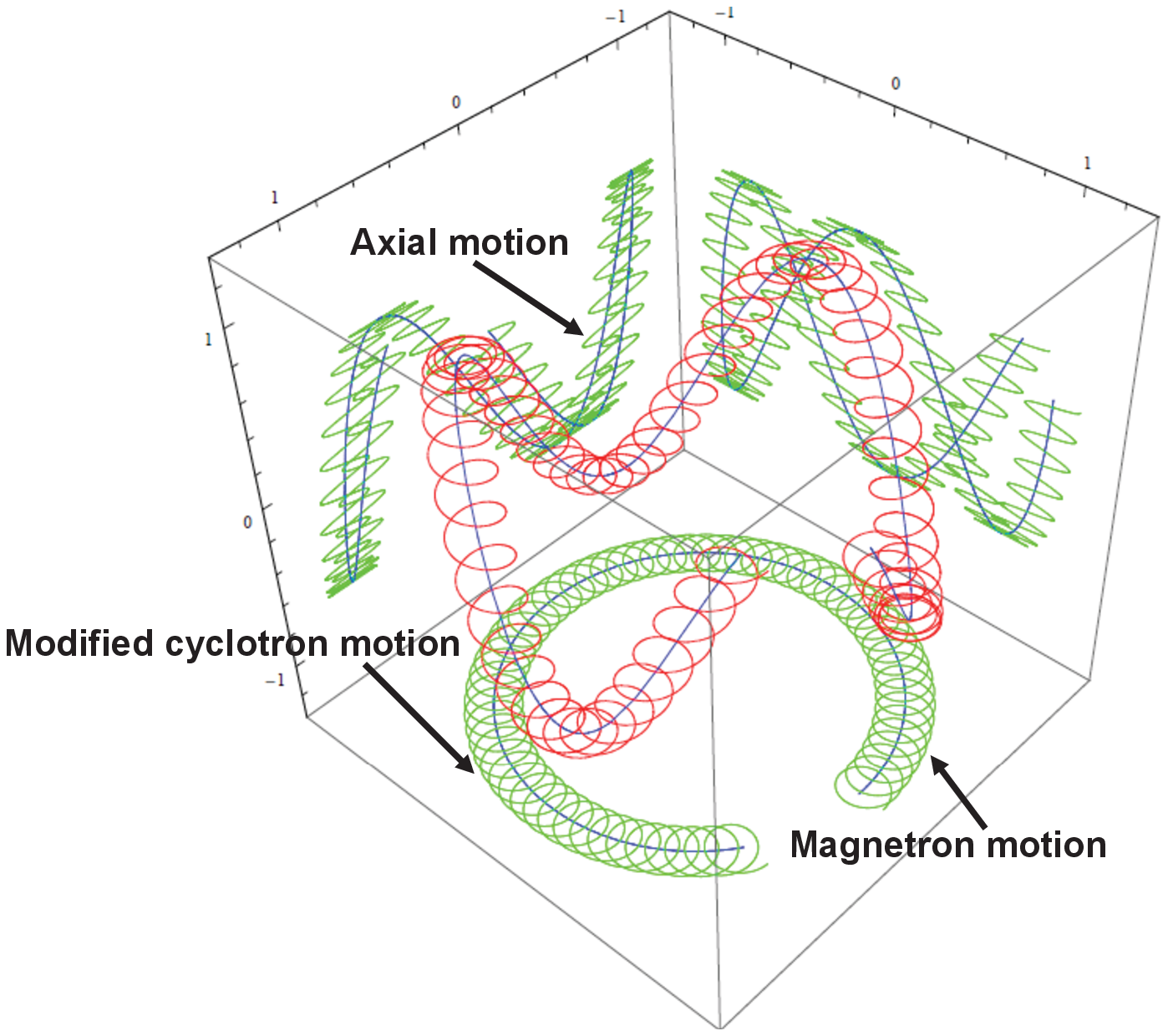}
\end{center}
\caption{Left: Schematic of a hyperbolic Penning trap system. For details see text. Right: (Colour online) The trajectory and three Eigenmotions of a charged particle in a Penning trap. The figure shows the combined motion and the three isolated projections: axial motion, magnetron motion, and modified cyclotron motion.}
\label{PT_schematic_traject}
\end{figure}

Penning traps store charged particles by the combination of a strong homogeneous magnetic field and a weak static (typically) quadrupolar electric field. Figure \ref{PT_schematic_traject} shows on the left the schematic configuration of a Penning trap system. The magnetic field points upwards and the electric field is generated by an applied voltage to the upper and lower end-cap electrodes, and one to the central ring electrode, leading to a potential difference of $V_{0}$. The typical electrode geometry is of hyperbolic nature, which allows one to achieve a harmonic potential. The trap-specific dimensions are $r_{0}$, the distance between trap center and ring electrode (in the minimum), and $z_{0}$; between trap center and end caps.\\ For the mass measurements, the electrodes are segmented to allow for either a differential read out of induced image currents (used for FT-ICR) or to apply RF excitations (for TOF-ICR). Both methods are explained below.

The Penning trap configuration leads to the storage of charged particles in three dimensions, and the motion of the particle is well understood and can be described classically \cite{Kretzschmar1992} and quantum mechanically \cite{Graeff1967,Abe2010}. An ion with a charge-to-mass ratio $m/q$ is exposed to the electric and magnetic forces and the resulting net effect is a cyclotron motion, which is composed of three Eigenmotions. These are the axial motion (up and down along the magnetic field axis) with frequency $\omega_{z}$, and two rotational motions, the magnetron ($\omega_{-}$) and modified cyclotron motion ($\omega_{+}$). The latter are both perpendicular to the axial motion. Figure \ref{PT_schematic_traject} shows on the right a calculated trajectory of an ion in a Penning trap and the deconvolution projection of the Eigenmotions.

The cyclotron frequency $\omega_{c}$, is proportional to the charge-to-mass ratio and the magnetic field strength $B$, and is given by
\begin{equation}
\omega_{c}= \frac{q}{m} \cdot B \, .
\label{eq:cyclotron_frequency}
\end{equation}%
Moreover, $\omega_{c}$ can be expressed as the square root of the quadratic sum of all Eigenmotion frequencies as
\begin{equation}
\omega_{c}^{2}=\omega_{z}^{2}+\omega_{+}^{2}+\omega_{-}^{2}.
\label{eq:B_G_Invariance}
\end{equation}%
This so-called Brown-Gabrielse-invariance theorem \cite{Brown1982,Gabr2009} cancels field imperfections of the Penning trap to first order. Employing this relationship allows one to determine the mass of the stored particle via the determination of the cyclotron frequency $\omega_{c}$. There are two main techniques applied in the field of radioactive beam Penning-trap mass measurements, the so-called Time-Of-Flight Ion Cyclotron Resonance (TOF-ICR) method and the Fourier-Transform Ion Cyclotron Resonance (FT-ICR) technique.

%\subsubsection{Mass Determination via Cyclotron-Frequency Measurement}
\subsubsection{Time-Of-Flight Ion Cyclotron Resonance (TOF-ICR) Technique}
\begin{figure}[tbh]
\begin{center}
\includegraphics[width=15cm]{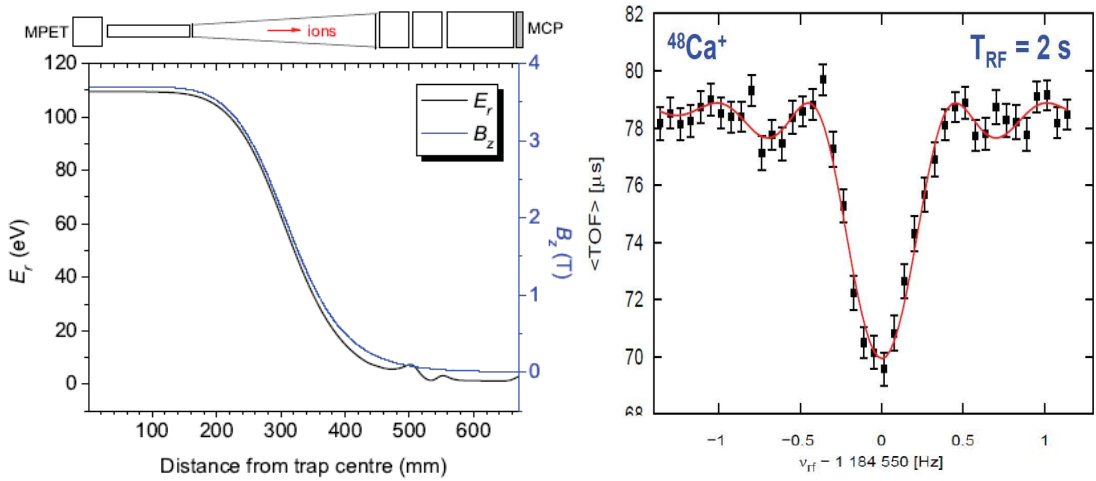}
\end{center}
\caption{(Colour online) Left: Radial energy conversion of the released ions in the changing magnetic field. The top of the figure shows the schematic of the trap, some adjustment ion optics elements and the MCP detector. The graph shows the longitudinal magnetic field gradient (scale on the right, blue curve), and the radial energy as a function of distance from the trap (scale on the left, black curve). Right:
A typical resonance TOF spectrum for $^{48}$Ca$^{+}$. This spectrum is taken with a $T_{\rm RF}=2s$ excitation time using the TITAN spectrometer at TRIUMF. The continuous line is a fit of the theoretical line shape to the data \cite{Koenig1995}.}
\label{TOF_mag_grad_plusTOF}
\end{figure}

%The cyclotron frequency  $\omega_{c}$ is proportional to the mass of the stored particle. Hence, a direct determination of the frequency leads to the mass, once the charge state and the magnetic field strength are known. There are two main techniques applied in the field of radioactive beam Penning-trap mass measurements, the so-called Time-Of-Flight Ion Cyclotron Resonance (TOF-ICR) method and the Fourier Transform Ion Cyclotron Resonance (FT-ICR) method.
For the Time-Of-Flight Ion Cyclotron Resonance TOF-ICR \cite{Graeff1980} technique, the ions are stored in the Penning trap and (typically) excited by an RF quadrupole field. This RF field at frequency $\nu_{\rm RF}$ is applied to a 4-fold or 8-fold segmented ring electrode for a period $T_{\rm RF}$. The RF excitation leads to a conversion from magnetron to cyclotron motion of the ions, which is accompanied by an increase in radial energy $E_{r}$. After the excitation, the ions are released from the trap and delivered to a detector system, typically a micro-channel plate or channeltron detector. The detector is installed outside the strong magnetic field region. The time of flight (TOF) from the trap to the detector is recorded. While traveling through the magnetic field gradient from high (on the order of 3-9 T) to the low magnetic-field region (typically 0.0001 T) radial energy $E_{r}$ is converted to longitudinal energy $E_{z}$. Figure \ref{TOF_mag_grad_plusTOF} on the left shows a schematic of the extraction from the trap to the detector on top of a graph that shows the magnetic field gradient and the corresponding radial energy as a function of distance from the trap. The minimal radial energy at the position of the detector indicates maximal longitudinal energy. By repeating this excitation, extraction, and TOF recording process for a set of frequencies around the expected cyclotron frequency $\omega_{c}$ a so-called TOF resonance spectrum is generated. The TOF is minimal for the maximal radial energy of the stored ions. This is reached when the excitation frequency $\omega_{\rm RF}$ corresponds to the cyclotron frequency $\omega_{c}$ of the ions and the conversion from magnetron motion to cyclotron motion is maximal. For this excitation frequency the ions have maximal radial energy in the trap. A typical TOF resonance spectrum is shown in the right part of figure \ref{TOF_mag_grad_plusTOF}. Since the cyclotron frequency $\omega_{c}$ depends on the charge-to-mass ratio $q/m$ of the ions as well as the magnetic field strength $B$, the derivation of the mass requires a determination of the charge state $q$ as well as $B$. The charge state $q$ is for most radioactive beam facilities (see P.v. Duppen's and T. Nilsson's contribution in this issue) either singly or doubly charged. Since the cyclotron frequency is directly proportional to $q$ the difference from $q=1^{+}$ to $q=2^{+}$ leads to a doubling of the frequency and is easily detected. For facilities with highly charged ions (HCI) capabilities, like TITAN \cite{Dilling2006} at TRIUMF or the HITRAP facility \cite{Quint2001} at GSI some charge selection technique is applied. HITRAP \cite{Herfurth2010} employs a selecting dipol magnet and TITAN makes use of separation via time-of-flight of a bunched beam with a Bradbury-Nilsen gate \cite{Brunner2012}. Both methods lead to unambiguous $q$ identification. The magnetic field strength $B$ is determined from calibration measurements using ions of very well-known mass such as $^{12}$C. To reach higher mass regions, one often uses carbon-clusters $^{12}$C$_{x}$. The mass of the ion of interest is then extracted from the frequency ratio of the two measurements, taking electron masses and binding energies into account.\\

%\subsubsection{Fourier Transform Mass Measurement Technique}
\subsubsection{Fourier-Transform Ion Cyclotron Resonance (FT-ICR) Technique}
\begin{figure}[tbh]
\begin{center}
\includegraphics[clip,width=7cm]{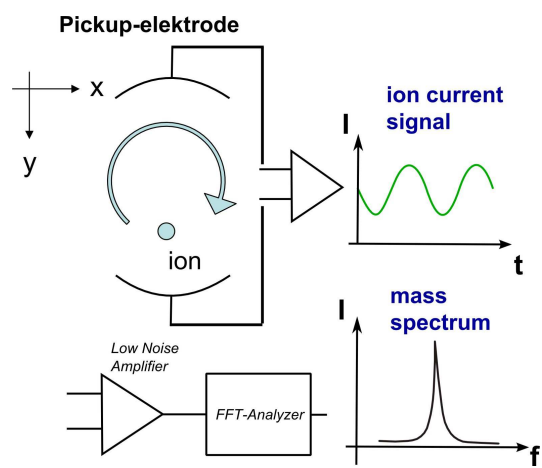}
\end{center}
\caption{(Colour online) Schematic of the FT-ICR mass measurement method. The top shows the circulating ion in the segmented Penning trap electrode configuration. Induced image currents are read out through a tuned tank circuit and a time dependent current spectrum is recorded. The lower part shows the current signal, which is amplified using a low-noise amplifier. A Fast Fourier Transform (FFT) is used to generate the mass spectrum.}
\label{FT_ICR}
\end{figure}

The method of Fourier-Transform Ion Cyclotron Resonance (FT-ICR) mass spectrometry was originally developed in the 1970's for mass determinations of molecular species in the field of analytical chemistry \cite{Comisarow1974} and has only recently been adapted for spectrometers for mass measurements of radioactive ions \cite{Ketelaer2009}. The schematic of the technique is shown in figure \ref{FT_ICR}. A stored circulating ion in the segmented Penning trap electrode configuration induces image currents. Those are read out through a tuned tank circuit and a time dependent current spectrum can be recorded. This current signal is amplified using a low-noise amplifier and a mass spectrum is generated by employing a Fast Fourier Transform (FFT).

The main advantage of the FT-ICR method over the otherwise very robust TOF-ICR approach is that a mass determination is in principle possible with a single ion. For the TOF-ICR a minimum of 200-300 ions are needed, however, with only one ion at a time in the trap. The downside of the FT-ICR is that
the induced signal current is very small (a few fA). With that the electronic noise becomes a major obstacle for practical applications. It can be overcome by using dedicated high quality tuned tank circuits (typical quality factor $Q \approx 5\,000$), reduction of the ambient Johnson-noise by going to cryogenic temperatures, or enhancing the induced signal by using higher charge states and/or longer accumulation times. Ideal applications for this method are mass measurements of super-heavy element (SHE) isotopes for nuclear structure studies as they are performed at SHIPTRAP \cite{Bloc2007,Minaya2012}. SHE are produced in minuscule quantities, and typically have half-lives of hundreds of milliseconds to a few seconds.
A proof-of-principle experiment is planned at the TRIGA reactor facility TRIGA-TRAP at Mainz University \cite{Ketelaer2008} and applications for HITRAP with highly charged ions are foreseen \cite{Klug2008}.

\subsubsection{Requirements for Mass Measurements of Radioactive Ions}
The specific requirements for the mass measurements of radioactive ions stem from the parameters of the ions themselves. For example the required sensitivity (depending on the production yield) and measurement speed (depending on the half-life) as well as the envisaged application which dictates the required precision. In order to be meaningful, all measurements should deliver reliable data, hence precise and accurate. The physics requirements can be categorized with the corresponding relative precision as follows:

\begin{itemize}
\item nuclear structure, $\delta m/m \approx 1 \times 10^{-7}$
\item nuclear astrophysics,  $\delta m/m \approx 1 \times 10^{-7 / -8}$
\item test of fundamental symmetries, neutrino physics,  $\delta m/m \approx 1 \times 10^{-8 / -9}$
\end{itemize}

\begin{figure}[tbh]
\begin{center}
\includegraphics[width=8cm]{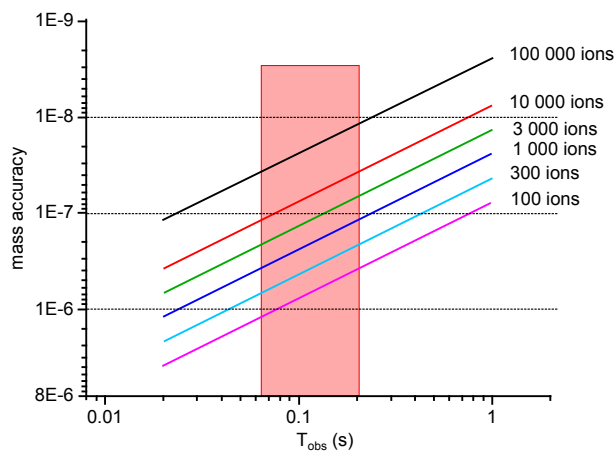}
\end{center}
\caption{(Colour online) Example of the achievable precision of a mass measurement using TOF-ICR for $^{74}$Rb with a half-life of $t_{1/2} = 65$~ms as a function of excitation time (the red shaded area indicates the longest possible excitation time for that half-life). The various statistical variations are indicated.}
\label{PT_precision}
\end{figure}

\paragraph{Accuracy and precision}
A major advantage of the Penning-trap mass measurement approach is the fact that Penning traps were developed for measurements of stable species. Penning traps behave exactly the same for stable and radioactive ions, hence the measurement method does not rely on the decay in any form. This in turn allows one to test, optimize, and calibrate the devices to a very high degree using stable isotopes. Moreover, the trapping and measurement principles have been theoretically investigated and are very well understood (see for example \cite{Brown1986,Bollen1990,Kretzschmar2007}). These facts lead to high confidence in reaching both accurate and precise data in the measurements. Virtually all operational Penning trap systems at radioactive beam facilities have studied their systematic uncertainties and developed new tools to further improve them.

One of the breakthroughs in the systematic studies was the introduction of the so-called carbon-cluster measurements.
Carbon clusters are molecular formations of $^{12}$C atoms in form of C$_{x}$, where $x$ can be between 2 and $\approx$ 60. Carbon clusters can therefore be generated in the entire mass range relevant for nuclear physics application, between mass $A=12$ and $A=240$ in mass steps of 12 units. Using such a grid it was possible to determine, quantify, and improve mass dependent systematic shifts. This was first developed for the ISOLTRAP system at CERN \cite{Blaum2002}, and is now also used at SHIPTRAP \cite{Chaudhuri2007}, TRIGA-TRAP \cite{Ketelaer2010} and JYFLTRAP \cite{Elomaa2009}. In addition, so-called compensation techniques have been developed (see for example \cite{Gabrielse1984}) which allows one to address a misalignment between electric and magnetic field axis. Misalignments lead to mass-dependent shifts, for example between species of interest and reference ions with a mass-to-charge difference of $\Delta (m/q)$. Again, this can be tested and determined very well with stable species by varying the frequency ratio $R$. Recently a new compensation method was developed reaching relative ratio precision (or mass dependent shift) of $\Delta R/R = - 4 \cdot 10^{-12} \times \delta(m/q) \times V_{0}$ \cite{Brodeur2012_MPET}, with $V_{0}$ being the applied potential between end-caps and center electrode at the Penning trap.

\paragraph{Speed of the mass measurements}
The Penning-trap mass measurement method which is essentially exclusively applied to radionuclides is the TOF-ICR method (with the exception of the above mentioned proof-of-principle experiment using FT-ICR). As can be seen in equation \ref{TOF_precision}, for the TOF-ICR method the achievable resolution (or mass uncertainty) is  inversely proportional to the excitation time $T_{\rm RF}$. Other factors are the charge state $q$, the magnetic field strength $B$, and a statistical factor $N$. Hence the required speed, or in other words the half-life of the radioactive isotope to be measured, dictates what duration can be used for the RF excitation $T_{\rm RF}$ and therefore governs the achievable  mass uncertainty of the measurement
\begin{equation}
\delta m \approx \frac{1}{\omega_{c}} \propto \frac{1}{T_{\rm RF} \times q \times B \times \sqrt{N}}.
\label{TOF_precision}
\end{equation}%
Figure \ref{PT_precision} shows the achievable precision in Penning-trap mass spectrometry as a function of the excitation time for the case of the short-lived radioactive isotope $^{74}$Rb with a half-life of $t_{1/2} = 65$~ms. Besides the actual time required for the RF excitation $T_{\rm RF}$, the preparation of the ions prior to the excitation takes time.

The initial condition of the ions before the RF fields are applied should be similar for all ions within one resonance spectrum. The preparation of these conditions has recently been dramatically improved by the invention of the so-called
Lorentz-steerer \cite{Ringle2006}, a device which introduces an electrical steering field within the magnetic field of the Penning trap. This preparation provides reliable injection parameters and initial conditions for the ions, while requiring virtually no time. Using such Lorentz-steerer devices, it was possible to reach isotopes as short-lived as $^{11}$Li \cite{Smith2008} with $t_{1/2} = 8.6$\,ms.

\paragraph{Improving the mass measurement precision}
The precision of the Penning-trap mass measurement method is given by equation \ref{TOF_precision}. One parameter that can not be manipulated is the half-life of the radioactive isotopes. In addition, the statistical factor $N$ only scales with the square-root and it is hence not very beneficial, particularly for experiments at radioactive beam facilities. The other two factors are the charge state $q$ and the magnetic field strength $B$.

\begin{figure}[tbh]
\begin{center}
\includegraphics[width=7cm]{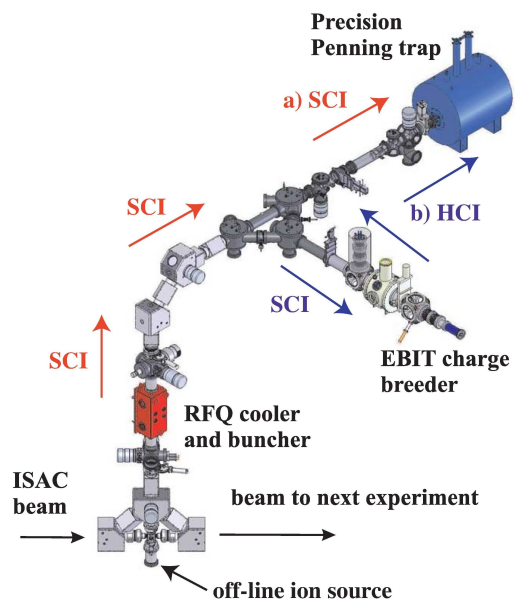}
\end{center}
\caption{(Colour online) The TITAN Penning trap facility \cite{Dilling2003} at TRIUMF where a charge breeder EBIT is employed to boost the achievable precision. SCI refers to singly charged ions and HCI to highly charged ions.} \label{titan}
\end{figure}

The precision is directly proportional to the charge state, and hence breeding the charge to a higher level will boost the precision accordingly (neglecting statistical losses due to inefficiencies and decay). Charge breeding has been applied for mass measurements of stable species with SMILETRAP \cite{Bergstroem2002} and has now also been adapted to radioactive species using an Electron Beam Ion Trap (EBIT) \cite{MSimon2012} system at TITAN. Figure \ref{titan} shows the set-up of the TITAN facility at TRIUMF, where the originally singly charged radioactive isotopes from ISAC are cooled, bunched in a linear Paul trap, and charge bred using the EBIT, before injecting them into the Penning trap for the mass determination. A number of experiments with highly charged ions were carried out \cite{Gallant2012_2,VSimon2012} and it was demonstrated that it is possible to increase the precision by up to two orders of magnitude at a constant excitation time, i.e., half-life of the isotope \cite{Ettenauer2011}.

The magnetic field strength also allows one to improve the precision, and nowadays Penning trap systems with magnetic field strengths of up to $9.6\,$T \cite{Bollen2006} are operational. This provides a gain in precision of about a factor of 2 over conventional systems with typical magnetic fields of 3 -- 5~T.

Both approaches reach the gain by a boost of the cyclotron frequency and hence the excitation frequency $\omega_{\rm RF}$.
This can also be reached by going to higher excitation modes. Typically quadrupole excitations are used, but recently it was
demonstrated at the LEBIT facility at MSU/NSCL \cite{Ringle2007}, the SHIPTRAP experiment at GSI \cite{Eliseev2007}, and ISOLTRAP \cite{Rosenbusch2012} that octupole excitations are possible.
Moreover, an improvement over conventional excitation and achievable precision of a factor 30 was achieved \cite{Eliseev2011}.

Another interesting new development is to enhance the reach of Penning traps at radioactive beam facilities by re-capturing decay products. This opens the door to mass measurements of species which would otherwise not be accessible due to limitations in the production mechanism, ionization, or extraction. This was recently demonstrated at the ISOLTRAP facility \cite{Herl2005,Herlerth2012} and is also planned for the TAMU-TRAP system \cite{Mehlman2012} at Texas A \& M University.

\subsubsection{Storage-Ring Mass Spectrometry}
The complementary devices to Penning traps for high-precision mass spectrometry on radionuclides are the high-energy heavy-ion storage rings, which are designed to store charged particles at relativistic energies. Presently, there are only two storage ring facilities in the world performing mass measurements of exotic nuclei, the Experimental Storage Ring ESR at GSI Helmholtzzentrum f\"ur Schwerionenforschung in Darmstadt (Germany) \cite{Fran1987} and the Experimental Cooler-Storage Ring CSRe at the Institute of Modern Physics of the Chinese Academy of Science in Lanzhou (China) \cite{Xia2002}. Both rings are installed at facilities with in-flight production of exotic nuclei (for the different production mechanisms of radioactive ion beams see the contribution by P. van Duppen in this issue).
Similar to the Penning-trap technique, the mass-to-charge ratios $(m/q)$ of the stored ions get related to their revolution frequencies $\nu$ according to \cite{Fran2008,Litv2011}
\begin{equation}
\frac{\Delta \nu}{\nu}=-\frac{1}{\gamma_t^2}\frac{\Delta (m/q)}{(m/q)}+\biggr(1-\frac{\gamma}{\gamma_t^2}\biggl)\frac{\Delta v}{v},
\end{equation}
with $\Delta v/v$ being the velocity spread of the ions, $\gamma$ the relativistic Lorentz factor and $\gamma_t$ the transition point of the storage ring, which is constant for a given ion-optical setting of the ring.

In order to minimize the term containing the velocity spread and hence to get a direct relation between the revolution frequency and the mass-to-charge ratio of the stored ions, two complementary storage-ring mass spectrometry techniques have been developed: Schottky and Isochronous Mass Spectrometry, SMS and IMS, respectively \cite{Fran2008}. While in SMS $\Delta v/v \rightarrow 0$ is realized by stochastic and electron cooling of the stored ion beams, which unfortunately takes a few seconds and thus limits the accessible half-life of the exotic species, in IMS the ring is tuned to a specific ion-optical mode where $\gamma_t=\gamma$ \cite{Stad2004}. In this mode the different velocities of the stored ions are compensated by the corresponding orbit lengths, such that all ions of a given species have the same revolution frequency. Nuclides with a half-life as short as a few ten microseconds can be addressed with this technique since no cooling is required \cite{Sun2010}.

The measurement of the revolution frequency is in the case of SMS non-destructive by detecting the induced image currents of the circulating ions on a Schottky pick-up and performing the Fourier transform of the periodic signals at typically the 30$^{\rm th}$ harmonic of the revolution frequency. In IMS operation a thin foil of a few $\mu$g/cm$^2$ is moved into the circulating beam. The penetrating ions produce secondary electrons, which are detected and allow for a measurement of the time of flight and thus the revolution frequency of the stored particles. As becomes apparent from the physics applications and results that will be discussed below, both storage-ring mass spectrometry techniques provide single ion sensitivity and can be applied to measure masses of nuclides with production rates as low as one ion per day.

\subsection{Ground-State Properties from Atomic Hyperfine Structure}

The properties of the atomic nucleus have a small but finite effect on the
appearance of their optical spectrum. They are summarized in the so-called
hyperfine structure of the transition lines. The properties that can be
extracted from the optical spectrum are the nuclear charge radius, the
magnetic dipole and the electric quadrupole moment as well as the nuclear
spin. Since the hyperfine structure is usually a 1~ppm to 1~ppb effect in
frequency when compared to the absolute transition frequency,
high-resolution techniques are required to observe the hyperfine structure.
Laser spectroscopy offers this resolution and was established as a unique
tool for the investigation of long chains of short-lived isotopes during the
last decades \cite{Cheal2010,Otten1989,Billowes1995,Kluge2003}.

\subsubsection{Isotope shift}
\label{sec:IS}
The isotope shift is the difference in the resonance frequency of an
electronic transition between two isotopes. It is usually defined as
\begin{equation}
\delta \nu _{\mathrm{IS}}^{AA^{\prime }}=\nu ^{A^{\prime }}-\nu ^{A},
\end{equation}%
i.e., the required change in frequency for tuning the light source that is
in resonance with the isotope $A$ at frequency $\nu ^{A}$ to the resonance
with the isotope of mass number $A^{\prime }$ at a frequency $\nu
^{A^{\prime }}$. One of the isotopes within the chain - in most cases a stable
isotope - is chosen as a reference. The contributions to the isotope shift
can be separated into two terms%
\begin{equation}
\delta \nu _{\mathrm{IS}}^{AA^{\prime }}=K_{\mathrm{MS}}\cdot \frac{%
M_{A^{\prime }}-M_{A}}{M_{A}M_{A^{\prime }}}+F~\delta \left\langle r_{%
\mathrm{c}}^{2}\right\rangle ^{AA^{\prime }}.  \label{eq:IS}
\end{equation}%
The first term is the mass shift that arises from the center-of-mass motion
of the nucleus and the fact that this motion changes with the mass of the
nucleus. It is obvious that a large relative change - as it is caused by
adding a single neutron to a light nucleus - has a much larger impact on the
shift than the addition of a neutron to a heavy nucleus. This is directly
reflected in the $1/M_{A}M_{A^{\prime }}$ dependence of the mass shift term.
The proportionality constant $K_{\mathrm{MS}}$ is usually divided once more: the
contribution expected for a two-body system
\begin{equation}
K_{\mathrm{NMS}}=m_{e}\nu _{0}
\end{equation}%
is called the normal mass shift (NMS) and always leads to a positive shift
for the heavier isotope. The remaining part is caused by the correlations
between the electrons and can have a positive or negative sign. It is called
the specific mass shift (SMS) and is notoriously difficult to evaluate because of the electron correlation integrals. Both contributions have the same mass dependence and thus it is possible to write
\begin{equation}
K_{\mathrm{MS}}=K_{\mathrm{NMS}}+K_{\mathrm{SMS}}.
\end{equation}%
Reliable ab-initio calculations with spectroscopic accuracy for the SMS are currently
only available for systems with up to three electrons as will be discussed
below.

The second term in equation \ref{eq:IS}, the so-called field shift, arises from the
change in the electron energy due to the finite size of the nucleus. For a
radially symmetric charge distribution, the field outside the nucleus is
equal to that of a point-like charge. But electrons that have a finite
probability to be found inside the nucleus probe the deviations of the
potential from the $1/r$ potential within the nuclear volume. It can be
shown that the change in energy of an electronic state is proportional to
the mean-square charge radius of the nucleus $\left\langle r_{\mathrm{c}%
}^{2}\right\rangle $\ and to the probability density of the electron inside the
nucleus $\left\vert \Psi _{\mathrm{e}}(0)\right\vert ^{2}$. Consequently,
the field shift contribution to the total transition energy between two
atomic states is proportional to the change in the probability density $%
\Delta \left\vert \Psi _{\mathrm{e}}(0)\right\vert _{i\rightarrow f}^{2}$
between the lower state $\left\vert i\right\rangle $ and the upper state $%
\left\vert f\right\rangle $. This offers, in principle, the possibility to
extract the absolute nuclear charge radius from the measurement of the total
transition frequency if all other contributions are sufficiently well known.
However, this is currently only possible for hydrogen \cite{Udem1997} and
hydrogenlike systems\footnote{Even though it is not an application of radioactive beams, it should be mentioned that absolute charge radii can be determined by laser spectroscopy on muonic atoms in combination with QED calculations. The recent result for muonic hydrogen \cite{Pohl2010} constitutes the so-called \emph{proton radius puzzle}, since it is 10 times more precise but $5\sigma$ smaller than the previous CODATA value \cite{Mohr2008}. Thus, it strongly disagrees with results from elastic electron scattering \cite{Bernauer2010} and the two-photon spectroscopy on hydrogen \cite{Parthey2010}.}. Already a second electron complicates the calculation
of mass-independent quantum-electrodynamical (QED) corrections so much that
the required accuracy cannot be achieved. With increasing number of electrons even the non-relativistic energies are not reliably calculable anymore. But the difference in the
transition energy between two isotopes can be measured and is proportional
to the change in the mean-square charge radius $\delta \left\langle r_{%
\mathrm{c}}^{2}\right\rangle ^{AA^{\prime }}$. As a result of a
nonrelativistic calculation, assuming a constant probability density of the
electron within the nuclear volume, one obtains
\begin{equation}
\delta \nu _{\mathrm{FS}}^{AA^{\prime }}=-\frac{Ze^{2}}{6\varepsilon _{0}}%
~\Delta \left\vert \Psi _{\mathrm{e}}(0)\right\vert _{i\rightarrow
f}^{2}~\delta \left\langle r_{\mathrm{c}}^{2}\right\rangle ^{AA^{\prime
}}=F_{i\rightarrow f}~\delta \left\langle r_{\mathrm{c}}^{2}\right\rangle
^{AA^{\prime }}.  \label{eq:FS}
\end{equation}%
The index at the field shift constant $F$, specifying the electronic transition, is usually suppressed.

\begin{figure}[tbh]
\begin{center}
\includegraphics[width=8cm]{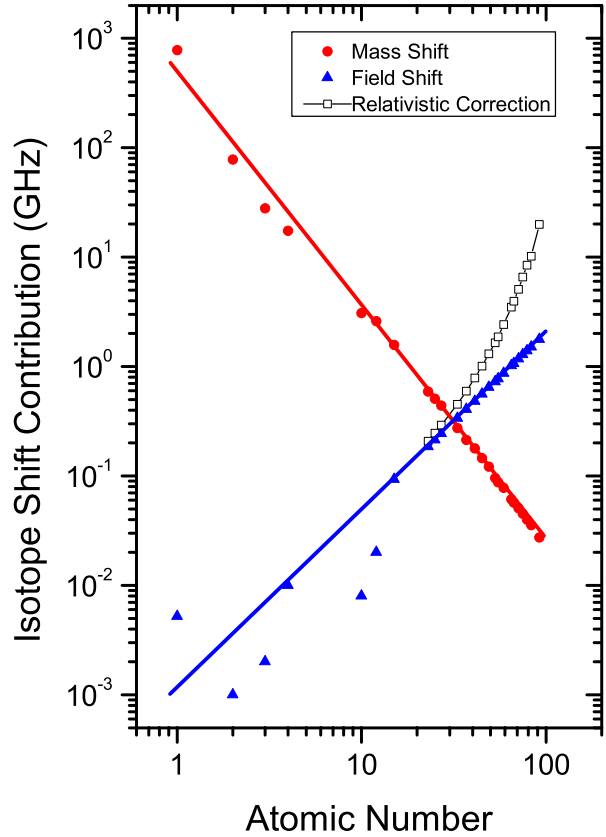}
\end{center}
\caption{(Colour online) Mass shift and field shift as a function of the atomic number $Z$. For the lightest elements (up to Na, $Z=11$) experimental values are plotted. The variation reflects the sensitivity of the respective transition used for spectroscopy. The values for the heavier elements are extrapolated based on the beryllium data using equations (\ref{eq:IS}) and (\ref{eq:FSDependence}). Relativistic contributions to the field shift are estimated on the values listed in \cite{Blundel1985} (see text).}
\label{fig:IS_vs_Z}
\end{figure}

An estimation of the field shift dependence on the atomic number $Z$ and the
mass number $A$ results in the rough approximation \cite{Otten1989}
\begin{equation}
\delta \nu _{\mathrm{FS}}^{AA^{\prime }}\varpropto \frac{Z^{2}}{\sqrt[3]{A}}.
\label{eq:FSDependence}
\end{equation}%
In figure~\ref{fig:IS_vs_Z} the mass-shift ($\circ $) and field-shift ($%
\triangle $) contributions are plotted as a function of the atomic number.
Both estimations are based on realistic numbers. For the mass shift, $%
K_{\mathrm{MS}}$, the measured mass shift in the $2s\rightarrow 2p$
resonance transition of Be$^{+}$ ions ($\lambda =313$~nm) was used and the
mass shift for a similar transition in heavier elements was calculated according to the mass dependence in equation (\ref{eq:IS}). The
field shifts for the heavier elements are extrapolated based also on the
field shift of beryllium using the simple relation given in equation (\ref{eq:FSDependence}).
For the lightest elements (up to $Z=11$, H -- Na) data was taken from experiments. The scatter indicates the variation of sensitivity with the respective transition. Especially in the light elements there is little choice in which transition to take for a given charge state -- particularly if the boundary conditions for the spectroscopy of short-lived isotopes and the availability of an appropriate laser system are considered. For He and Li the sensitivity is strongly reduced compared to hydrogen because the $2s$ electron does not probe the nuclear interior as strong as the 1s electron. Beryllium on the other hand uses also the 2s electron but now in a singly charged ion -- hence the electron is much tighter bound.
It is instructive to realize that the binding energy of an
electron decreases with increasing nuclear charge radius. Assuming an
increasing charge radius towards heavier isotopes and a transition with a
reduction of the electron density at the nucleus, will result in opposite
effects in the isotopes shift: The field shift decreases the transition
frequency whereas the mass shift increases it. If both are approximately of
the same size, the effect cancels and the isotope shift becomes very small.
This is indeed what is observed around the elements $Z\approx 38$ in
accordance with the data in figure~\ref{fig:IS_vs_Z}.

For heavier elements, the radial change in the electron density inside the
nucleus and relativistic effects cannot be neglected anymore. The nuclear
factor has then to be replaced by a power series of radial moments%
\begin{equation}
\lambda ^{AA^{\prime }}=\delta \left\langle r_{\mathrm{c}}^{2}\right\rangle
^{AA^{\prime }}+\frac{G(Z)}{F(Z)}\delta \left\langle r_{\mathrm{c}%
}^{4}\right\rangle ^{AA^{\prime }}+\frac{H(Z)}{F(Z)}\delta \left\langle r_{%
\mathrm{c}}^{6}\right\rangle ^{AA^{\prime }}+\cdots
\end{equation}%
where the fractional coefficients are the Seltzer coefficients
tabulated in \cite{Seltzer1969}. The nonrelativistic change in the
probability density of the electron is corrected by a factor that is
obtained from a realistic solution of the Dirac equation for a finite-size
nucleus with radius of liquid-drop dimension $R_{0}=r_{0}\sqrt[3]{A}$. The
effect has been included for illustration in figure~\ref{fig:IS_vs_Z} ($\square $) based on the
table provided in \cite{Blundel1985}. It is obvious that the relativistic
contributions significantly contribute for heavy elements.

\subsubsection{Magnetic Hyperfine Structure}

The magnetic interaction of an atomic nucleus with spin $I\neq 0$ and the
electron shell can be approximated in first order as that of a point
magnetic dipole interacting with the magnetic field that is produced by the
electrons at the center of the nucleus $B_{e}(0)$. It leads to a coupling of the total
angular momentum $J$ of the electron shell and the nuclear spin $I$ to the
atomic total angular momentum $F=J+I$. According to the angular momentum
coupling rules, $F$ can take any value between $I+J$ and $\left\vert
I-J\right\vert $ and the energy of the corresponding level is proportional
to the scalar product $I\cdot J$, which can be written
\[
\Delta E=a\,I\cdot J=a\frac{1}{2}\left( F^{2}-I^{2}-J^{2}\right)
\]%
and leads to a hyperfine splitting energy%
\begin{equation}
\Delta \nu _{\mathrm{mag}}=\frac{A}{2}C=\frac{A}{2}\left[ F\left( F+1\right)
-J\left( J+1\right) -I\left( I+1\right) \right]
\end{equation}%
with
\begin{equation}
A=\frac{\mu _{I}B_{e}(0)}{hIJ}.
\end{equation}%
The magnetic interaction is dominated by the contact term, i.e., the interaction of the electron's intrinsic magnetic moment with the magnetic moment of the nucleus. Since this term contributes only if there is a non-vanishing electron-spin density inside the nucleus, it is largest for $s$-electrons. \\
The ratio of the $A$ factors of different levels
is -- besides the \textquotedblleft trivial\textquotedblright\ dependence
on the quantum numbers $I$ and $J$ -- governed by the ratio of the magnetic
field at the nucleus. If we look at the $A$ factors of different isotopes
along a chain, it is obvious that - in first order - the ratios should be
determined by the ratios of the magnetic moments. We obtain for the
ratio of the magnetic dipole hyperfine structure constant in two electronic
levels $i$ and $f$ for two isotopes 1 and 2 the relation
\begin{equation}
\frac{A_{1}^{f}}{A_{1}^{i}}=\frac{A_{2}^{f}}{A_{2}^{j}}
\end{equation}%
which is often used as a constraint in fitting spectra with low statistics.
Similarly, the unknown nuclear moment of one isotope can be connected to the
known moment of a reference isotope via%
\[
\mu =\frac{A}{A_{\mathrm{Ref}}}\frac{I}{I_{\mathrm{Ref}}}\mu _{\mathrm{Ref}%
}.
\]%
These relation neglect finite-size contributions of the nucleus,
which arise from the change of the electron wavefunction due to the spatial
extension, the distribution of the nuclear charge (Breit-Rosenthal effect,
BR) and the distribution of the nuclear magnetic moment inside the nucleus
(Bohr-Weisskopf effect, BW). They are usually accounted for by two factors
\begin{equation}
A=A_{\mathrm{point}}~\left( 1+\epsilon _{\mathrm{BW}}\right) \left(
1+\epsilon _{\mathrm{BR}}\right) .
\end{equation}%
These modifications from the point-like nucleus give rise to the so-called
hyperfine anomaly in the ratio of the $A$ factors between two isotopes%
\begin{equation}
\frac{A_{1}}{A_{2}}\approx \frac{g_{I}(1)}{g_{I}(2)}\left( 1+\,^{1}\Delta
^{2}\right)
\end{equation}%
where $g_{I}$ is the nuclear gyromagnetic ratio $\mu _{I}= g_{I}\mu _{N} I$ and
$^{1}\Delta ^{2}$ the differential hyperfine anomaly. In extreme cases, $%
^{1}\Delta ^{2}$ can be of the order of a few percent but in most cases it
is on the 10$^{-4}$ level or even smaller. A recent compilation can be found
in \cite{Pearson2012}. The effect is usually too small to be extracted from
optical spectra but microwave spectroscopy is a useful tool for its
determination \cite{Werth1995}.

\subsubsection{Electric Hyperfine Structure}

A non-spherical nucleus with spin posseses also higher electromagnetic multipole
moments up to the order $2I$. Besides the monopole term -- accounted for
by the Coulomb term -- the quadrupole moment is the next possible higher
order of an electric moment. It arises from the energy of the orientation
of the nuclear charge distribution in the inhomogeneous electric field
of the electron shell and is given by
\begin{equation}
\Delta \nu _{\mathrm{el}}=B\frac{\frac{3}{4}C\left( C+1\right) -I\left(
I+1\right) J\left( J+1\right) }{2I\left( 2I-1\right) J\left( 2J-1\right) },
\end{equation}%
where%
\begin{equation}
B=\frac{eQ_{s}}{h}\left. \frac{\partial ^{2}V}{\partial z^{2}}\right\vert
_{r=0},
\end{equation}%
with $Q_{s}$ being the spectroscopic quadrupole moment of the nucleus, and $%
V_{zz}(0)$ the electric field gradient at the nucleus. The electric
hyperfine structure appears only for nuclei and electronic states with $%
I>1/2 $ and $J>1/2$, respectively. It shifts the magnetic hyperfine levels
in a characteristic way, depending on the prolate or oblate form of the
nucleus.

The observation of the atomic hyperfine structure allows us to determine
also the spin of a nucleus if it is unknown. As long as $I<J$, the spin can
directly and unambiguously be determined from the number of observed
hyperfine components. Otherwise, the relative distances between the
hyperfine components as well as the relative intensities are signatures of
the spin. The theoretical line strength $S_{FF^{\prime }}$ of a hyperfine
transition between the hyperfine levels $F$ and $F^{\prime }$ arising from
the fine structure levels with total angular momentum $J$ and $J^{\prime }$
are related to the line strength in the underlying fine structure transition
$S_{JJ^{\prime }}$ by%
\begin{equation}
S_{FF^{\prime }}=\left( 2F+1\right) \left( 2F^{\prime }+1\right) \left\{
\begin{array}{ccc}
F & F^{\prime } & 1 \\
J^{\prime } & J & I%
\end{array}%
\right\} ^{2}~S_{JJ^{\prime }},
\end{equation}%
where $\{\cdots \}$ denotes the 6-j coefficient that can be taken from
standard books on angular momentum theory. However, these intensities must
be handled with care since they are only correct for the excitation
with unpolarized light and isotropic detection efficiency. Otherwise the
detection geometry and the intensity distribution must be taken into
account. Moreover, optical pumping can significantly alter the intensity
ratios and lineshapes. In such cases, detailled studies of the lineshape
have to be carried out.

\subsubsection{Application of Laser Spectroscopy for Short-Lived Isotopes}

\begin{figure}[tbh]
\begin{center}
\includegraphics[width=15cm]{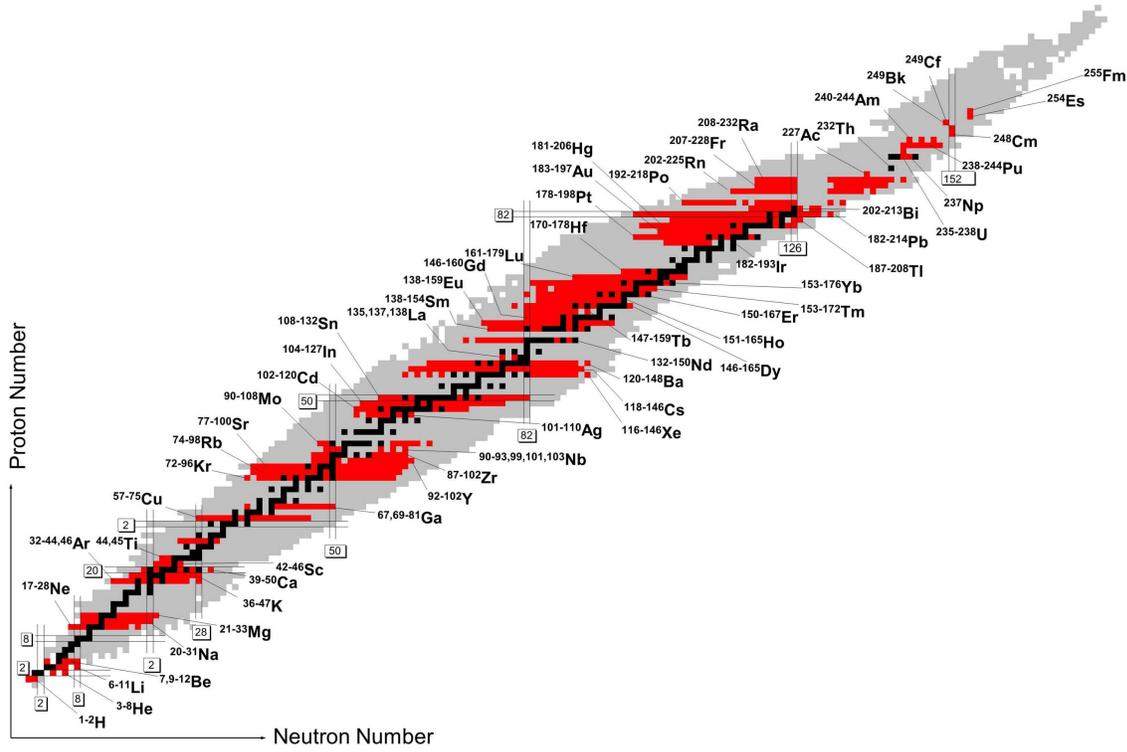}
\end{center}
\caption[Laser spectroscopy in the nuclear chart]{(Colour online) Laser spectroscopy in the nuclear chart. Black squares indicate the stable and very long-lived nuclei, while the known radioactive isotopes are plotted in grey. Those radioactive isotopes for which laser spectroscopy was performed online - in a few cases with long-lived isotopes off-line - are marked in red.}
\label{fig:NuclearChart}
\end{figure}

Figure~\ref{fig:NuclearChart} shows a nuclear chart with all radioactive
isotopes that have been studied with laser spectroscopy at on-line facilities -- and in a few cases off-line -- to extract properties of the nuclear ground state or
sufficiently long-lived isomers.
In recent years the region of the lightest nuclei, the region of the island of
inversion around $N=20$ and the region of the refractive elements above Sr,
where a sudden-onset of deformation at $N=60$ is observed, were studied
intensively. Furthermore, the filling of the neutron $pfg$-shell in isotopes
above the doubly-magic $^{56}$Ni have lately been of much interest
and was studied in copper and gallium. Further work in progress includes the region at
the $N=28$ shell closure above potassium and calcium and the $N=50,82$ shell
closures for nuclei in the tin region. Some of these developments have been
enabled by new experimental techniques as will be discussed in the following.

\subsubsection{Laser Spectroscopic Techniques}

The majority of laser spectroscopic data at on-line facilities has been
obtained using two techniques: collinear laser spectroscopy (CLS)
\cite{Kaufman1976,Anton1978} and resonance ionization (mass) spectroscopy
(RIS/RIMS) \cite{Letokhov1977,Hurst1977} in various modifications. Both
techniques are rather complementary to each other: generally CLS provides
higher resolution but is less sensitive, while RIS is usually performed with
pulsed lasers and is therefore limited in resolution but can be extremely
sensitive. However, in special cases the sensitivity of collinear
spectroscopy can be strongly increased by particle detection and RIS/RIMS
can provide high-resolution if cw lasers are applied. Also mixed forms of
CLS/RIS have already been demonstrated. Examples will be given below.
Besides these techniques, dedicated methods were developed for the lightest
isotopes in order to meet the challenges for a charge radius determination
with extremely small field shifts and low production rates. This will be
one focus of this proceeding.

\paragraph{Collinear Laser Spectroscopy (CLS)}

\begin{figure}[bth]
\begin{center}
\includegraphics[width=8cm]{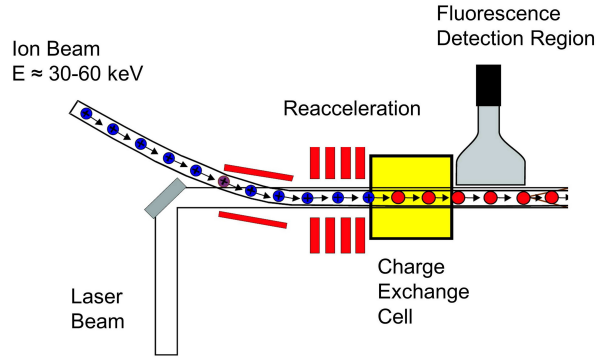}
\end{center}
\caption{(Colour online) General setup of collinear laser spectroscopy (CLS). An ion beam is superimposed with a laser beam in an electrostatic deflector. The ion beam can be neutralized in a charge-exchange cell (CEC) filled with a vapor of an alkaline metal at low-pressure. The fluorescence light is detected with photomultipliers in the optical detection region (ODR). An additional potential is applied either to the CEC or to the ODR to avoid interaction between the laser and the ion beam before the ODR is reached.}
\label{fig:CLSPrinciple}
\end{figure}
In CLS the laser beam is superimposed with a beam of fast ions or atoms as
shown in figure \ref{fig:CLSPrinciple} and the
resonance fluorescence is detected with a photomultiplier perpendicularly to
the flight direction. Since the atoms are propagating parallel or
antiparallel to the laser beam with velocity $\beta =\upsilon /c$, the
resonance frequency $\nu _{0}$ of the atom is shifted in the laboratory
system by the relativistic Doppler effect according to%
\begin{equation}
\nu _{\mathrm{p,a}}=\nu _{0}\gamma (1\pm \beta ),  \label{eq:relDoppler}
\end{equation}%
with the time dilation factor $\gamma =(1-\beta ^{2})^{-1/2}$. CLS utilizes
the fact that the acceleration of an ion ensemble with a static electric
potential compresses the longitudinal velocity distribution. Thus, the
Doppler width is considerably reduced \cite{Kaufman1976,Wing1976}. In a
simplified way, the kinematic Doppler compression can be deduced using the
differential variation of energy with velocity $\mathrm{d}E=m\upsilon \,%
\mathrm{d}\upsilon $. The energy uncertainty $\delta E$ in the source,
caused by the Doppler distribution $\delta \upsilon _{\mathrm{D}}$ of the
ions, stays constant under electrostatic acceleration with a potential $U$.
From $\delta E=\delta (m\upsilon ^{2}/2)=m\upsilon \,\delta \upsilon =%
\mathrm{const}$., it follows that any increase in velocity must be
compensated by a corresponding reduction of $\delta \upsilon _{\mathrm{D}}$%
\begin{equation}
\delta \upsilon =\frac{\delta E}{m\upsilon }=\frac{\delta E}{\sqrt{2meU}}.
\label{eq:DopplerReduction1}
\end{equation}%
Since the remaining Doppler width $\Delta \nu _{\mathrm{D}}=\nu _{0}(\delta
\upsilon /c)$ for a resonance transition at frequency $\nu _{0}$ is
proportional to $\delta \upsilon _{\mathrm{D}}$ it follows that
\begin{equation}
\Delta \nu _{\mathrm{D}}=\nu _{0}\frac{\delta E}{\sqrt{2eUmc^{2}}}.
\label{eq:DopplerReduction2}
\end{equation}%
Hence the Doppler width decreases proportionally to $1/\sqrt{U}$ and beam
energies of about 50~keV are usually sufficient to reduce the Doppler width
to some 10~MHz and therefore into the order of the natural line width for
allowed dipole transitions. CLS has been applied on-line first at the TRIGA reactor in
Mainz \cite{Schinzler1978} and soon thereafter at ISOLDE \cite{Neugart1981,Mueller1983}.

The collinear geometry allows for sufficiently long interaction times and
high resolution. The velocity dependence of the laser frequency in the rest
frame of the ion according to equation (\ref{eq:relDoppler}) can be utilized to keep
the laser frequency in the laboratory system fixed and to tune instead the
resonance frequency of the ion across the laser frequency by changing the
ion velocity. This Doppler-tuning is usually realized by applying an
additional post-acceleration potential to the optical detection region (ODR) in which the
resonance fluorescence is detected. This has the additional advantage that
optical pumping into dark states cannot occur before the ions enter the
detection region.

Spectroscopy of fast neutral atoms can also be carried out if a charge
exchange cell, filled with a thin vapor of an alkaline element, is mounted
in front of the detection region. The cross sections for charge exchange
collisions of a beam with ions $\mathrm{M}^{+}$ on, e.g., Cs vapor
\begin{equation}
\mathrm{M}^{+}+\mathrm{Cs}\longrightarrow \mathrm{M}+\mathrm{Cs}^{+}+\Delta E
\end{equation}
is large ($\approx $ $10^{-14}$ cm$^{2}$) at typical beam energies and leads
preferably to the charge exchange into states with $\Delta E$ close to 0
(resonant charge exchange). This provides also the opportunity to populate
metastable higher lying levels if the electron donator is properly chosen.
Doppler-tuning must in this case be realized by applying the
post-acceleration voltage to the charge exchange cell (CEC) and the optical
detection region should be mounted as close as possible behind the CEC. Technical details and a comparison of two CECs -- a horizontal and a vertical design as they are in use at COLLAPS and at TRIUMF, respectively -- were recently presented in \cite{Klose2012}.

While the direct optical detection is the most straightforward detection
scheme, its sensitivity is limited by the small solid angle, the quantum
efficiency of the detectors, background from scattered laser light and
several other factors. During the last decades CLS was therefore combined
with various other detection techniques with the aim to increase sensitivity
and accuracy of the technique \cite{Cheal2010,Geithner2000,Neugart2006}. Some of the recent
developments will be discussed below.

\paragraph{Resonance Ionization Spectroscopy (RIS)}

RIS combines the large cross section of resonant optical excitation with the
very sensitive charged-particle detection by exciting an electron in the
atom stepwise along dipole-allowed transitions until the electron is
detached from the atom and either the electron, the ion, or both are
detected. There is a multitude of arrangements for RIS. Efficient excitation
at medium to low resolution can be obtained using pulsed lasers with
bandwidths adjusted to the Doppler width of the atomic ensemble that has to
be ionized. It has been first applied for nuclear charge radii measurements
of short-lived isotopes at Gatchina \cite{Alkhazov1983}. Here, $^{145-149}$%
Eu isotopes were produced on-line, mass separated and collected. After the
transfer into a hot atomic beam source, laser spectroscopy was applied on
the atomic beam with the laser beams perpendicular to reduce
Doppler-broadening. For shorter-lived isotopes of europium the mass
separated ions were not collected but introduced into a hot oven were they
were neutralized and evaporated. Spectroscopy was then again performed on
the atomic beam \cite{Fedoseyev1984}. The concept of a hot-cavity laser ion
source for on-line use was suggested by Kluge et al.\ \cite{Kluge1985} as
well as by Andreev et al.\ \cite{Andreev1986} and first realized at
Gatchina \cite{Alkhazov1989}. In this modification it it now widely used as
an ion source for element selective ionization. The resonant laser ion
source RILIS at ISOLDE \cite{Marsh2010}\ is now the mostly used of all ion
sources at ISOLDE. It has also been well established at TRIUMF\ \cite{Lassen2010} and was demonstrated at the Holyfield Facility at Oak Ridge
National Laboratory (ORNL) \cite{Gottwald2008}. Developments towards a laser ion source at SPIRAL2 are going on at ALTO and at GANIL. The combination of a resonant laser ion source with a radiofrequency trap, the so-called LIST (laser ion source trap) scheme to improve the selectivity and the suppression of isobaric contaminations due to surface ionization has been proposed in \cite{Blau2003b} and its performance demonstrated in \cite{Schwellnus2010}. Soon the first on-line coupling at ISOLDE will be demonstrated.

A laser ion source based on a gas cell has been proposed by van Duppen and
coworkers \cite{vanDuppen1992} and demonstrated at the Leuven Isotope
Separator on-line (LISOL) \cite{Vermeeren1994,Kudryatsev1996}. A development
of an on-line gas-cell laser ion source is going on at Jyv\"{a}skyl\"{a}
\cite{Kessler2006} and is also foreseen at RIKEN (Rikagaku Kenkyusho,
Institute of Physical and Chemical Research, Japan) \cite{Wada2010} and for
the next-generation facility FAIR at GSI Darmstadt (Facility for Antiproton
and Ion Research).

Resonance ionization spectroscopy is also being used to measure nuclear
moments and charge radii: It has been applied combined with collinear laser
spectroscopy \cite{Schulz1991,Billowes2008}, in a gas cell \cite{Lauth1992},
after deposition and laser ablation from a beam catcher as applied by the
COMPLIS collaboration \cite{Sauvage2000}, e.g. for the tin isotopes \cite%
{LeBlanc2002,LeBlanc2005}, or directly inside the hot-source (in order of
decreasing resolution). The last method offers the lowest resolution but is
extremely sensitive and the resolution is often still sufficient for the
heavy isotopes that exhibit large field shifts and large hyperfine
splittings. A recent review on RIS for nuclear physics has been given by
Fedoseev, Kudryatsev and Mishin \cite{Fedoseev2012}.

It should be noted that the technique can also provide ultrahigh isotopic
selectivity if cw lasers and a multi-step excitation scheme is used.
This approach has been chosen for the trace detection of many
ultra-low abundant or radiotoxic isotopes, for example $^{41}$Ca, $^{89,90}$Sr,
$^{135,137}$Cs \cite{Wendt1999,Lu2003}. Continuous-wave RIMS was also used for the
investigation of the charge radii of the lithium isotopes as discussed below.

\section{Probing the Nuclear Interaction through Ground-State Properties}

\subsection{Isotope Chains with Halo Nuclei}

During the last decades, great progress has been made in the ab-initio modeling
of the nuclear structure of the lightest elements. A forerunner in this
field was the Green's function Monte-Carlo technique \cite{Pieper2001}:
Starting from phenomenological two-body ($NN$) potentials with parameters
fitted to the phase shifts observed in nucleon-nucleon scattering, it was
found that three-body interactions ($3N$) are required to get the binding
energies and the level structure right. These calculations have now been
carried out up to $^{12}$C. Recently, there were additional attempts to
bridge the gap between the fundamental theory of QCD and the
phenomenological interactions using effective field theories \cite%
{Epelbaum2010} as being discussed in the contribution by C. Forssen in this
issue. Other theories based on $NN$ interactions are, e.g., the No-Core
Shell Model \cite{Navratil1998}\ and the Fermionic Molecular Dynamics model
\cite{Feldmeier1998,Roth2010}. Alternative descriptions of light nuclei are based
on their strong clusterization which is expressed experimentally by the
small cluster thresholds -- it is for example easier to remove a\ deuteron
from $^{6}$Li than a single neutron or proton -- and is also obtained in
ab-initio calculations of light nuclei, e.g. in Green's Function Monte Carlo
calculations. A number of cluster models, for example the stochastic
variational multi-cluster approach \cite{Varga1994}\ have been developed
that treat small clusters like $\alpha $ particles, deuterons, tritons, $%
^{3} $He and individual neutrons and protons as the building blocks.
Calculations of the nuclear ground states can then be performed based on
effective potentials between these clusters.

\begin{figure}[tbh]
\begin{center}
\includegraphics[width=10cm]{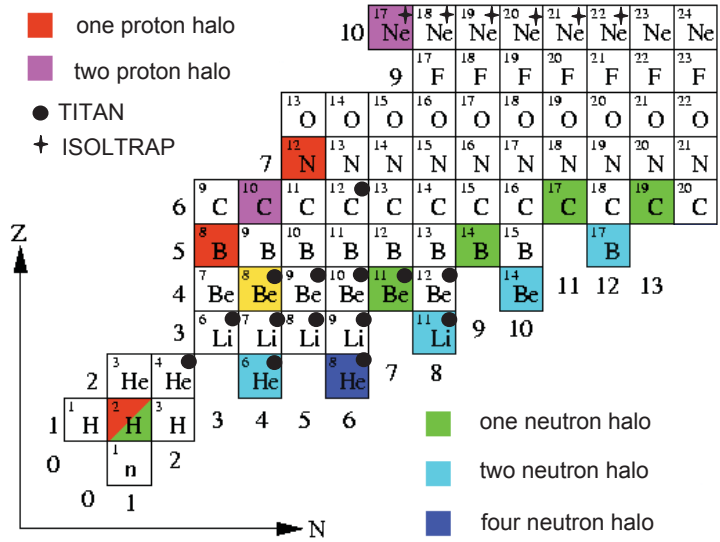}
\end{center}
\caption{(Colour online) A section of the chart of nuclides where found or anticipated neutron and proton halos are color coded. Mass measurements carried out by the ISOLTRAP and TITAN facility are labeled.} \label{halo_chart}
\end{figure}

Nuclear masses, radii and moments along the isotopic chains of the lightest
elements provide benchmarks to test these theories as well as the cluster
and three-particle dynamics of the intriguing halo nuclei that appear in
this region of the nuclear chart.
Halo nuclei (for more details see contribution by I.\ Tanihata in this issue)
are a challenge for precision measurements due to their short half-life
and their low production yield at radioactive beam facilities.
They can be characterized by three fingerprints: a very low binding
energy of the last nucleon(s), a large
interaction cross section and a narrow momentum distribution of the
fragments in breakup reactions. All three characteristics are intimately connected with observables of high-accuracy atomic physics measurements. The first point clearly indicates that the mass is one of the key properties to describe and understand the halo phenomenon. In particular, the mass gives access to the nucleon separation energies, hence the one, two, or four proton or neutron separation energies, which defines how tightly the last nucleon(s) are bound to the core. Moreover, in order to extract the charge radius of nuclei from precision isotope shift measurements (see section \ref{sec:IS}) the mass shift needs to be known. Mass shift calculations for the lightest nuclei require a determination of the nuclear mass with accuracy $\delta m \approx 1$~keV or even below \cite{Yan2008}. But for a long time, masses of halo isotopes and other radioactive light nuclei were typically determined in reactions or fragmentation processes coupled to time-of-flight spectrometers. Since these techniques are limited in accuracy, very accurate mass measurements were not available for these exotic nuclei. Considerable progress has been made in the last few years to obtain such data and the development of Penning trap techniques played an important role.
Figure \ref{halo_chart} shows the lower section of the Segre chart of nuclei. Color-coded are the various identified or anticipated halo nuclei and those light isotopes for which mass measurements were carried out at ISOLDE or TRIUMF are labeled.

Concerning the second fingerprint, it was a laser spectroscopic measurement of the magnetic dipole \cite{Arnold1987}\ and electric
quadrupole moment of $^{11}$Li \cite{Arnold1992}\ that clearly disproved a strong deformation of the $^{11}$Li nucleus as a possible explanation for the large interaction cross section measured by Tanihata and coworkers \cite{Tanihata1985} and therefore strongly supported the halo picture as it was developed by Hansen and Jonson \cite{Hansen1987}. In recent years, the nuclear moments \cite{Borremans2005,Neugart2008} of this nucleus have been measured with increased accuracy.

Finally, the charge radius of a neutron-halo nucleus compared to that of the core nucleus provides information about the correlation between the halo neutrons, which can be combined with the results of breakup reactions, e.g.\ in \cite{Simon1999} or \cite{Nakamura2006a} in order to get a detailed picture of the relative motions between core and halo nucleons and a possible contribution of core excitation. Hence, charge radius measurements of halo nuclei were considered important from the very beginning of their investigations. However, isotope shift measurements -- as the most reliable source of charge radii changes so far -- are challenging for the lightest elements because
the field shift contribution to the isotope shift is very small, typically a
$10^{-5}$ part of the total isotope shift as discussed above. In absolute numbers: The mass shift is on the order of
a few 10~GHz and, thus, an accuracy of the isotope shift measurement of the
order of 100~kHz is required to obtain the charge radius with an uncertainty
of about 1\%. A photon in the blue region is already Doppler-shifted by
about 100 kHz in the rest frame of an ion or atom moving at just 5~cm/s
and the absorption or emission of a photon leads to a recoil velocity of
about 20~m/s. Thus, it is obvious that the external degrees of freedom must
be very well under control during the experiment or the technique must be
insensitive to the thermal motion of an ion. For the lightest elements with
halo isotopes, He, Li, and Be, three dedicated approaches were
developed.
It should be noted that a similar accuracy as in the experiment is required in the theoretical calculation of the atomic mass shift contribution. For three-electron systems like Li and Be$^+$, this has been achieved for the first time in \cite{Yan2000}. Since then, the accuracy of these calculations has been improved by about a factor of 100 and is much more accurate than all experimental isotope shifts \cite{Yan2008,Puchalski2006,Puchalski2008,Noertershaeuser2011a}. This increase is partially due to improved mass measurements of halo isotopes.

Similar measurements were performed for many other halo nuclei and will be discussed in the following. They now provide sets of accurate numbers to obtain a conclusive picture of the structure of these nuclei. We will in the following concentrate on the techniques and the results of the measurements. A deeper theoretical discussion and analysis of the consequences for nuclear structure will be provided in the other contributions of this proceeding.

\subsubsection{Helium}

\begin{figure}[tbh]
\begin{center}
\includegraphics[width=12cm]{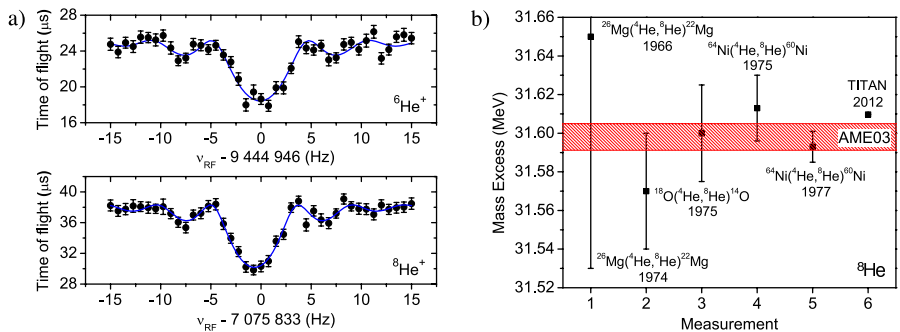}
\end{center}
\caption{(Colour online) Left: TOF spectra for $^6$He (top) and $^8$He (bottom) (taken from \cite{Brodeur2012}). Right: Mass exccess for $^{8}$He as measured with TITAN \cite{Ryjkov2008} compared to results from previous indirect mass determinations and to the value from AME 2003 \cite{AME2003} (figure modified from \cite{Ryjkov2008}).}
\label{helium_8}
\end{figure}

The isotopic chain of helium has 4 bound nuclei, the stable $^{3,4}$He, the two-neutron halo $^{6}$He, and the four-neutron halo $^{8}$He. Table \ref{Helium} shows the isotopes and the basic properties. The mass measurement requirements for the short-lived isotopes included fast preparation and measurements cycle while maintaining accuracy and precision on the sub-keV level. The measurements were carried out at the TITAN facility \cite{Dilling2006, Dilling2003} at the TRIUMF-ISAC radioactive beam facility in Vancouver, Canada. For helium, the isotopes were produced using the 500-MeV proton cyclotron. A silicon-carbide target coupled to a forced electron beam ion arc discharge  (FEBIAD) source \cite{Bricault2008} was bombarded with $70$\,$\mu$A of protons. The singly charged ions were extracted from the target, mass separated using the ISAC dipole isotopes separator with a typical resolution of $R \approx 3\,000$, and delivered as a continuous beam to the TITAN experiment.
The TITAN experiment (see figure \ref{titan}) consists of three ion traps in series, a linear Paul trap \cite{Brunner2012,Smith2006}, for cooling and bunching, an electron beam ion trap \cite{Lapierre2010}, for charge breeding (not used for these experiments), and a mass measurement Penning trap apparatus \cite{Brodeur2012_MPET, Brodeur2009}. For the mass measurements the TOF-ICR method is applied, and a fast excitation cycle in combination with the Lorentz-steerer for injection into the Penning trap and preparation of initial condition was employed. Figure \ref{helium_8} (right) shows the mass of the $^{8}$He isotopes as measured with TITAN \cite{Ryjkov2008} in comparison to previous indirect mass determinations. The resonance spectra are shown on the left. The uncertainty for $^{8}$He was improved by more than an order of magnitude and a final uncertainty of $\delta m = 110$~eV was reached \cite{Brodeur2012}. For $^{6}$He the new mass measurements led to a deviation from the AME2003 \cite{AME2003} of $4\sigma$ and an uncertainty of $\delta m = 54$~eV was achieved\cite{Brodeur2012}, corresponding to a 14-fold improvement over previous measurements.

\begin {table}[tp]%
\caption {Helium isotopes and their basic properties.}
\label{Helium}
\centering %
\begin {tabular}{|c|c|c|}
%\toprule %
\hline
isotope & $T_{1/2}$ & $\delta m$(AME2003)\\
\hline
$^{3}$He & stable & 0.0024 \, keV \\
$^{4}$He& stable & 0.00006 \, keV \\
$^{6}$He& 806.7 (1.5) \, ms  & 0.8 \, keV \\
$^{8}$He& 119.0 (1.5) \, ms  & 7.0 \, keV \\
\hline
\end {tabular}
\end {table}

\begin{figure}[tbh]
\begin{center}
\includegraphics[width=8cm]{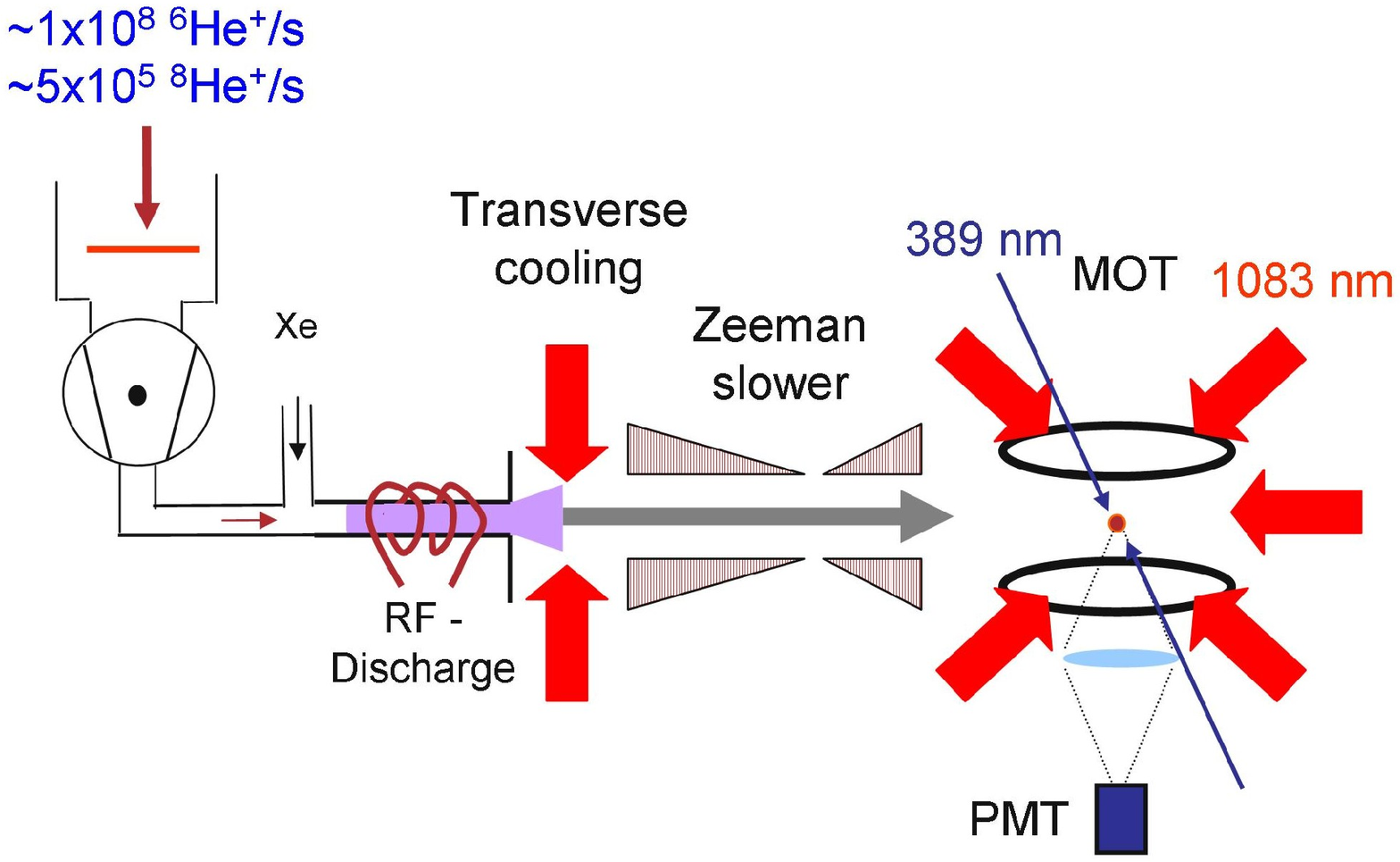}
\includegraphics[width=7cm]{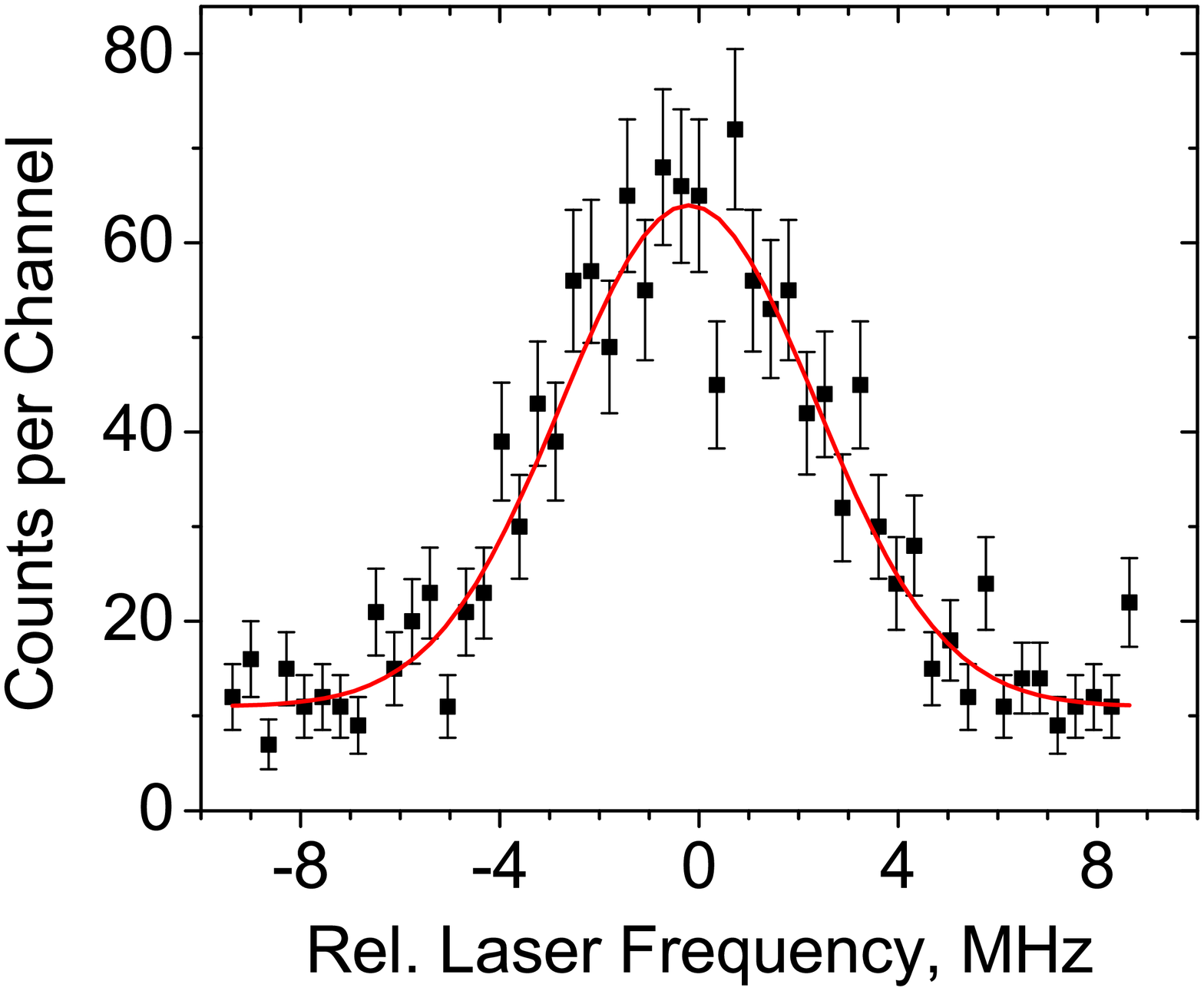}
\end{center}
\caption{(Colour online) Experimental setup for the isotope shift measurements of He isotopes (left) as explained in the text and a resonance signal of $^8$He (right) obtained within a measurement time of 2 hours and about 60 atoms of the 100-ms isotope $^8$He.}
\label{fig:HeSetup}
\end{figure}

The isotope shift of the helium isotopes was measured on trapped and cooled
helium atoms in a magneto-optical trap (MOT) \cite{Wang2004,Mueller2007}.
A schematic of the setup is shown in figure~\ref{fig:HeSetup}. The halo nuclei $^{6}$He and $^{8}$He were produced in the stripping
reaction $^{7}\mathrm{Li} + \, ^{\mathrm{12}}\mathrm{C} \longrightarrow
\,^{\mathrm{6}}\mathrm{He} + \, ^{13}\mathrm{N}$ at the Argonne Tandem
Linear Accelerator System (ATLAS) and $^{13}\mathrm{C}+\,^{12}\mathrm{C} \longrightarrow \,^{6}\mathrm{He,} \,^{8}\mathrm{He} + \mathrm{X}$ at GANIL. Laser Spectroscopy of helium atoms is
only possible from an excited metastable atomic state since excitation from
the ground state would require laser light at 59~nm. Thus, the ortho-helium $%
1s2s\,^{3}\mathrm{S}_{1}$\ metastable state has been populated in a gas
discharge cell (efficiency $\approx 10^{-7}$) and laser cooling and spectroscopy was performed on the $%
1s2s\,^{3}\mathrm{S}_{1}\rightarrow 1s2p\,^{3}\mathrm{P}_{2}$\ (1083~nm) and
$1s2s\,^{3}\mathrm{S}_{1}\rightarrow 1s3p\,^{3}\mathrm{P}_{J}$\
(389~nm) transitions, respectively. A simplified experimental setup is shown in figure~\ref{fig:HeSetup}.
Excited helium atoms leaving the RF discharge cell are first transversely
cooled before they enter a Zeeman slower in which they are slowed down by
repetitive absorption of counter propagating photons until they are
sufficiently slow to be captured in the shallow potential of a
magneto-optical trap located at the end of the Zeeman slower. Even though
the total efficiency of the apparatus is only on the order of $10^{-7}$, it
was still possible to perform laser spectroscopy with production rates as low as $10^{5}$ $^{8}$He atoms/s. The capture rate of only 30 atoms per hour is compensated
by the large amount of photons that can be scattered by a single atom even
within the short half-life of $^{8}$He (119~ms). The atom trap is operated
in the capture modus, i.e., only the cooling lasers are on until an atom is
trapped. Such an event is observed by a significant increase of scattered
photons on the photodetector. Then, the MOT\ operation is switched to the
spectroscopy modus in which cooling and spectroscopy is alternately
performed: The spectroscopy laser is switched on for typically 2~$%
\mu $s and this period is followed by 8~$\mu $s of laser cooling. The laser
beam intensities have to be carefully balanced in order to reduce any
systematic shift as much as possible and the spectroscopy was performed on
all three fine-structure transitions in $^{6}$He and on the transitions to
the $^{3}\mathrm{P}_{J=1,2}$ states in $^{8}$He. The results for
the field shift are consistent within their uncertainties.

\begin{figure}[tbh]
\begin{center}
\includegraphics[width=9cm]{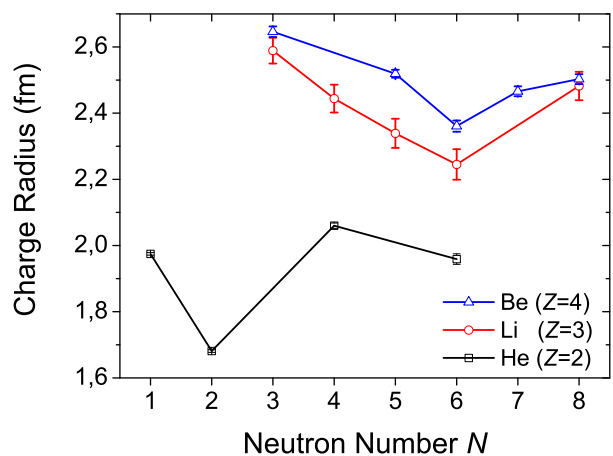}
\end{center}
\caption{(Colour online) Nuclear charge radii of helium, lithium and beryllium isotopes as a function of neutron number as obtained from isotope shift measurements. The error bars are dominated by the uncertainty of the reference radius determined by elastic electron scattering ($^4$He, $^6$Li and $^9$Be). In all cases a considerable increase in charge radius is observed for the neutron-halo isotopes. This is to a large part driven by the center-of-mass motion of the core nucleus.}
\label{fig:HaloChargeRadii}
\end{figure}

The results of the isotope shift measurements presented in \cite%
{Wang2004,Mueller2007} were reevaluated in \cite{Brodeur2012} based on the new
mass measurements of $^{6,8}$He at TITAN. Charge radii are shown in figure~%
\ref{fig:HaloChargeRadii}. The charge radius increases from $^{4}$He to $%
^{6}$He but decreases then again towards $^{8}$He. The Gamov Shell Model
has recently been used to analyze the differences in charge radii \cite%
{Papadimitriou2011}. It was concluded that the change in the charge radius
between $^{4}$He and $^{6,8}$He is dominated by the center-of-mass motion of
the $\alpha $-core, Smaller, but not negligible, are contributions from the
spin-orbit term of the halo neutrons and a swelling of the $^{4}$He core due
to core polarization. The estimated size of the latter effect, taken from
Green's function Monte-Carlo calculations, amounts to about 5\% of the $%
^{4}$He point-proton size \cite{Pieper2001}. The decrease from $^{6}$He to $%
^{8}$He is accordingly caused by the difference in the recoil term and the
spin-orbit contribution, which is about twice as large and negative for $^{8}$He compared to $^{6}$He.

\subsubsection{Lithium}
\begin{figure}[bth]
\begin{center}
\includegraphics[width=7cm]{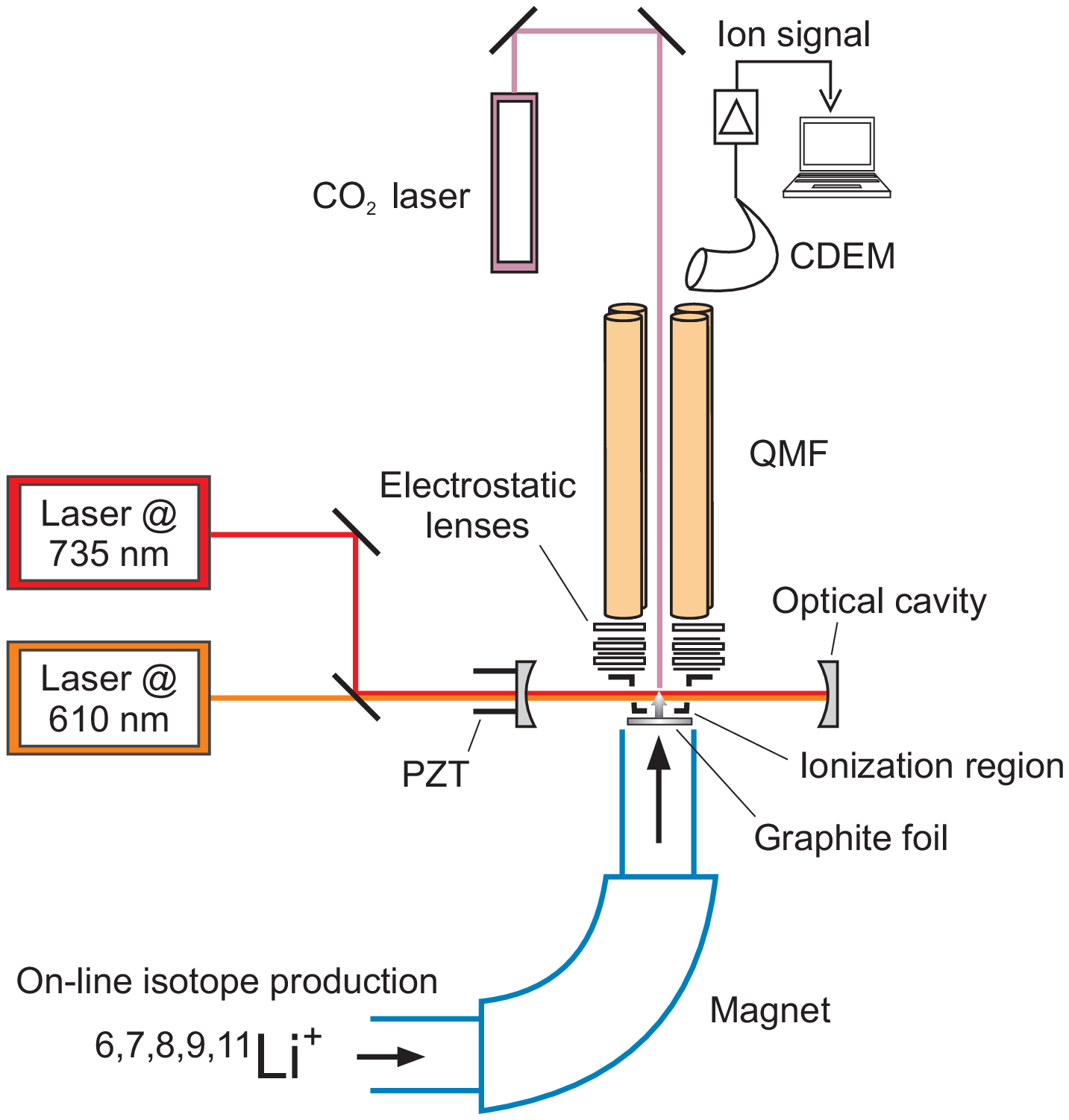}
\hspace{0.5cm}
\includegraphics[width=7cm]{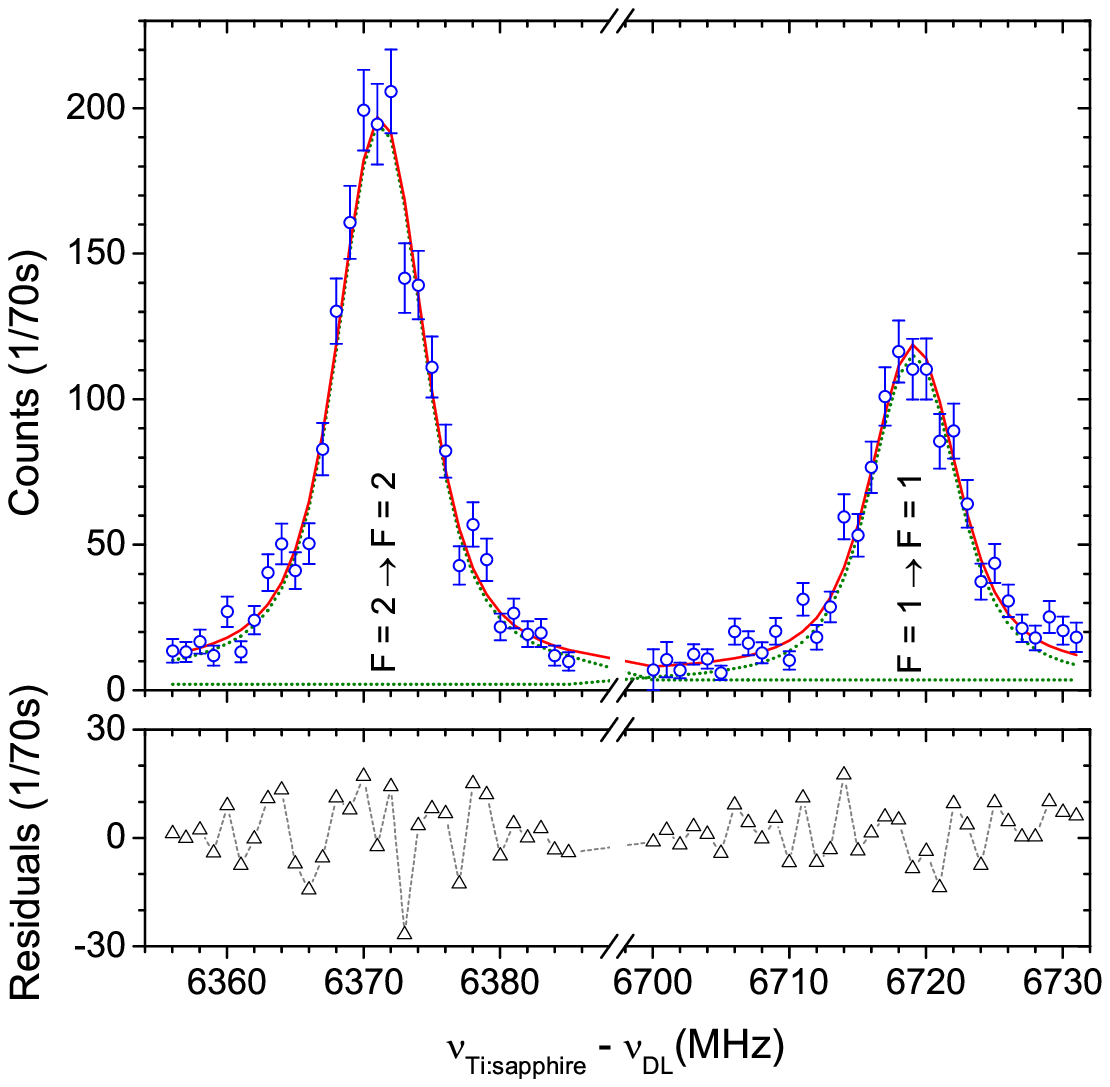}
\end{center}
\caption{(Colour online) Simplified experimental setup for the lithium spectroscopy (left) and a typical spectrum of $^{11}$Li (right) as a function of the two-photon transition frequency. Two hyperfine lines are observed due to the $\Delta F=0$ selection rule for such this transition. Voigt profiles are fitted to the experimental data points and the residuals of the fit are shown below the spectrum. (Reprinted with permission from Phys.\ Rev.\ A \textbf{83} 012516 Copyright 2011 American Physical Society)}
\label{fig:LiSetupSpectrum}
\end{figure}

In the case of the lithium isotopes, spectroscopy was performed using
Doppler-free two-photon excitation on neutral lithium atoms, followed by
resonance ionization and mass separation \cite%
{Noertershaeuser2011a,Ewald2004,Ewald2005,Sanchez2006}. The technique was developed
 to achieve an accuracy of about 100~kHz in the isotope shift for the low production rate
 ($\approx 10\,000$~ions/sec) and the short lifetime ($t_{1/2} = 8.75\,(14)$~ms) of the isotope $^{11}$Li.
 As it is shown on the left in figure~\ref{fig:LiSetupSpectrum}, the beam of lithium ions,
 produced either at the GSI on-line mass separator beamline in Darmstadt ($^{6-9}$Li)
 or at the ISAC  mass separator facility at TRIUMF ($^{6-11}$Li), is first stopped in a
 thin graphite foil (about 300~nm thickness, $\approx 80~\mu$g/cm$^{2}$). The foil is
 heated by a CO$_{2}$ laser to about 2000$^{\circ}$C, such that the implanted
 and neutralized atoms quickly diffuse out of the
foil and drift into the laser interaction region which is placed a few mm in
front of the foil inside the ionization region of a quadrupole mass
spectrometer.
The neutral atoms are first excited by two photons at 735~nm
from the $2s\,^{2}\mathrm{S}_{1/2}$ ground state into the $3s\,^{2}\mathrm{S}%
_{1/2}$ level. In this transition the atom absorbs two photons
simultaneously. In order to obtain the two counterpropagating beams and the
relatively high intensity that is required
for this type of non-linear spectroscopy, the 735-nm laser-beam intensity is enhanced in a
resonator by reflecting the photons back and forth between two highly
reflective mirrors. The (non-relativistic) Doppler shift
\begin{equation}
\Delta \nu _{\mathrm{D}} = \nu _{\mathrm{L}} \left( 1 + \frac{\upsilon _{\Vert }}{c} \right)
\end{equation}%
of the laser frequency $\nu _{\mathrm{L}}$ in the rest frame of an atom with
velocity component $\upsilon _{\Vert }$ along the wave vector $\vec{k}$ of
the laser beam has equal amplitudes but opposite signs for two
counterpropagating beams with identical frequency. The sum frequency of two
photons is then independent of the atom's velocity if one photon is taken
from each beam. This results in a (first-order) Doppler-free signal to which
all atoms of the thermal ensemble contribute. After excitation to the
$3s\,^{2}\mathrm{S}_{1/2}$ level
the atoms will relax via the $2p$ level into the $2s$ ground-state. A
second laser is used to transfer the population of the $2p$ level resonantly
into the $3d$ state out of which the atoms can undergo non-resonant
ionization by absorbing another photon either from the 735-nm or the 610-nm
beam. The power of the 610-nm laser beam is also enhanced in the same cavity as the
735-nm laser beam. This excitation scheme leads to very selective and
efficient ionization only of those atoms which have previously been excited
in the $2s\rightarrow 3s$ transition. The ionized atoms are extracted
from the ionization volume with the ion optics of the quadrupole mass
spectrometer and mass separated before being detected with a continuous
dynode electron multiplier (CDEM). A typical spectrum of $^{11}$Li is shown
on the right in figure~\ref{fig:LiSetupSpectrum}. It shows the two hyperfine structure
components from which the center of gravity (cg) can be obtained. Hyperfine-induced fine-structure mixing could in principle affect
the center of gravity but is expected to be far smaller than the accuracy of
these measurements.
The charge radii within the lithium chain, as shown in figure~\ref%
{fig:HaloChargeRadii}, exhibit a decrease of the radius from $^{6}$Li to $%
^{9}$Li, followed by a significant increase to $^{11}$Li. This behavior is
caused by the strong clusterization of the light nuclei \cite{Noertershaeuser2011b}. The increasing
charge radius for the halo isotopes is again dominated by the center-of-mass
motion of the core nucleus, caused by the halo neutrons. Core polarization
can also contribute to the larger charge radius, but its contribution is
expected to be small. The amount of core polarization cannot be extracted
from laser spectroscopic investigations alone, but by combining measurements
of matter radii, charge radii and the low-lying dipole strength $B(\mathrm{E1%
})$ distribution as it is discussed e.g. in \cite{Noertershaeuser2011b,Esbensen2007}. However, the quantification is hampered
by other contributions like, e.g., the spin-orbit contribution that must be
determined theoretically.

Due to the short half-life of $^{11}$Li, a mass measurement in a Penning trap is also extremely challenging. It was also carried out at TITAN, using a measurement cycle of 50~Hz, allowing a mass measurement precision of $\delta m = 690$~eV \cite{Smith2008}. It improved the uncertainty $\delta m (\mathrm{AME2003}) =19$~keV of the AME 2003 \cite{AME2003} by almost a factor 30. The measurement practically eliminated the uncertainty arising from the mass for the derivation of the charge radius \cite{Noertershaeuser2011b}. $^{11}$Li is now the shortest-lived isotope for which a mass measurement in a Penning trap was performed, with almost an order of magnitude shorter half-life than the previous record, set by ISOLTRAP for $^{74}$Rb \cite{Kellerbauer2004}. Measurements of the lighter lithium isotopes were also carried out in order to extract the two-neutron separation energies along the lithium chain and to verify the masses used for the mass-shift calculations required to determine their charge radii \cite{Smith2008, Brodeur2009, Brodeur2012_Li}.

\subsubsection{Beryllium}
\begin{figure}[tbh]
\begin{center}
\includegraphics[width=6cm]{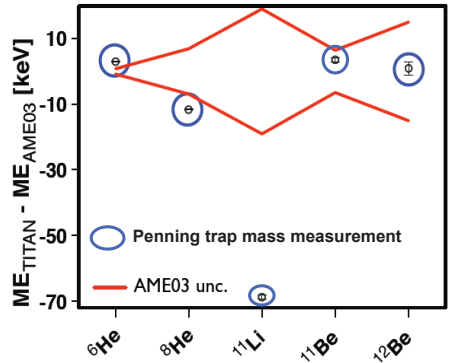}
\end{center}
\caption{(Colour online) Mass measurements of the neutron halo nuclei using the TITAN Penning trap. The error band shows the mass uncertainties as given in the AME 2003 \cite{AME2003}. The significant improvement in the precision is clearly visible. The masses of $^{6,8}$He deviate clearly from the previous values, the discrepancy for $^{11}$Li is huge.}
\label{halo_harvest}
\end{figure}

Beryllium mass measurements were carried out for $^{9-12}$Be \cite{Ringle2009, Ettenauer2010}. The results for the short-lived isotopes are included in figure \ref{halo_harvest}. The reached precision was sufficient for the mass-shift calculations to determine the charge radius from isotope shift measurements. The mass measurement of the two-neutron halo candidate $^{14}$Be remains a challenge for the future. With a half-life of  $t_{1/2} = 4.35\,(17)$~ms it would establish a new Penning-trap mass measurement record.

Laser spectroscopy on beryllium had to be performed on the three-electron system Be%
$^{+}$ because sufficiently accurate calculations of the mass shift term are
still not available for four-electron systems. Singly
charged beryllium ions offer an alkaline excitation spectrum with the
resonance transition being in the near ultraviolett at 313~nm. Due to the strong
binding, the valence electrons wavefunction has a relatively high probability
density at the nuclear site, providing a
sensitivity to changes in the mean-square charge radius that is
approximately 10 times larger than in the case of helium or lithium \cite{Yan2008,Puchalski2008}.
Consequently, an accuracy of about 1~MHz in the determination of the isotope
shift is sufficient to extract the nuclear charge radius with a relative precision of
about 1\%. This was achieved by performing collinear laser spectroscopy
quasi-simultaneously in collinear and anticollinear geometry \cite%
{Noertershaeuser2009, Krieger2012}. In the usual approach of collinear laser
spectroscopy -- performing fluorescence spectroscopy just collinearly or
anticollinearly -- the available relative precision of 10$^{-4}$ in the voltage
determination of the 50~kV acceleration potential, leads to an uncertainty
of about 40~MHz for the isotope shift between $^{12}$Be and the stable
reference isotope $^{9}$Be. This is disastrous for a determination
at the level of accuracy that is required to extract the small field shift.
To become independent from the exact knowledge of the ion velocity,
the absolute transition frequencies required for excitation in collinear
and anticollinear geometry $\nu _{c,a}$ according to equation (\ref{eq:relDoppler})
are determined simultaneously. Their product can then be easily evaluated
\begin{equation}
\nu _{\mathrm{a}}\,\nu _{\mathrm{c}}=\nu _{0}^{2}\,\gamma ^{2}\,(1+\beta
)\,(1-\beta )=\nu _{0}^{2}  \label{eq:FreqProd}
\end{equation}%
and provides the rest-frame transition frequency $\nu _{0}$ independently of
the acceleration voltage. This collinear-anticollinear approach
was previously used for precision tests of QED
on helium-like ions of stable isotopes from helium to fluorine
\cite{Riis1994,Thompson1998,Myers1999} but was never applied to radioactive isotopes.
The prize one has to pay for this advantage is
the determination of the absolute laser frequencies with accuracy better
than 10$^{-9}$, whereas
standard collinear laser spectroscopy requires only that the laser frequency
is sufficiently stable and known on a 10$^{-5}$ level. The new approach was
therefore strongly facilitated by the invention of the frequency comb \cite%
{Udem1999}, a device that allows to directly link radiofrequencies with
optical frequencies and has been demonstrated to be able to determine
transition frequencies with accuracy at the 10$^{-17}$ level \cite%
{Rosenband2008}.
\begin{figure}[bth]
\begin{center}
\includegraphics[width=12cm]{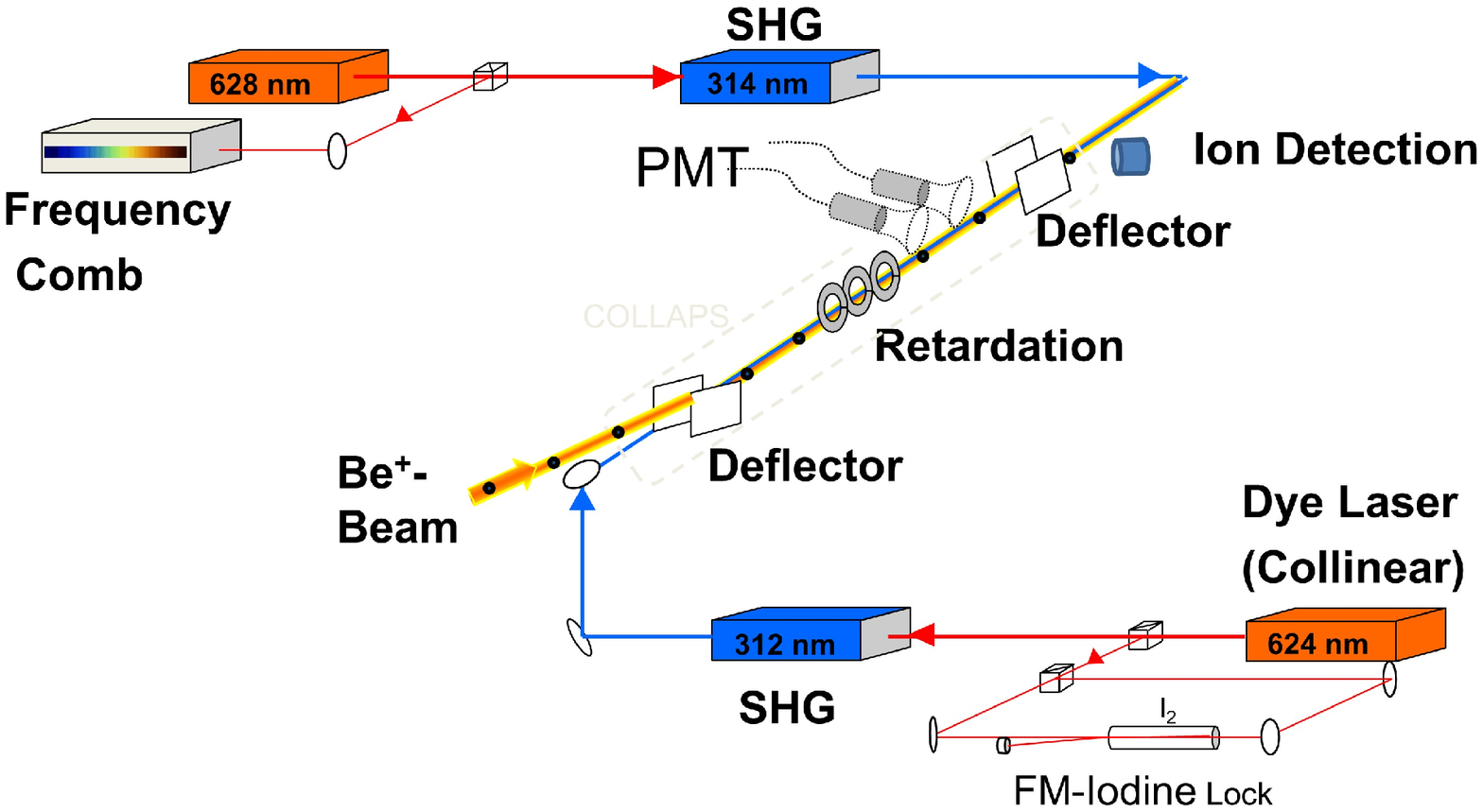}
\includegraphics[width=12cm]{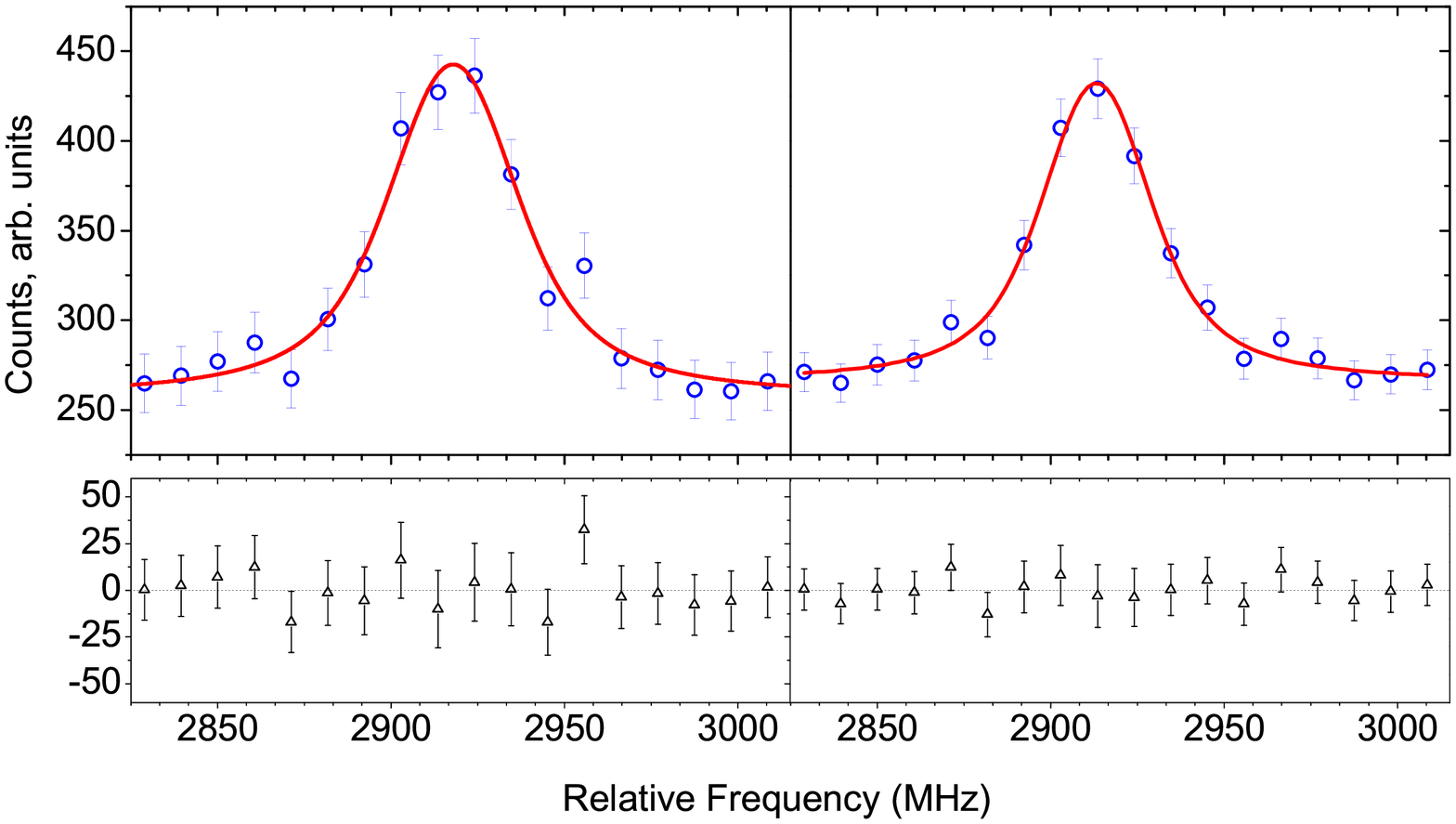}
\end{center}
\caption{(Colour online) Top: Simplified experimental setup (top) for the quasi-simultaneous collinear and anticollinear spectroscopy of beryllium ions at ISOLDE. Two frequency-doubled laser systems providing absolute frequency information are used, one in collinear and one in anticollinear geometry. Bottom: A resonance of the most exotic isotope in this chain $^{12}$Be obtained with a yield of about 8,000~ions/s.}
\label{fig:BeSetupSpectrum}
\end{figure}

At ISOLDE\ the approach was realized using two dye lasers at 628~nm and
624~nm, respectively \cite{Noertershaeuser2009}. One laser was locked to
well-known reference transitions in molecular iodine and the other one
directly to a frequency comb. Both laser beams were frequency doubled and
sent into the COLLAPS laser spectroscopy beamline in collinear (312~nm) and
anticollinear (314~nm) configuration. Using fast shutters, spectra in both
directions were taken alternately and integrated until a sufficiently large
signal-to-noise ratio was observed. A typical spectrum of $^{12}$Be, the
isotope with the shortest half-life (20~ms) and production yield
($\approx 8\;000$~ions/s) is shown in figure~\ref{fig:BeSetupSpectrum}.
For this isotope it was necessary to implement additionally an ion-photon
coincidence detection in order to reduce the background from
scattered laser light \cite{Krieger2012}. The data points are the
experimentally observed count rates normalized by the number of ions as a
function of the Doppler-tuned laser frequency in the rest-frame of the ion.
The solid line is a fit to the spectrum using a Voigt profile.
Beryllium isotopes with an odd mass number exhibit hyperfine structure
that was analyzed by fitting the complete spectra with similar profiles for
each peak and adjusting the hyperfine parameters, the center-of-gravity,
peak intensities, and an offset as well as the Doppler width for all Voigt
profiles. In these cases it was required to include a small secondary peak
that arises 4~V below the main peak. This is caused by non-elastic collisions
of the ions in which they are excited into the $2p$ excited state during the
encounter of the residual gas atom. The energy for the transition is taken
from the ions kinetic energy and subsequently irradiated in the form of
a 4-eV photon.

The charge radii obtained in these experiments are included in figure~\ref%
{fig:HaloChargeRadii}. The trend is quite similar to that observed for the
lithium isotopes and the underlying physics as well. The cluster basis in
this chain is $\alpha +\,^{3}\mathrm{He}$ for $^{7}$Be and $\alpha +\alpha
+xn$ for $^{9-12}$Be. The center-of-mass motion in the one-neutron halo
isotope $^{11}$Be is again the dominant contribution to the increase of the
charge radius. Since $^{12}$Be is not believed to be a halo isotope and has
a significantly smaller matter radius than $^{11}$Be it is (at first glance)
surprising that the measured charge radius of $^{12}$Be is larger than that
of $^{11}$Be. The observation can be explained theoretically by a strong
admixture of $(sd)^{2}$ to the ground-state wavefunction of $^{12}$Be \cite%
{Krieger2012}. It has been shown in Fermionic Molecular Dynamics calculations
that this actually increases the $\alpha$\,-$\alpha$  distance in the
underlying $^{10}$Be cluster structure, such that the $^{12}$Be charge
radius is again comparable to that of $^{9}$Be.

\subsubsection{Neon}
The proton-rich isotopes of the neon isotopic chain presented a difficult challenge for mass measurements, and even though laser spectroscopy had been done at ISOLDE in the late 1990s, the charge radius extraction was hindered by a mass shift determination. This was recently accomplished using results from high precision experimental mass measurements performed at ISOLTRAP \cite{Geithner2008}.

\begin{figure}[tbh]
\begin{center}
\includegraphics[width=10cm]{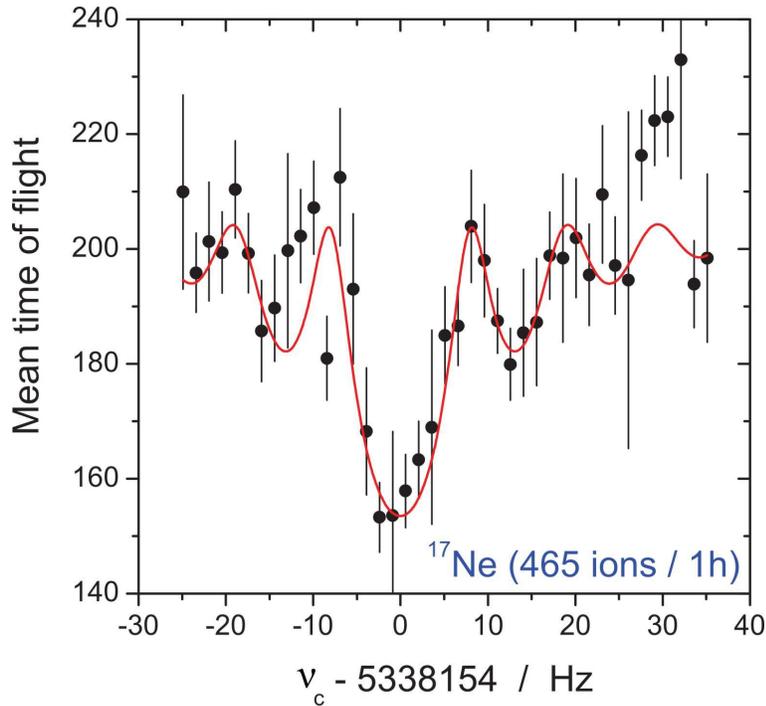}
\end{center}
\caption{(Colour online) TOF resonance spectrum of $^{17}$Ne taken with the ISOLTRAP Penning-trap mass spectrometer.}
\label{Neon}
\end{figure}

The mass measurements were carried out using the ISOLDE facility for the production of the neon isotopes, where 1.4-GeV protons impinge on CaO or MgO targets to produce the isotopes. A plasma source was used to ionize the isotopes, which were then mass selected and delivered as a pulsed
60-keV beam to the ISOLTRAP setup \cite{Mukherjee2008}. ISOLTRAP consists of 4 main devices in series, a linear RFQ cooler and buncher \cite{Herfurth2001}, a recently added multi-reflection time-of-flight separator \cite{Wolf2012}, a buffer-gas filled isobar separator trap, and a mass measurement Penning trap. The measurement cycle was adjusted for this measurement such that the short half-life of $^{17}$Ne with $t_{1/2} = 109.2(6)$\,ms could be accommodated within a $400$\,ms measurement cycle from proton pulse to ion detection. The uncertainty was reduced by a factor 50 compared to AME 2003 and reached a level of $\delta m = 570$~eV for $^{17}$Ne. Figure \ref{Neon} shows a resonance spectrum of the measurement \cite{Geithner2008} from ISOLTRAP. These mass measurements provided the required precision for the extraction of the charge radius.

\begin{figure}[bth]
\begin{center}
\includegraphics[clip,width=15cm]{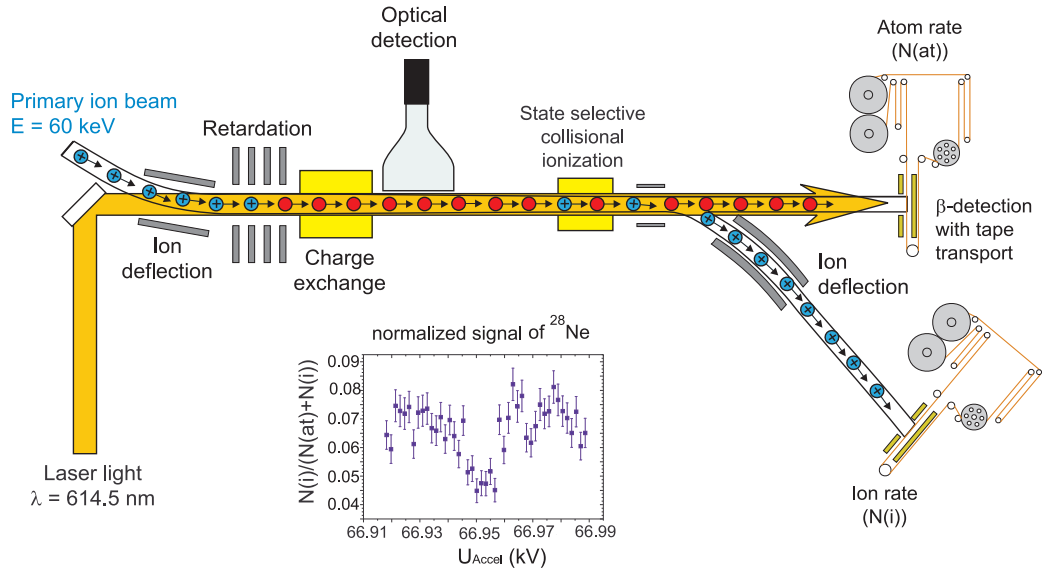}
\end{center}
\caption{(Colour online) Simplified experimental setup for the neon spectroscopy with collinear laser spectroscopy and a spectrum of $^{28}$Ne. Ne$^+$ ions provided from the plasma ion source at ISOLDE were first neutralized into a metastable state after superimposing the ion beam with a laser beam. On resonance the metastable state is depopulated by pumping it into the ground state and the decreased population is probed by state selective ionization in a second gas cell. Recording the ratio of ions and atoms as a function of Doppler-tuning voltage results in spectra like the one shown on the bottom for $^{28}$Ne. This spectrum was taken with a yield of only 50 atoms per proton pulse at ISOLDE, i.e., about 25 atoms/s. (Reprinted with permission from Phys.\ Rev.\ C \textbf{84} 034313 Copyright 2011 American Physical Society)}
\label{fig:SetupNeon}
\end{figure}

The isotope $^{17}$Ne is the only \textit{proton-halo} candidate that has so far been
studied using laser spectroscopy. Collinear laser spectroscopy was also
applied in this case, and state-selective ionization was used for efficient
detection of the induced resonant transition \cite{Geithner2008,Geithner2005}.
The neon ions produced at ISOLDE were ionized in a plasma ion
source and sent to the COLLAPS apparatus. Here, resonant charge exchange
into the $[2p^{5}(^{2}\mathrm{P}_{3/2}^{\mathrm{o}})3s]_{2}$ metastable
level was induced by passing the ion beam through the charge exchange cell
operated with sodium. Subsequent laser excitation into the metastable $%
[2p^{5}(^{2}\mathrm{P}_{3/2}^{\mathrm{o}})3p]_{2}$ state leads to the
decay into the $[2p^{5}(^{2}\mathrm{P}_{3/2}^{\mathrm{o}})3s]_{1}$ level
which is quickly relaxing into the $2s^{2}\,2p^{6}$ ground state of the neon
atom. On resonance, the populated metastable level will be quickly
depopulated. Electrons bound in the $2s^{2}\,2p^{6}$ ground state have
ionization energies of about 20\,eV and are much less susceptible to
collisional ionization than the metastable state with a binding energy of
only about 5\,eV. Hence, the ratio of neutralized to ionized atoms after
transmission through a second gas cell is used to detect the transition as
it is shown in figure~\ref{fig:SetupNeon}. This detection technique was also
applied for the isotopes of other noble gases (Ar, Kr, Xe, Rn) \cite%
{Klein1996,Blaum2008,Keim1995,Borchers1989,Neugart1988}. A speciality
is again the voltage calibration for the neon measurements.
Since the ionization occurs in a plasma ion source, the starting potential
is not as well defined as for a
surface ion source. Therefore a peculiarity of the neon excitation scheme
was exploited: The transition frequencies for the
%\begin{eqnarray*}
$\lbrack 2p^{5}(^{2}\mathrm{P}_{3/2}^{\mathrm{o}})3s]_{2} \rightarrow
[2p^{5}(^{2}\mathrm{P}_{1/2}^{\mathrm{o}})3p]_{2}$ and the
$\lbrack 2p^{5}(^{2}\mathrm{P}_{3/2}^{\mathrm{o}})3s]_{2} \rightarrow
[2p^{5}(^{2}\mathrm{P}_{1/2}^{\mathrm{o}})3p]_{1}$
%\end{eqnarray*}%
transitions conincide in the laboratory frame if the first transition is
excited anticollinearly and the second one collinearly at a beam energy of
61,758.77~eV. Hence, a single laser beam can be used to excite both
transitions if it is retroreflected at the end of the beamline. This
approach was used in the neon measurements for the beam energy
determination \cite{Marinova2011}.

The charge radii obtained from that analysis are plotted
together with the neighboring isotopes of Na and Mg in figure~\ref%
{fig:Rc_Ne-Mg}. The FMD approach has been used to describe the neon and the
magnesium isotopes and the results reproduce the trend of the charge radii
very well. For neon, especially the strong increase in charge radius from $%
^{18}$Ne to $^{17}$Ne is remarkable. This reflects a structural change in
the isotopic chain and is correctly reproduced in the FMD calculation from
which an amount of approximately 40\% $s^{2}$ contribution is extracted.
This clearly indicates at least the onset of a proton halo \cite{Geithner2008}.
\begin{figure}[bth]
\begin{center}
\includegraphics[width=15cm]{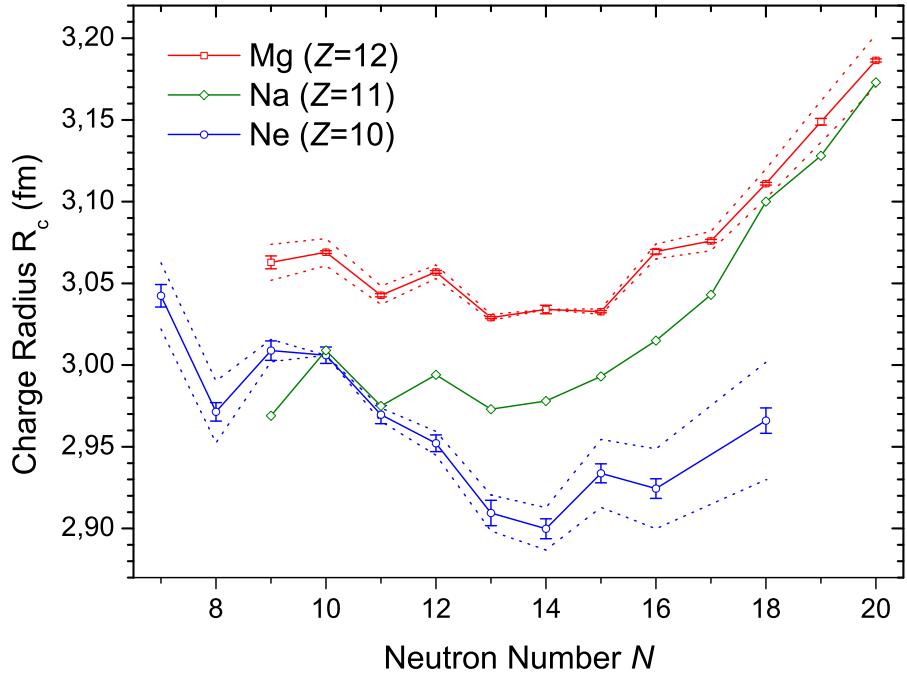}
\end{center}
\caption{(Colour online) Nuclear charge radii of neon, sodium and magnesium isotopes (details see text).}
\label{fig:Rc_Ne-Mg}
\end{figure}

\subsubsection{Nuclear Moments of Light Nuclei}

After the discovery of the halo isotopes, the question arose whether the
large radii are due to a strong nuclear deformation or the long tails of a
halo state. Particularly the nuclear moments of $^{11}$Li were considered to
be a key property of this nucleus. While magnetic moments and quadrupole
moments can usually be extracted from optical spectra as discussed above,
the quadrupole interaction in very light elements can hardly be resolved.
Furthermore, nuclear magnetic resonance (NMR) is often able to provide
more accurate values. NMR requires a beam of
spin-oriented (polarized) nuclei that can be produced by optical pumping
with $\sigma ^{\pm }$-light in collinear geometry. The NMR\ technique can
then be applied for radioactive beams if the isotopes are afterwards stopped
inside a suitable crystal and the $\beta $-asymmetry is detected \cite%
{Neugart2006}. Since detection of charged particles is much more efficient
than photon detection, the $\beta $-asymmetry detection can be applied together
with CLS to enhance the sensitivity. Hence, it seems
only natural to combine these two techniques. In addition, the combined
measurement of the hyperfine structure from which $\mu $ can be extracted
and of the $g$-factor allows one to\ directly measure the nuclear spin as it
will be discussed below for the case of Mg. Another technique with high
sensitivity and accuracy is the RF-optical double resonance. This method has
been applied to the radioactive beryllium isotopes.

\paragraph{Nuclear Magnetic Resonance with $\protect\beta $-asymmetry
detection ($\protect\beta $-NMR)}

\begin{figure}[bth]
\begin{center}
\includegraphics[clip,width=15cm]{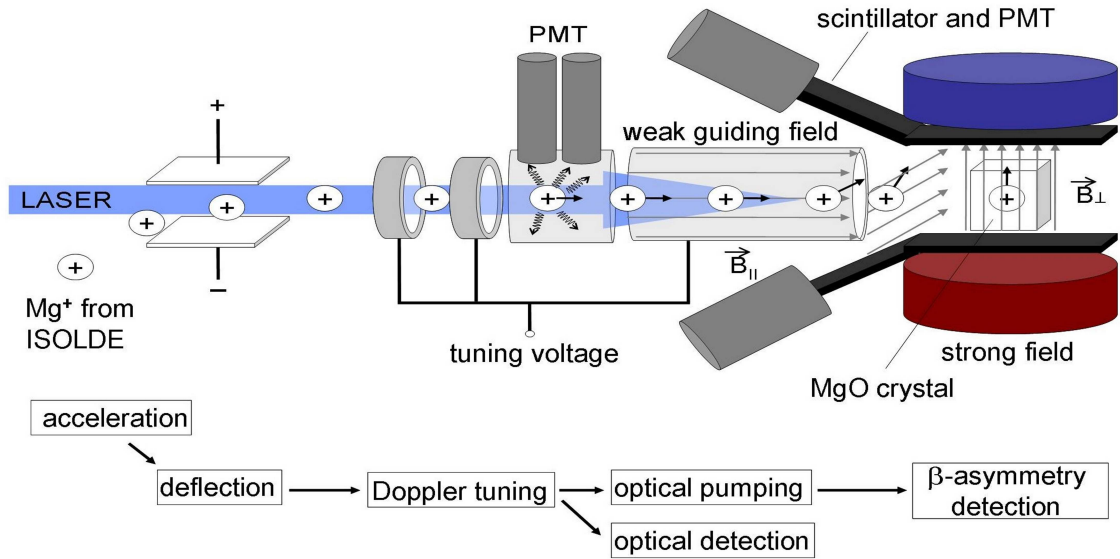}
\end{center}
\caption{(Colour online) Collinear laser spectroscopy in combination with optical pumping and $\beta$-NMR. The laser-ion interaction polarizes the ions on resonance and leads to a $\beta$-asymmetry that is detected after implantation into a crystal inside a strong magnetic field. RF-coils around the crystal (not shown in this figure) can then be used to destroy the orientation produced by the laser beam and to obtain a nuclear magnetic resonance (NMR) signal.}
\label{fig:CLS_NMR}
\end{figure}

At ISOLDE, a first measurement of the nuclear magnetic dipole moment and the
spin of $^{11}$Li was performed already in 1987 \cite{Arnold1987}, followed by a
measurement of the electric quadrupole moment in 1992 \cite{Arnold1992}.
%Optical pumping in collinear geometry followed by $\beta $-asymmetry
%detection and combined with $\beta $-NMR is a very sensitive technique for
%accurate determinations of nuclear moments especially of light elements \cite{Neugart2006}.
Later, the nuclear magnetic moment of $^{11}$Be has also been measured
with this technique \cite{Geithner1999}.
The experimental setup for these measurements is presented in figure~\ref{fig:CLS_NMR}: The
laser light is circularly polarized ($\sigma ^{+}$, $\sigma ^{-}$) and is
used to optically pump the atoms into the $m_{F}$-state state with the
maximum (absolute) value of $m_{F}$. Then both, the electronic and the
nuclear moments are maximally aligned along the quantization axis provided
by the direction of a weak magnetic guiding field
produced by a set of coils along the beamline. In order to use $\beta $%
-asymmetry detection, the optically pumped and oriented ions or atoms must
be implanted into a crystal inside a strong magnetic field. Then, the $\beta
$-asymmetry is measured as the difference of $\beta $-particles emitted
parallel and antiparallel to the magnetic field normalized to their total
number%
\begin{equation}
a=\frac{N^{\uparrow }-N^{\downarrow }}{N^{\uparrow }+N^{\downarrow }}.
\end{equation}%
Recording $a$ as a function of laser frequency (or Doppler tuning voltage)
allows to record the hyperfine structure of the atom with higher sensitivity
than with pure optical detection. The total size of the asymmetry $a$
depends on the nuclear transition but also on the host crystal in which the
ion is implanted, its structure and its purity, and often this has to be
clarified in  systematic pilot investigations.

Once the laser resonance has been found that provides the maximum asymmetry,
NMR\ can be used to measure the nuclear $g$ factor or the quadrupole moment
with high accuracy. For a measurement of the $g$ factor a crystal with a
cubic lattice structure is chosen since there is no electric field gradient
at the regular lattice positions. Consequently, the measurement becomes
independent from a possibly existing quadrupole moment and the magnetic
substates $m_{I}$ of the nucleus remain degenerate. If an external magnetic
field is applied, this degeneracy is lifted and the additional energy of the
substates is proportional to $m_{I}\mu _{I}B.$ The laser-induced
polarization -- i.e.\ the predominant population of the $m_{I}=\pm I$
substates for $\sigma ^{\pm }$ pumping -- can then be destroyed by inducing
transitions between neighboring magnetic substates with the Larmor frequency
\begin{equation}
\omega _{L}=\frac{g_{I}\mu _{N}I}{\hbar }B_{0}  \label{eq:LarmorFrequency}
\end{equation}%
which is directly connected to the nuclear $g$-factor provided that the
magnetic field is known. The latter is usually calibrated using a second
species with a well-known nuclear magnetic moment like, e.g., $^{8}$Li.

Contrary, if the quadrupole moment is the property of interest, a crystal
with a non-cubic lattice structure has to be uesd, where an electric field
gradient is present at the regular lattice position. If an implanted ion
inhabits such a regular position, the quadrupole interaction leads to a
non-equal splitting between the different magnetic substates. In the
NMR spectrum, the corresponding transitions occur at equidistant
radiofrequencies. In the easiest case (axially symmetric electric
field gradient along the $z$-axis of the principle axis system),
the frequency difference $\Delta \nu$ between neighboring lines is
directly related to the quadrupole interaction constant and given by %
\begin{equation}
\Delta \nu_{Q} = \nu(m\leftrightarrow m-1)-\nu(m-1 \leftrightarrow m-2) = \frac{3eQ_{s}V_{zz}/h}{2I(2I-1)}.
\label{eq:QuadrupoleSplitting}
\end{equation}

Already the early results for the spin and the magnetic moment of $^{11}$Li
supported the halo picture and excluded a strong deformation:
The spin was detemined to be $I=3/2$ and the
magnetic moment of $\mu =3.6673(23)$~$\mu _{N}$ \cite{Arnold1987} to be very
close to the Schmidt-value for a spherical $1p_{3/2}$ proton. Finally, the
result of the measurements of the spectroscopic quadrupole moments of $^{9}$%
Li and $^{11}$Li $\left\vert Q_{s}(^{11}\mathrm{Li})/Q_{s}(^{9}\mathrm{Li}%
)\right\vert =1.14(16)$ excluded again a large deformation as the source of
the large matter radius and also implied that the $^{9}$Li core is
essentially unchanged by the presence of the halo neutrons \cite{Arnold1992}%
. The accuracy of the $^{11}$Li moments has been increased by an order of
magnitude since then, resulting in $\left\vert Q_{s}(^{11}\mathrm{Li})/Q_{s}(^{9}\mathrm{Li}%
)\right\vert =1.088(15)$. The currently most accurate values for the nuclear moments of $^{11}$Li are $\mu =3.6712(3)$%
~$\mu _{N}$ and $Q_{s}(^{11}\mathrm{Li})=-33.3(5)$~mb compared to $Q_{s}(^{9}%
\mathrm{Li})=-30.6(2)$~mb \cite{Borremans2005,Neugart2008}. An even more
accurate measurement of the quadrupole moment of $^{9}$Li relative to $^{8}$%
Li has been reported from TRIUMF \cite{Voss2011}: using $\beta $-detected
nuclear quadrupole resonance (NQR) it was measured to be $\left\vert
Q_{s}(^{9}\mathrm{Li})/Q_{s}(^{8}\mathrm{Li})\right\vert =0.96675(9)$. The
ion beam of $^{9}$Li was neutralized in a charge exchange cell, optically
pumped, reionized in a helium gas jet and implemented in a SrTiO$_{3}$
crystal. Contrary to the measurements at ISOLDE, the magnetic field at the
implantation site in SrTiO$_{3}$ is actively compensated in order to obtain
a pure quadrupole resonance spectrum. The same technique will soon be
applied for a measurement of the quadrupole moment of $^{11}$Li.

The small increase of $Q_{s}$ from $^{9}$Li to $^{11}$Li of 8.8(1.5)\% can
in principle arise either from a $^{9}$Li core polarization or from the
change of the charge distribution of the core due to the recoil effect
caused by a $d$ component in the halo-neutron wavefunction. In \cite%
{Shulgina2009} a pure three-body wavefunction was designed that was
optimized to reproduce experimentally known properties of the halo isotope
consistently. The model wavefunction indeed reproduces the quadrupole
moment, magnetic moment and the nuclear charge radius as well as the binding
energy in excellent agreement with experiment. Also all other properties are
within the measurement uncertainties. For microscopic theories this halo
nucleus is still a challenge but the set of high-precision data from atomic
physics low-energy experiments that has now been established along the
complete chain of isotopes provides excellent benchmarks for ab-initio,
cluster, and shell-model theories \cite{Noertershaeuser2011b}.

\paragraph{RF-optical Double Resonance in an Ion Trap}

\begin{figure}[bth]
\begin{center}
\includegraphics[clip,width=7.5cm]{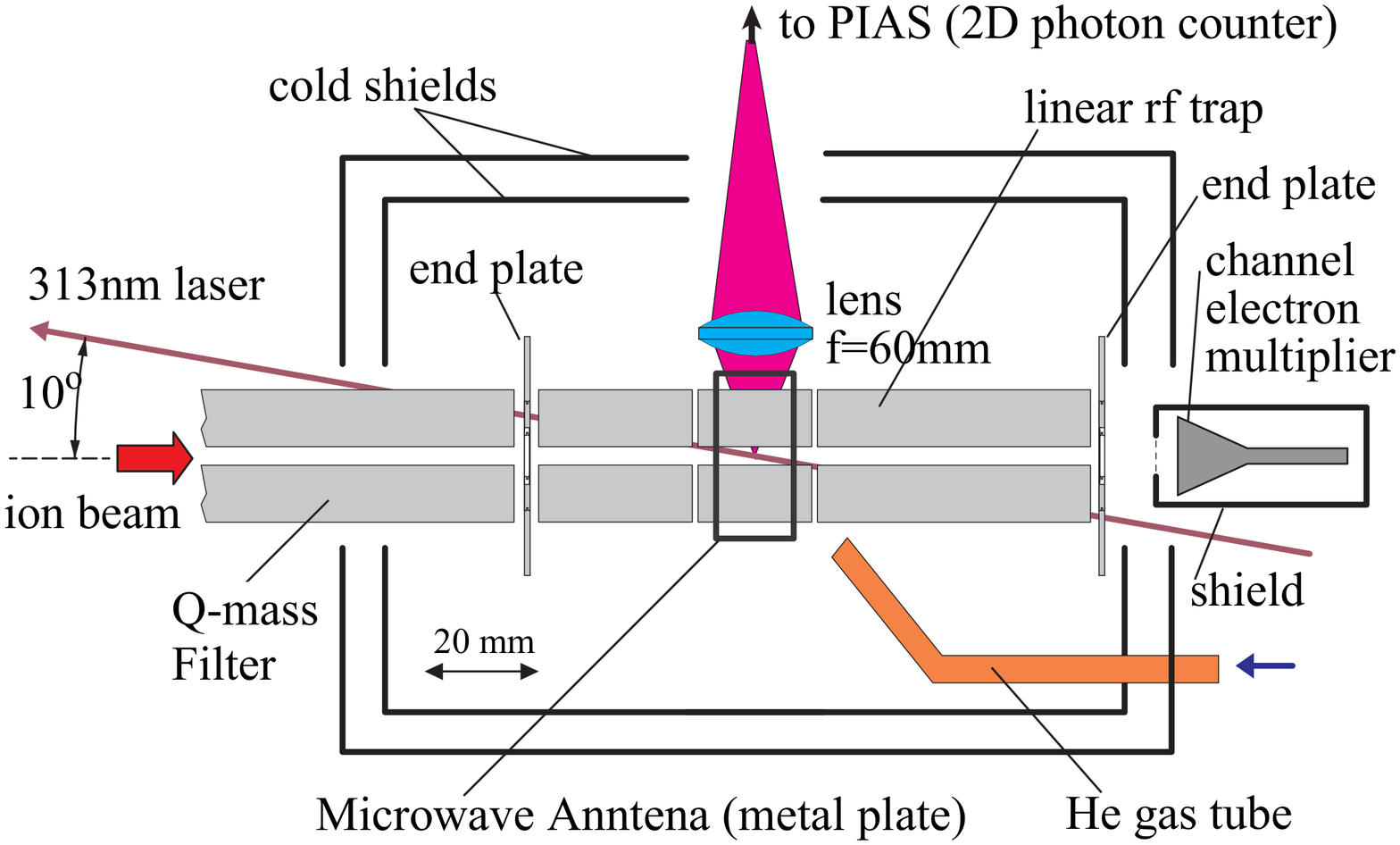}
\includegraphics[clip,width=7.5cm]{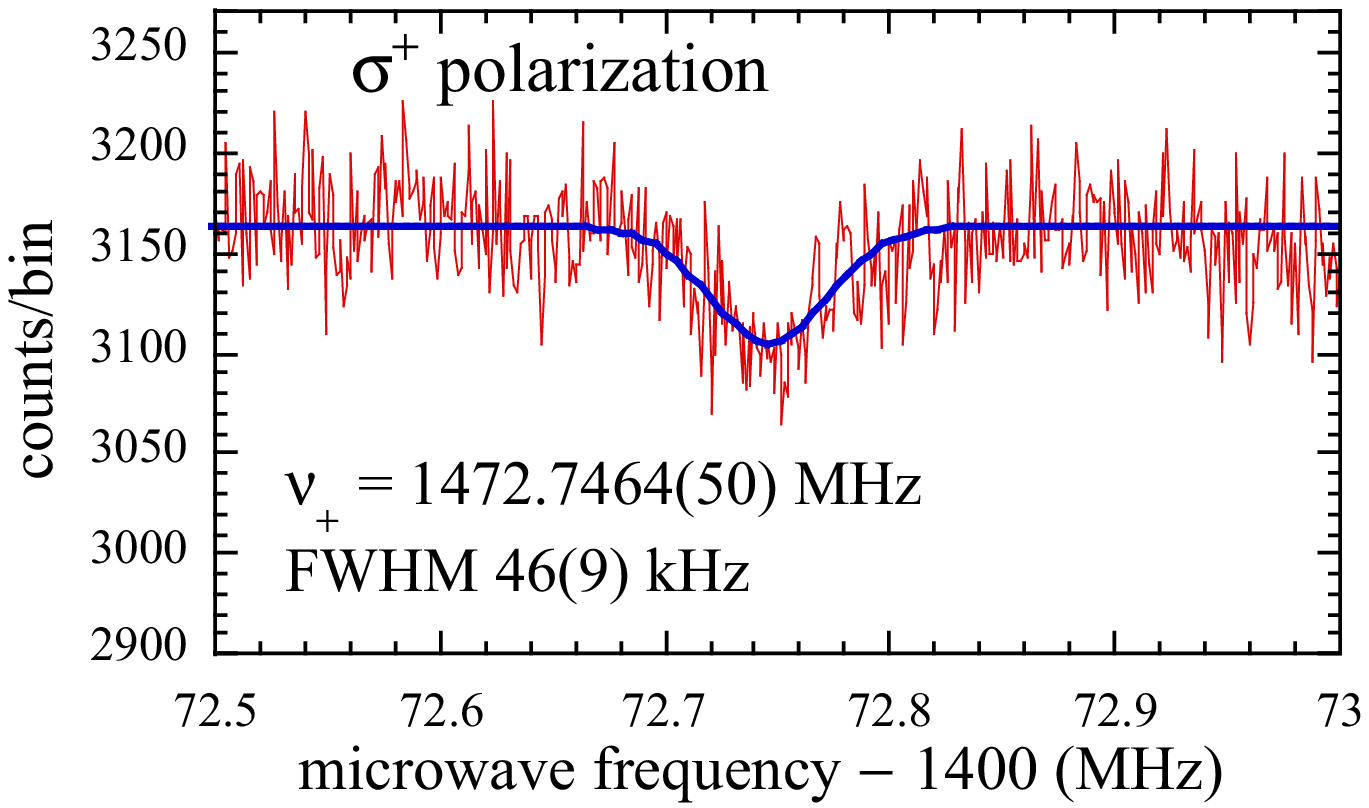}
\end{center}
\caption{(Colour online) Setup (left) and signal (right) for mirowave-optical double-resonance spectroscopy on laser cooled $^{7}$Be ions performed at RIKEN \cite{Okada2008}. Laser resonance fluorescence induced by the $\sigma^+$-polarized cooling laser on the $2s_{1/2} (F=2,$ $m_{F}=+2) \rightarrow 2p_{3/2} (F=3,$ $m_{F}=+3)$ transition is recorded as a function of the microwave frequency coupling the $F=2,\; m_{F}=+2$ and the $F=2,\;m_{F}=+ 1$ states introduced via the microwave antenna. At resonance, the microwave removes population from the cycling cooling transition, reducing the amount of resonance fluorescence. (Reprinted with permission from Phys.\ Rev.\ Lett.\ \textbf{101} 212502 Copyright 2008 American Physical Society)}
\label{fig:RF_DoubleRes7Be}
\end{figure}

The nuclear moments of beryllium isotopes are also the subject of an
experiment at the SLOWRI facility at RIKEN. Here, radioactive beryllium
isotopes are produced using projectile fragmentation of a 100~MeV/u $%
^{13}$C beam. The fragments are stopped in a gas cell, extracted as a
2-eV low-energy beam and transferred into a cryogenic (10~K) Paul trap. The central region of the segmented Paul trap is shown on the left in figure~\ref{fig:RF_DoubleRes7Be} \cite{Okada2008}. During the ion accumulation time, a small amount of cold He gas is introduced for buffer gas cooling. Laser Doppler-cooling on the $2s^{~2}\mathrm{S}_{1/2}\rightarrow 2p~^{2}\mathrm{P}_{3/2}$
transition is performed using a red-detuned laser introduced at a small angle relative to the trap axis in order to cool all degrees of freedom simultaneously and a final temperature
of the ions of the order of 10~mK is reached. Hence, the energy of the ions
is reduced by about 15 orders of magnitude \cite{Nakamura2006b}. In order to
obtain a closed two-level system on the transition, a magnetic field is employed,
to lift the degeneracy of the $m_{F}$ levels. During the
cooling process the ions are optically polarized and pumped with $\sigma\pm$ light
into one of the
$2s_{1/2} (F=2,$ $m_{F}=\pm 2) \rightarrow 2p_{3/2} (F=3,$ $m_{F}=\pm 3)$
closed two-level transitions. The fluorescence is observed with a 2D photon-counting detector.
The separation between the $F=2,$ $m_{F}=\pm 2$ and the $%
F=2,\;m_{F}=\pm 1$ states is then be measured by introducing a
radio-frequency via the microwave antenna that couples the respective substates and therefore removes population from the laser transition. Thus, the observed amount of
fluorescence decreases and a dip in the RF spectrum is observed as it is
shown on the right in figure~\ref{fig:RF_DoubleRes7Be} \cite{Okada2008}. Measuring the RF
transition frequencies for the two transitions at two magnetic field
strengths and fitting the Breit-Rabi formula to the results allows the
extraction of the $A$ factors of $^{7}$Be and -- assuming no significant
hyperfine anomaly -- the determination of the magnetic moment of this
isotope to $\mu (^{7}\mathrm{Be})=1.399\,28(2)$~$\mu _{\mathrm{N}}$. The $A$
factor in the $^{2}\mathrm{S}_{1/2}$ ground state of $^{11}$Be has also been
measured with high precision \cite{Wada2012} but has not been published yet.
In order to extract the hyperfine anomaly, an independent measurement of the
magnetic moment with higher accuracy than the previous CLS measurement in \cite{Geithner1999} is required.

The trapped ions were also used for isotope
shift measurements, applying an optical-optical double-resonance technique
\cite{Takamine2009}. Final results have not been published so far, but reasonable
agreement with the collinear data has been obtained in the preliminary data
analysis \cite{Wada2012}.

\subsection{The Island of Inversion}

\begin{figure}[bth]
\begin{center}
\includegraphics[clip,width=12cm]{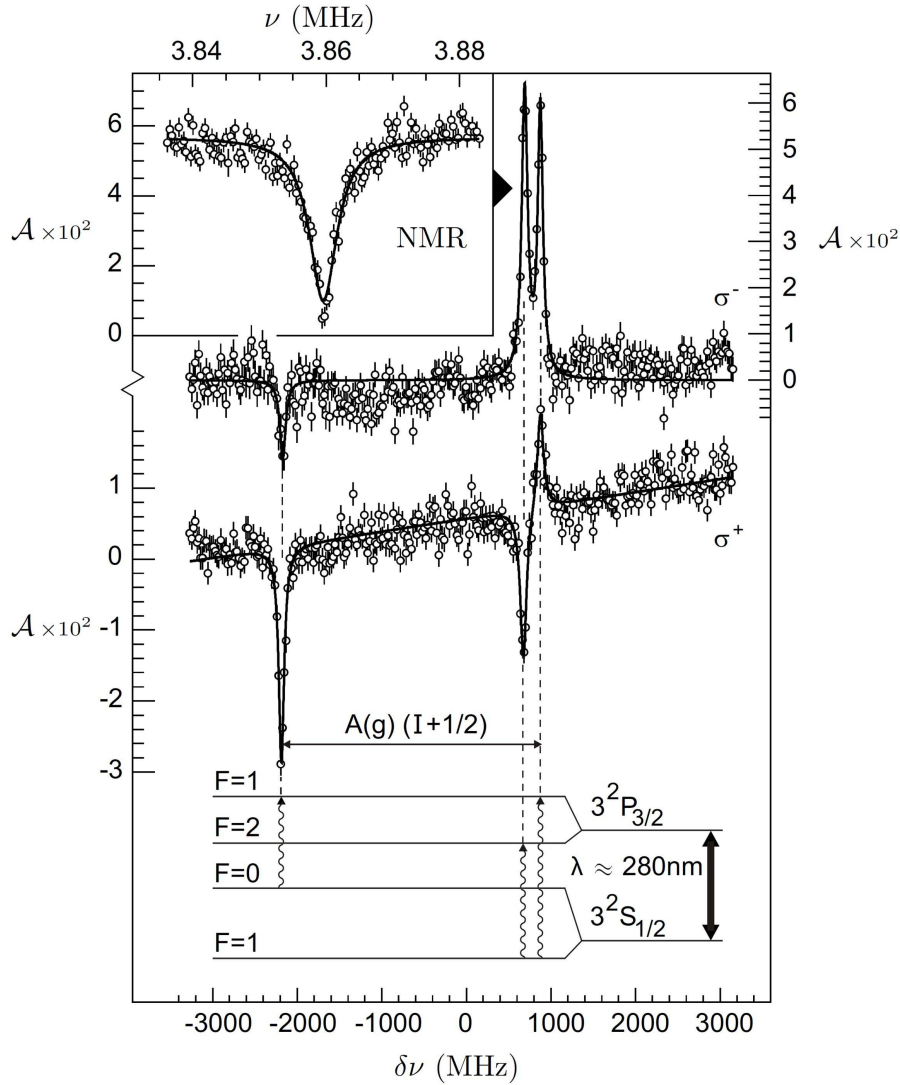}
\end{center}
\caption{Hyperfine spectrum of $^{31}$Mg in the D2 line detected using $\beta$-asymmetry detection recorded as a function of Doppler-tuned laser frequency \cite{Yordanov2007a}. The respective transitions are indicated on the level scheme below the resonances. The inset on the top shows the $\beta$-NMR signal as a function of the applied radiofrequency after optical pumping with $\sigma^-$ on the transition with the highest $\beta$-asymmetry ($F=1 \rightarrow F=2$). (Please note that the labels $\sigma^\pm$ are reversed compared to those shown in reference \cite{Neyens2005}, where they are wrongly assigned.)}
\label{fig:BetaAsymSpec31Mg}
\end{figure}

The so-called island of inversion around the $N=20$ shell closure of the
isotopes of Ne, Na, Mg and Al has now been investigated for more than 35
years after the observation of anomalous ground-state properties of $^{31}$%
Na: An unusual increase in binding energy was seen for $N=20,21$ ($^{31,32}$%
Na) instead of the normal behavior of a sudden drop after a shell has been
completed \cite{Thibault1975}. It was from the beginning interpreted as
being a sign of sudden deformation, which is unusal for a magic nucleus.
Laser spectroscopic studies of the sodium isotopes supported this picture by
finding a spin of $I=3/2$ for $^{31}$Na instead of the expected $I=5/2$ for
a $d_{5/2}$ proton and isotope shift measurements also indicated an
additional volume effect by a prolate deformed nucleus \cite{Huber1978}. At
ISOLDE,\ the neighboring isotopic chains of Mg and Ne were intensively
investigated during the last decade by laser spectroscopy at COLLAPS \cite%
{Neyens2005,Kowalska2008,Yordanov2007a,Yordanov2007b,Kraemer2009,Yordanov2012}.
During these investigations, the spin and magnetic
moment of $^{31}$Mg came as a surprise \cite{Neyens2005}: optical pumping
combined with $\beta $-asymmetry detection and $\beta $-NMR was used to
determine the spin and magnetic moment of this nucleus as described above
for the case of lithium \cite{Kowalska2008}. The spins in the Mg$^{+}$ ion
beam were oriented with a circularly polarized laser beam at 280~nm in the $%
2s\,^{2}\mathrm{S}_{1/2}\rightarrow 2p\,^{2}\mathrm{P}_{3/2}$ transition.
The $\beta $-asymmetry spectrum as a function of the laser frequency, as
shown in figure~\ref{fig:BetaAsymSpec31Mg} \cite{Yordanov2007a}, exhibits only
three resonances. The vertical lines along the peaks guide the eye to the
respective transition in the level scheme shown below the spectrum. The
observation of only three lines is already a clear signature for a $I=1/2$
nucleus since any other half-integer spin would result in six allowed
transitions. A \emph{direct spin measurement} was also performed by an
independent determination of the magnetic moment from the optical spectrum
and the $g$-factor from the $\beta $-NMR measurement. The ground-state
splitting obtained in the optical spectrum in a $^{2}$S$_{1/2}$ state is $%
\Delta E=A\left( I+1/2\right) $ with $A=\frac{g_{I}\mu _{N}B_{0}}{IJ}$while
the Larmor frequency measured in $\beta $-NMR is $\Delta \nu
_{L}=g_{I}\mu _{N}B_{0}$. If the ratio $A/g_{I}$ is known from a reference
isotope, the only unknown is the spin $I$ which can then be unambigiously
determined. In this way the spin $I=1/2$ was determined for $^{31}$Mg and
similarly the ground-state spin of $^{33}$Mg was determined to be $I=3/2$.
These findings were explained by two-particle two-hole excitations across
the $N=20$ shell closure being dominant in both nuclei \cite{Yordanov2007b}.

The nuclear moments of $^{31}$Mg and $^{33}$Mg are in the Nilsson model
connected to strongly prolate deformed levels as they are observed for all
nuclei in the island of inversion. Measurements of the quadrupole moments
along the isotopic chain would therefore be desirable. However, the key
isotopes $^{30,31,32}$Mg are either spinless or have spin 1/2 and do
therefore not exhibit a
spectroscopic quadrupole moment. Instead, the charge radii along the
complete $sd$-shell were determined by isotope shift measurements as an
observable that is also sensitive to deformations. Since fluorescence
spectra are not available for $^{21,31,33}$Mg because the production yields
are too low, the $\beta $-asymmetry spectra must be exploited for such a
measurement \cite{Yordanov2012}.

\begin{figure}[bth]
\begin{center}
\includegraphics[clip,width=12cm]{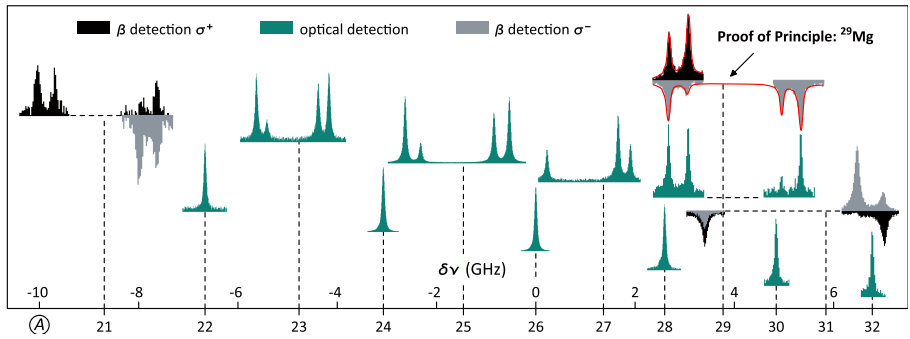}
\end{center}
\caption{(Colour online) Spectral lines of Mg isotopes on the D1 transition in singly charged ions  \cite{Yordanov2012}. The (Doppler-tuning) frequency scale is relative to the isotope $^{26}$Mg. Resonance spectra for $^{22-30}$Mg and $^{32}$Mg ar recorded by standard fluorescence detection, where for $^{30,32}$Mg ion-photon coincidence detection was used. Only $\beta$-asymmetry detection was possible for the isotopes $^{21,31}$Mg. The proof-of principle that analysis of these spectra is compatible with data obtained from fluorescence spectra was shown with the isotope $^{29}$Mg, for which both spectra were obtained. (Reprinted with permission from Phys.\ Rev.\ Lett.\ \textbf{108} 042504 Copyright 2012 American Physical Society)}
\label{fig:ISMg}
\end{figure}

Therefore a quantitative description of the peak shape and the line
intensities in the $\beta $-asymmetry spectrum based on rate equations of
the optical pumping along the path of flight is required. A quantitative agreement was found
\cite{Kowalska2008} for Mg, which was the basis for the determination of the
isotope shift of $^{21,31}$Mg \cite{Yordanov2012}. Figure~\ref{fig:ISMg}
shows the spectra obtained for the Mg isotopes along the complete $sd$%
-shell. The isotopes $^{22-29}$Mg were investigated using CLS with standard
optical fluorescence detection. The exotic even isotopes $^{30,32}$Mg could
also be observed by fluorescence detection but it was necessary to use the
ion-photon coincidence technique and the observation time was limited to
about 100~ms after each proton pulse. Detection of the resonances of $%
^{21,31}$Mg was only possible applying the $\beta $-asymmetry detection. To
ensure that the resonance positions obtained by this technique are in
accordance with those from the optical fluorescence, both techniques were
applied to the isotope $^{29}$Mg. This proof of principle was successful and
the measurements were combined to obtain the change of the charge radius
along the Mg chain.

For magnesium, ab-initio calculations of the atomic mass shift constants with
spectroscopic accuracy are currently not feasible. Instead, the mass shift and the field
shift coefficient were obtained with a King-Plot procedure, using the charge
radii of the stable isotopes as input. Results are in excellent agreement
with the best theoretical calculations currently available \cite{Sahoo2010}
and the charge radii extracted with this
technique are plotted in figure \ref{fig:Rc_Ne-Mg} together with those of
the neighboring isotopes of neon and sodium. The charge radii of Na are taken from
\cite{Fricke2004} based on the measurements reported in \cite{Huber1978}.
Sodium has only a single stable isotope and a King-plot procedure is therefore
not possible to extract the specific mass shift contribution.
Touchard and coworkers \cite{Touchard1982} estimated the SMS using the empirical
observation that the change in charge radius without deformation
$\delta \langle r^2 \rangle_{\rm Sph}$ is usually only half of that expected
for a uniform-density nucleus $R=r_0 A^{1/3}$ with  $r_0=1.2$~fm and assuming
a negligible change of deformation between $^{25}$Na and $^{27}$Na.
%$\delta \langle r^2 \rangle_{\rm unif.} = \frac{2}{5} r_0^2 A^{-1/3} \delta A$.
The extracted charge radii are very sensitive to the estimated SMS and uncertainties
are difficult to quantify. To improve the quality of this data, high-precision
calculations of the mass shift contributions are highly desirable.
It appears that the slope of the sodium chain is slightly too steep compared to the
chains of neon and magnesium. It is striking that all all three elements exhibit a shallow
minimum of the charge radius around $N=14$, i.e., after filling the $s_{1/2}$
and the $d_{3/2}$ orbital in the $sd$ shell.

\subsection{Studies of Medium Heavy Nuclei using Cooled and Bunched Beams}

Studies of the medium heavy radioactive nuclei have strongly profited from
the application of gas-filled radio-frequency quadrupoles (RFQ) for cooling
and bunching \cite{Herfurth2001}. A segmented linear radio-frequency ion
trap is mounted on a high-voltage platform, such that the ion beam is
decelerated and the ions
are stopped inside the electrode structure. While the RF field at the rods
stores the ions radially, a longitudinal DC potential is applied along the
individual segments to form a potential well. Since the RFQ is filled with
buffer gas at room temperature with typical pressures around $10^{-5}$ to $%
10^{-4}$~mbar, the ions loose kinetic energy in the collisions with the gas
and are cooled into the bottom of the potential well. After a preselected collection
time, the potential barrier at the last electrode is rapidly pulsed down and
all ions accumulated during the collection time are extracted in a short
bunch and reaccelerated to an energy that is determined by the high-voltage
platform of the cooler. Accumulation times can be varied depending on the
ion yield, the lifetime of the isotope of interest or other requirements.
Usual values are on the order of a few 10~ms and typical temporal
bunch lengths are a few $\mu$s. The accumulation time is often limited by the
amount of isobaric contamination since the number of ions that can be stored
in the trap before space charge effects become dominant is of the order
of $10^5$--$10^7$.
This technique was first applied at ISOLDE for Penning-trap mass
measurements \cite{Herfurth2001} and its advantages for laser spectroscopy
were impressively demonstrated at the Jyv\"{a}skyl\"{a}
IGISOL (ion guide isotope-separation on-line) facility \cite{Nieminen2002}. The
principal idea behind the bunching is to suppress the unavoidable background
from laser light scattered at apertures and at the entrance windows. This
leads even in favorable cases to a background of the order of a few 100
counts per second per mW of laser light. The statistical fluctuation of
this background is the limit for the sensitivity to small resonance
fluorescence signals. If the observation time of the photomultiplier
background can be limited to those times where indeed an ion or atom of the
species of interest is inside the observation region, this background can be
strongly suppressed and the signal-to-noise ratio improved. This is the
advantage of the cooling and bunching technique. The background suppression can in this case
be easily calculated from the ratio of the bunch length and the collection
time. A typical example is a collection time of 50~ms and a bunch length of
5~$\mu $s, resulting in a background suppression of $10^{4}$. The background is
now limited by ion-beam related background which cannot be suppressed with this
technique.

Ion beam cooler-bunchers for collinear laser spectroscopy are now in
operation at JYFL \cite{Nieminen2002}, ISCOOL at ISOLDE \cite%
{Franberg2008,Mane2009} and at ISAC/TRIUMF \cite{Mane2011a}. At JYFL work
was performed in the region of the $N=28$ shell closure above Ca and around the deformation onset at $N=60$ from Y to Mo as discussed below. At ISCOOL work has been concentrated on studies of the
monopole migration in Cu and Ga isotopes but most recently it was also used
for the spectroscopy of K, Ca and Cd isotopes. A speciality of the ISAC beam
cooler is the possibility to extract the cooled ions in forward and in
backward direction \cite{Mane2011a}. In forward direction the ions are
delivered to the Penning-trap mass spectrometer TITAN, whereas
the ions extracted backwards are used for collinear laser spectroscopy. This
was applied for laser spectroscopy on the $N=Z$ nucleus $^{74}$Rb \cite%
{Mane2011c} as will be discussed in section \ref{sec:CKMUnitarity}. In the
following we will mention some of the recent results using the cooling and
bunching technique at ISOLDE\ and JYFL.

\subsubsection{The N=28 Shell Closure around Calcium}

\begin{figure}[tbh]
\begin{center}
\includegraphics[width=7cm]{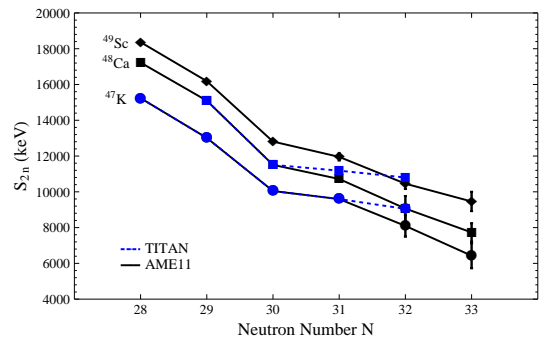}
\end{center}
\caption{(Colour online) Two-neutron separation energies ($S_{2n}$) of neutron-rich K, Ca, and Sc isotopes. The Penning-trap mass measurement values \cite{Gallant2012} are shown together with literature values from the AME 2003 \cite{AME2003}. (The figure is taken from \cite{Gallant2012}).}
\label{Ca_S2n}
\end{figure}

This shell closure is in-depth discussed in the contribution by O.\ Sorlin in these proceedings. Mass measurements provide excellent insight into the complex nuclear interaction as all effective forces are reflected in the nuclear binding energy. Hence, by providing precise and accurate binding energies (or total ground state properties) it is possible to refine and ultimately improve theoretical predictions.  Neutron-rich isotopes are particularity suited for such tests, as can for example be seen from the studies of halo nuclei. Another example, where new phenomena appear in very neutron-rich isotopes, is the chain of oxygen isotopes (see R.\ Kanungo's contribution in this issue and \cite{Janssens2009,Kanungo2009}). Predictions for changes or evolution of the magic numbers of the nuclear shell model now also exists for isotopes of calcium (see contribution of M.\ Hjort-Jensen). In order to test such predictions, ground state mass measurements are well suited, but reliable data are often difficult to obtain due to the low production yields and short half-lives, for example $t_{1/2}(^{54}$Ca$)=86(7)$\,ms.
Recently mass measurements of the Ca-chain and nearby K-chain were carried out. In a first campaign $^{47-50}$K and $^{49,50}$Ca were measured with the TITAN facility. Surprisingly large deviations from literature were found ranging from 4  to 10 standard deviations \cite{Lapierre2012}. A second campaign made use of a UC$_{x}$ target and even more exotic nuclei were accessible. The measurements of $^{51}$K and $^{52}$Ca revealed again large deviations from previous AME2003 \cite{AME2003} values and a strong change in the general trend of the two-neutron separation energies $S_{2n}$ appeared. The results together with AME2003 data are shown in figure \ref{Ca_S2n}. The observed separation energies and the general trend along the isotopic chain are in good agreement with theory if three-nucleon ($3N$) interactions are taken into account. This has been shown, for example, using a coupled cluster method with phenomenological $3N$ forces \cite{Hagen2012} and using chiral effective field theory \cite{Gallant2012}.

Laser spectroscopy in this region was recently performed at JYFL with investigations on Sc \cite{Avgoulea2011}, Ti \cite{Gangrsky2004}, and Mn \cite{Charlwood2010}. Laser measurements have also been completed for K and Ca isotopes at COLLAPS and are currently being analyzed.

\subsubsection{Studying the Monopole Interaction in the $pfg$ Shell}

With ISCOOL at ISOLDE the isotopic chains of copper and gallium have been
studied \cite%
{Flanagan2009,Cheal2010a,Cheal2010b,Vingerhoets2010,Vingerhoets2011,Mane2011b}.
The main objective of these works was the investigation of the behavior of
the proton $\pi p_{3/2}$ and $\pi f_{5/2}$ single-particle levels, while
the $\nu g_{9/2}$ orbital is filled with neutrons. The monopole interaction
between protons and neutrons in these states changes their relative
energies and leads to a lowering of the $\pi f_{5/2}$ level while the energy
of the $\pi p_{3/2}$ state is increased when neutrons are filled in the $\nu
g_{9/2}$ shell. This behaviour is understood in terms of the tensor force
\cite{Otsuka2005,Otsuka2010} which is repulsive when protons and neutrons
both occupy orbitals that are either $j_{>}=\ell +1/2$ or $j_{<}=\ell -1/2$
single-particle states (see also the contribution of T. Otsuka in this issue).
Contrary, the tensor force is attractive when
neutrons and protons occupy states of opposite coupling between orbital
angular momentum and spin. In copper and gallium this attraction lowers the
energy of the $\pi f_{5/2}$ ($j_{<}$) state when filling neutrons in the $%
\nu g_{9/2}$ shell and it was expected that this interaction would lead to an
inversion of the $p_{3/2}$ and the $f_{5/2}$ single-particle states
somewhere between $^{73}$Cu and $^{79}$Cu. While the copper isotopes up to $%
A=73$ exhibit an $I=3/2$ ground-state spin as expected from the simple shell
model, the isotope $^{75}$Cu was found to have an $I=5/2$ ground state. The spin
has been tentatively assigned already after first low-resolution in-source
measurements at ISOLDE and was then confirmed by collinear laser
spectroscopy, both reported in \cite{Flanagan2009}. In the gallium isotopic
chain the equivalent transition from an $I=3/2^{-}$ to an $I=5/2^{-}$ ground
state occurs between $^{79}$Ga and $^{81}$Ga. The ground state of $^{73}$%
Ga, previously assigned to be $3/2^{-}$ was surprisingly proven to be a $%
I=1/2^{-}$ state \cite{Cheal2010a}. This was concluded from the number of
observed resonances in both investigated transitions: the $4p\,^{2}\mathrm{P}%
_{3/2}\rightarrow 5s\,^{2}\mathrm{S}_{1/2}$ exhibits only three instead
of six lines and the $4p\,^{2}\mathrm{P}_{1/2}\rightarrow 5s\,^{2}%
\mathrm{S}_{1/2}$ three instead of four resonances. Additionally, the peak
intensity ratio agreed to that expected for a $I=1/2$ ground state. The
corresponding $1/2^{-}$ excited state in the neighboring isotopes shows a
strong variation of energy along the chain. In shell-model calculations using the
effective jj44b \cite{BrownUnpub} and the JUN45 \cite{Honma2009}
interactions, it dives down in energy, until a minimum is reached around $%
^{73,75}$Ga but in the calculations it does not become the ground state.
The $1/2^{-}$ state seems to be of strongly mixed character and its behaviour
cannot easily be understood in terms of the tensor force. Analysis of the magnetic and quadrupole
moments of the gallium isotopes revealed further details of the ground-state
composition, e.g., the sign inversion of the spectroscopic quadrupole moment
around $^{73}$Ga indicates a transition from a dominant $\pi (p_{3/2}^{3})$
state ($Q_{s}>0$) to a $\pi (p_{3/2}^{1})$ configuration ($Q_{s}<0$) with
two protons excited to higher levels \cite{Cheal2010}. The high-precision
mass data required for the isotope-shift analysis were obtained with
the Penning-trap mass spectrometer ISOLTRAP \cite{Roos2004,Blau2004,Guen2007}.

It should also be noted that the moments of various copper isotopes were addressed by
resonance ionization spectroscopy with limited resolution. In-source spectroscopy was applied at
ISOLDE in order to study the moments of $^{68}$Cu$^{g,m}$, $^{70}$Cu$%
^{g,m1,m2}$ \cite{Weissman2002}, and $^{58,59}$Cu \cite{Stone2008}, while a
higher resolution for these isotopes and sensitivity even for $^{57}$Cu was
obtained using gas-cell RIS at LISOL \cite{Cocolios2009,Cocolios2010}. $^{57}
$Cu was considered a key isotope since an earlier $\beta $-NMR measurement
reported a magnetic moment that was considerably smaller than expected from
shell-model calculations \cite{Minamisono2006} and its explanation required
strong excitation across the closed $N=28$ shell. However, this was disproven by
the laser spectroscopic measurements which
were substantially in agreement with shell-model calculations \cite{Cocolios2009,Cocolios2010}.

%The Cu and Ga isotopes do not provide a possibility to use a King plot
%procedure to separate mass shift and field shift part, since the only two
%stable isotopes occur in these chains. The intermediate element zinc ($Z=30$%
%), however, has 5 stable isotopes. Laser spectroscopy on these isotopes
%provides a possibility to check the consistency of the charge radii
%extracted for Cu and Ga and has been proposed (and accepted) to be studied
%at ISOLDE.

\subsubsection{Region of Deformation at $N=60$}

\begin{figure}[bth]
\begin{center}
\includegraphics[clip,width=12cm]{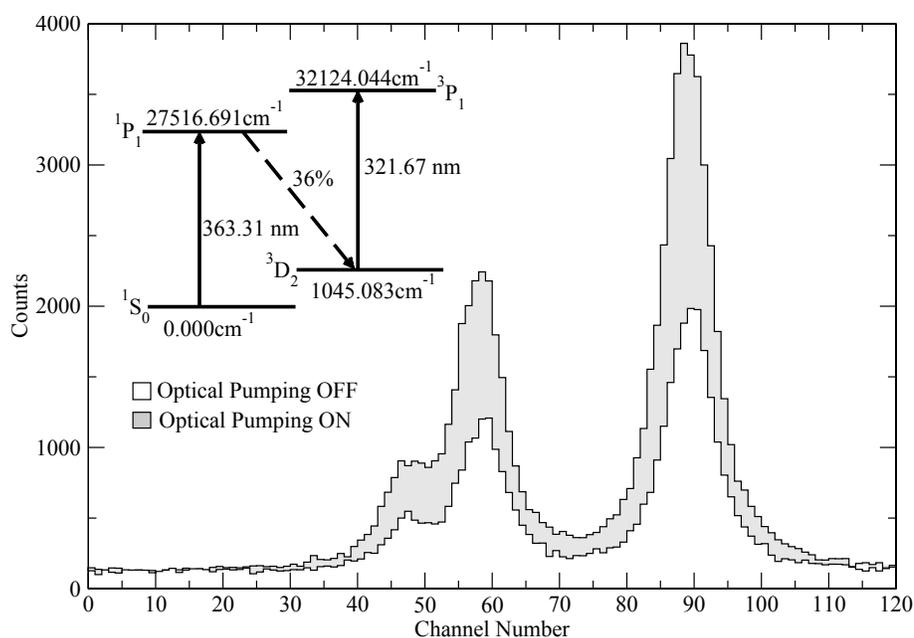}
\end{center}
\caption{Demonstration of signal enhancement in the 321.67 nm transition of $^{89}$Y$^{+}$ by optical pumping on the 363.31~nm transition in the RFQ cooler and buncher \cite{Baczynska2010} as shown on the level scheme in the inset. (Reprinted with permission from J. Phys. G \textbf{37} 105103 Copyright 2010 Institute of Physics Publishing)}
\label{fig:OpticalPumpingY}
\end{figure}

At JYFL the RFQ cooler-buncher technique was applied for the first time for
studies of the isotope shift and hyperfine structure of $^{175}$Hf \cite%
{Nieminen2002}. It was then used to systematically study the
region of deformation at $N=60$ above Sr.
A sudden onset of deformation above $N=59$
was observed at ISOLDE for Rb \cite{Thibault1981} and Sr \cite{Buchinger1990}
but not in Kr \cite{Keim1995}. Above Sr a region of
refractory elements starts, which cannot be studied at ISOL facilities. The
IGISOL at JYFL can produce these isotopes but only in small amounts. The
bunching technique was therefore the breakthrough for laser spectroscopy
with IGISOL. During the last decade the chains of Y \cite%
{Bissel2007,Cheal2007,Baczynska2010}, Zr \cite%
{Campbell2002a,Campbell2002b,Forest2002}, Nb \cite{Cheal2009} and Mo \cite%
{Charlwood2009} were studied. An additional technique facilitated the
spectroscopy of Y, Zr and Nb: Optical pumping in the cooler-buncher was
applied to prepare the ions in states that are more favorable for laser spectroscopy than the
respective ground state. For example, the $5s^{2}$~$^{1}\mathrm{S}_{0}$
ground state of Y$^{+}$ has a resonance transition into a $4d5p~^{1}\mathrm{P%
}_{1}$ state. This $J=0\rightarrow J=1$ transition is not well suited for
nuclear spin determinations. The $4d5s$ $^{3}\mathrm{D}_{2}\rightarrow 4d5p$
$^{3}\mathrm{P}_{1}$ transition is more advantageous but the population of
the metastable $^{3}\mathrm{D}_{2}$\ state is small. Therefore, a pulsed
laser was applied inside the RFQ cooler-buncher to excite ions from the
ground into the $4d5p~^{1}\mathrm{P}_{1}$ level, which has a $\approx 30\%$
branching ratio into the metastable $4d5s$ $^{3}\mathrm{D}_{2}$ state. The
signal was significantly enhanced by the optical pumping as it is shown in
figure~\ref{fig:OpticalPumpingY} \cite{Baczynska2010}.

Optical pumping in the RFQ for state preparation before collinear laser
spectroscopy has further possible applications. It can be used to pump the
population into states that have transitions easier to excite with cw laser
light. While pulsed lasers -- as they are used for the state preparation --
can be easily frequency-doubled, tripled and even quadrupled, these
processes require much more effort with cw lasers as they are used for the
high-precision spectroscopy. Metastable ionic states might in other cases
offer stronger transitions, better knowledge of the atomic factors for mass
and field shifts, larger hyperfine splitting, or the possibility to
distinguish the fluorescence light from the laser excitation. Depopulation
of the metastable state after cw excitation might also be a possible
detection scheme.

\begin{figure}[bth]
\begin{center}
\includegraphics[clip,width=8cm]{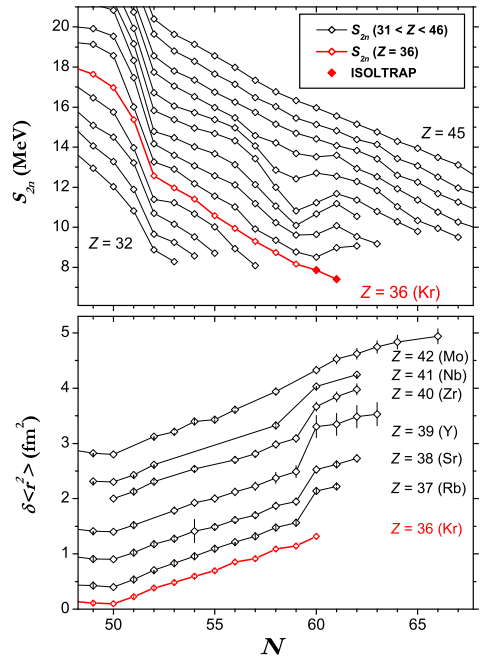}
\end{center}
\caption{(Colour online) Two-neutron separation energies ($S_{2n}$, top) and differences in the mean-square charge radii (bottom) in the $N=60$ region \cite{Naimi2010}. The charge radii clearly exhibit a sudden onset of deformation at $N=60$ from Rb to Nb. This is correlated with the unusual behavior of the binding energies for the respective isotopes of these elements. The increase in the $S_{2n}$ values indicates increased binding due to deformation. (Reprinted with permission from Phys.\ Rev.\ Lett.\ \textbf{105} 032502 Copyright 2010 American Physical Society)}
\label{fig:ChargeRadiiN60}
\end{figure}

The results of the investigations in this region at JYFL are shown in figure~%
\ref{fig:ChargeRadiiN60} in combination with previous results from ISOLDE. For Y and Zr the deformation is strong and similar
to that observed in Sr. In the chain of Nb the transition already weakens
and for Mo it is practically absent. It is interesting to see that this behavior seems to be also reflected in mass measurements on these isotopes, which were mainly performed at Jyv\"askyl\"a and partially with ISOLTRAP as it is shown in the upper part in figure~\ref{fig:ChargeRadiiN60}, taken from reference \cite{Naimi2010}.

\subsection{Studies of Heavy Nuclei}

\subsubsection{In-Source Laser Spectroscopy: Shape Coexistence around Lead }

The region of the heaviest stable elements has early been studied by optical
spectroscopy and later by laser spectroscopy. One of the highlights was the
discovery of the shape coexistence of mercury which leads to a large
odd-even staggering for the isotopes lighter than $^{184}$Hg \cite%
{Kuehl1977,Ulm1986}. Many of the isotopic chains around mercury
were investigated using resonance ionization laser spectroscopy (RIS).
The observation of shape coexistence in $^{186}$Pb \cite%
{Andreyev2000} revived the field.
Recent studies at ISOLDE were performed, e.g., on lead \cite%
{deWitte2007,Seliverstov2009}, the light even isotopes of Po from $^{192}$Po
up to $^{210}$Po and on the neutron-rich isotopes $^{216,218}$Po at ISOLDE
\cite{Cocolios2011}. Compared to the neighboring elements, Po showed a large
deviation from sphericity already at less exotic neutron numbers.

\subsubsection{Superheavy Elements}
Superheavy Elements (SHEs) are chemical elements beyond the actinides, with atomic numbers numbers $Z>103$. Even though there were some speculations about the existence of very heavy elements already in the 1950's \cite{Hermann2003}, the year 1966 is usually regarded as the birth of superheavy element research. Three theoretical papers were published in this year that discussed the existence of an "`Island of Stability"' due to shell-effects that stabilize the nuclei of these elements against spontaneous fission \cite{Myers1966,Meldner1966,Sobiczewski1966}.  Since then it was shown that the shell effects can indeed provide significant stability giving some of the very heavy isotopes half-lives in the range of seconds to hours. During the first period of SHE research, the term SHE was solely coined for the expected island of stability around $Z=114$. However, it is now widely used for all those nuclei, which are stabilized by nuclear-shell effects, starting around rutherfordium ($Z=104$). To date the periodic table has been extended up to element 118 \cite{Oganessian2006}. New elements are typically discovered and identified in fusion-evaporation experiments and subsequent decay identification of the daughter chains. The production of all elements from 113 up to 118 has been claimed in the meanwhile, but according to the last report from the IUPAC/IUPAP Joint Working Party on Discovery of Elements (JWP) \cite{Barber2011} only the papers on element 114 and 116 meet the strong criteria of the working group to really announce the discovery an element. These elements have in the meanwhile been named Flerovium (Fl) and Livermorium (Lv) (for more details see the contribution by P.~Greenlees in these proceedings).

Ideally, one can identify the proton number of the product, and then link the subsequent decay chain(s) to well known nuclei, such that no unambiguous event is falsely counted. However, the production cross-sections are extremely low, and typical production rates are around a few atoms per hour to few atoms per days, going down to even a few atoms/month. An additional potent path forward towards a 'clean' identification as well as to carry out systematic studies, is via direct mass measurements of the isotopes that act as the anker points of the decay chains and to map out where the shell closures occur. This would be of particular interest for the chains of hot fussion, where the decay chains can yet not be linked to known isotopes, since the $\alpha$-decay chains end in nuclei that are not investigated so far and decay by spontaneous fission.

The approach of direct Penning-trap mass measurements has been realized with the SHIPTRAP system at GSI Darmstadt. SHIP is a velocity filter system for fusion evaporation reaction products, and is used for the generation and identification for SHEs. This in-flight production and separation system is coupled via a gas-filled stopper cell to a mass measurement Penning trap (SHIPTRAP). Here, binding energies of isotopes of trans-uranium elements were obtained from direct mass measurements for the first time \cite{Block2010}. Masses of the $^{252-254}$No isotopes have been measured \cite{Block2010} and confirmed in combination with measurements of $^{255}$No and the lawrencium isotopes $^{255,256}$Lr \cite{Minaya2012} the deformed neutron-shell closure at $N=152$. With these mass measurements it has now been possible to map out the shell gap at $N=152$ and therefore providing information where it will be
most promising to investigate SHE and find the predicted island of stability. With production cross sections as low as tens of nano-barns the SHIPTRAP measurements constitute the Penning-trap mass measurement of isotopes
with the lowest cross section performed so far.

The small production are also an obstacle for laser spectroscopic studies in the region of superheavy elements. Moreover, laser spectroscopy must first be applied in order to identify atomic transitions
that can then be used for hyperfine structure investigations, because nothing is known about the atomic structure for elements with $Z>100$. This requires to scan huge
frequency spans since the predictive power of corresponding atomic structure calculations
is still limited and the results have rather large uncertainties. These boundary conditions make laser spectroscopic studies of super-heavy elements extremely challenging.
Gas-cell spectroscopy
with pulsed lasers was shown to be the technique of choice to study these nuclei
\cite{Backe1998,Sewtz2003}. The heaviest element studied so far by laser spectroscopy is the
element fermium ($Z=100$). Off-line studies with the 20.5-hours isotope $^{255}$Fm with
low-resolution resonance ionization spectroscopy in a buffer-gas cell
resulted in the observation of two transitions that can in the future be
used for higher resolution spectroscopic studies \cite{Sewtz2003}. The
resonances were observed by extracting the ions formed by the laser pulse
from the gas cell through a quadrupole ion guide system, subsequent mass
separation in a quadrupole mass spectrometer and ion detection with a
channeltron detector. The technique has been further developed towards
on-line application at the GSI SHIP installation for first spectroscopy of
the element nobelium ($Z=102$). The RADRIS (Radiation Detected Resonance
Ionization Spectroscopy) method is based on stopping the separated
fusion-evaporation product in a buffer gas cell at pressures of about
100~mbar. The fraction of the ions that are not neutralized can be guided
by suitable electric fields to a filament, where they are collected. The
remaining neutralized atoms in the gas can be laser ionized and the
laser-produced ions are guided to a particle detector where the ion is
detected by its $\alpha $-decay. Alternatively, in the so-called ICARE
technique (Ion Collection and Atom Re-Evaporation), the catcher filament is
heated by a short current pulse and the collected atoms are evaporated.
Spectroscopy is then performed by a laser pulse triggered shortly after the
evaporation and detection after resonance ionization as before. The
technique has been demonstrated on-line with $^{155}$Yb, a chemical homologue of
nobelium \cite{Backe2007}. Recently, in-gas jet, high-repetition, high-resolution laser resonance ionization spectroscopy of superheavy elements, also stopped previously in a buffer gas cell, has been proposed \cite{vanDuppen2012}. Besides the possibility to study nuclear
structure by hyperfine interactions, the identification of atomic
transitions as a first step is a very interesting subject in itself due to
the strongly relativistic effects in the electron shell.

\section{Nuclear Astrophysics Studies}

\begin{figure}[h!]
\begin{center}
\includegraphics[width=\textwidth]{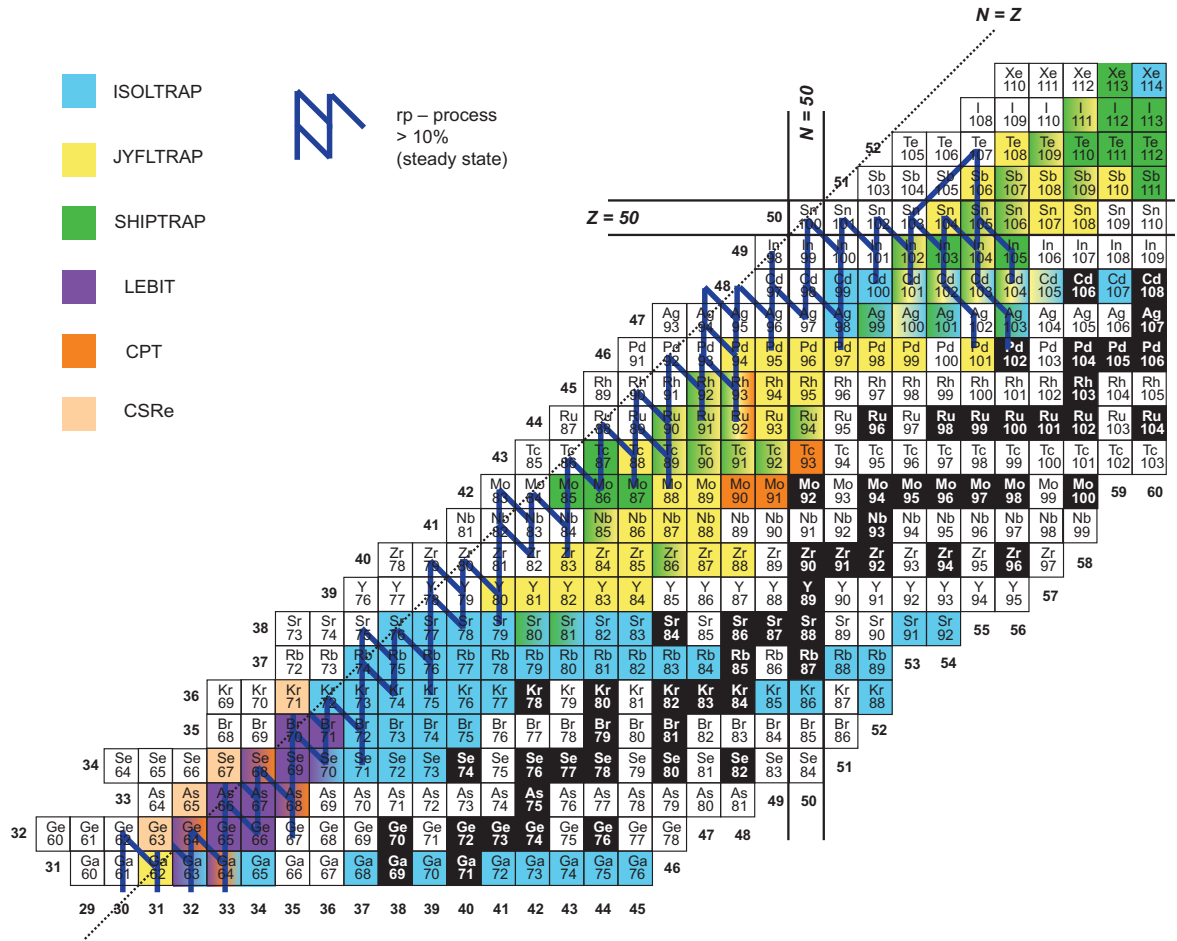}
\end{center}
\caption{(Colour online, updated figure from \cite{Herf2011}). Part of the chart of nuclei showing mass measurements relevant for the $rp$-process at different Penning-trap and storage-ring facilities. Black boxes indicate stable isotopes, the thick line the $rp$-process path, containing at least 10\% of the overall flow in the steady state solution, and the dashed line the $N=Z$ nuclei. Added to the masses measured with the Penning-trap technique and already shown in figure~5 of reference \cite{Herf2011} are the masses recently obtained at the CSRe storage ring in Lanzhou \cite{Tu2011}.}
\label{rpprocess}
\end{figure}

In the last few years a large amount of new high-precision mass data was produced in the context of nuclear astrophysics studies using the Penning-trap and storage-ring techniques, especially on neutron-deficient nuclei for the investigation of the $rp$- (rapid-proton capture) \cite{Scha2001} and the $\nu p$- (rapid-neutrino capture) processes \cite{Froe2006}. More than 2/3 of all nuclides on the $rp$-process path have been measured by now, as shown in figure~\ref{rpprocess}. Only in the mass region around the doubly-magic nucleus $^{100}$Sn experimental mass data is still scarce due to the low production rates of these species at the existing radioactive beam facilities. The data support the modeling of these two nucleosynthesis pathways, which aims at understanding the final elemental abundances and energy production and at comparing model results with the growing number of astronomical observations in a quantitative way. In this context tremendous progress in theory has been made in recent years, see e.g.\ \cite{Froe2012}.

The $rp$-process in type I x-ray bursts starts, for example, from a breakout of the
hot CNO cycle by a sequence of $\alpha$-induced reactions
and proceeds then via rapid proton-capture reactions and subsequent $\beta^+$ decays close
to the $N = Z$ line on the chart of nuclides. In particular in hydrogen-rich bursts it can
reach up to tellurium, where the predominant $\alpha$-instability, $^{107}$Te and $^{108}$Te,
returns the flow into the closed SnSbTe cycles, although recent mass results from JYFLTRAP have
shown that the branching into the closed SnSbTe cycle is weaker than expected \cite{Elom2009}.

In the Ge to Kr mass region mass excesses of short-lived $A=2Z-1$ nuclei have been addressed
for the first time with the CSRe storage ring employing isochronous mass measurements
\cite{Tu2011b}. Mass uncertainties of a few keV allowed to
deduce the proton-separation energy of $^{65}$As to be $S_p=-90(85)$\,keV \cite{Tu2011}, thus being slightly proton unbound.
%As a result of this new x-ray burst model calculations have shown that the majority of the reaction
%flow passes through $^{64}$Ge via proton capture for most relevant temperature-density conditions,
%thus removing this nuclide from the list of prominent $rp$-process waiting points. Combined with
%the Penning-trap mass results with impressively small uncertainties of only a few keV on the
%$rp$-process waiting points $^{68}$Se \cite{Clar2004} and $^{72}$Kr \cite{Rodr2004}, the
%uncertainties due to unknown nuclear masses on the x-ray burst light curves and on the final
%abundances of x-ray burst ashes have been reduced almost entirely.
%
Including the new data into x-ray burst model calculations showed that the majority of the reaction flow passes through $^{64}$Ge via proton capture for most relevant temperature-density conditions, thus, removing $^{64}$Ge from the list of prominent $rp$-process waiting points. Combined with the Penning-trap mass results with impressively small uncertainties of only a few keV on the $rp$-process waiting points $^{68}$Se \cite{Clar2004} and $^{72}$Kr \cite{Rodr2004}, the uncertainties on the x-ray burst light curves and on the final abundances of x-ray burst ashes caused by unknown nuclear masses was almost entirely removed.

In the mass region from Rb up to Rh the most exotic neutron deficient nuclides on the $rp$-process path have been addressed by the Penning-trap facilities JYFLTRAP \cite{Kank2006,Webe2008}, SHIPTRAP \cite{Webe2008}, and ISOLTRAP \cite{Herf2011}. Among them the $N=Z$ nuclides $^{74}$Rb \cite{Kellerbauer2004,Kell2007} and $^{76}$Sr \cite{Sikl2005} and the $N=Z+1$ nuclides $^{85}$Mo and $^{87}$Tc \cite{Haet2011}. Recently the nuclide $^{74}$Rb was remeasured because of its importance as a superallowed beta emitter (see below). This was performed at the TITAN Penning-trap mass spectrometer using for the first time highly charged ions \cite{Ettenauer2011}. Its mass was confirmed with a similar uncertainty of about 6 keV.

Strong mass deviations up to 1.6~MeV from literature values \cite{AME2003} were found for many of these $rp$-process nuclides. This caused significant changes of the elemental abundance in the ashes of astrophysical x-ray bursts in model calculations. This demonstrates once more the importance of accurate and precise mass data for nuclear astrophysics studies. More new $rp$- and $\nu p$-process relevant masses on less exotic isotopes of Nb, Mo, Tc, Ru, and Rh have recently been published by the Canadian Penning Trap collaboration \cite{Fall2011}.

The picture is by far not as complete for $r$-process nucleosynthesis studies \cite{Arno2007} as it is for the $rp$-process. Although the $r$-process is responsible for the origin of about half of the heavy elements beyond iron \cite{Cowa1991}, it is not even known with certainty where in the universe this process can occur. To understand the $r$-process is thus still one of the greatest challenges of modern nuclear astrophysics. This endeavor is hampered by the fact that hardly any of the extremely unstable participating nuclei could be studied, mainly due to the inaccessibility of the involved neutron-rich nuclides at present radioactive beam facilities. Thus, reliable mass data are scarce. Until new facilities start operating, the masses of many key nuclides required for nucleosynthesis calculations far away from the valley of stability can only be extrapolated with nuclear mass models from mass data closer to stability. In order to test the predictive power of these models, mass measurements of less-exotic nuclei with high-precision are still of importance \cite{Lunn2003}.

The $r$-process abundances show pronounced peaks at $A\approx80$ (zinc), 130 (tellurium), and 195 (platinum), which are assigned to possible waiting point nuclei. The first one has been addressed by ISOLTRAP and JYFLTRAP and their high-precision data has improved theoretical descriptions of various separation energies \cite{Haka2008,Baru2008}. Among others, the mass of $^{81}$Zn has been determined for the first time in 2008, making $^{80}$Zn the first of the few major waiting points along the path of the $r$-process where neutron-separation energy and neutron-capture $Q$-value are determined experimentally. Only recently, after several attempts by different facilities, ISOLTRAP succeeded to measure the unknown mass of the even more exotic short-lived nuclide $^{82}$Zn \cite{Wolf2012b} with an uncertainty of only about 1~keV, despite the difficulty of having an isobaric contamination of $^{82}$Rb being eight orders of magnitude more abundant. This became only possible due to the use of the newly installed multi-reflection time-of-flight mass separator \cite{Wolf2012}, which allows to suppress isobaric contaminations by about a factor of 10.000. The mass or, to be more specific, the binding energy of the beta-emitting $^{82}$Zn provides a key information to answer the question about the composition of a neutron-star crust, which is assumed to be the birthplace of many chemical elements.

Another region of interest that has been addressed around a shell closure is at $N=82$, where one takes advantage of the fact that the $r$-process gets closer to the stable nuclei. Here, beta-decay endpoint measurements on $^{130}$Cd gave hints that the $N=82$ shell closure might be quenched at $Z=48$ thus leading to the conclusion that $^{130}$Cd is not a waiting point \cite{Dill2003}. However, isomeric decay studies on $^{130}$Cd conclude in contrast that $N=82$ is not quenched at $Z=48$ \cite{Jung2007}. To resolve this controversy, the shell closure should be probed by direct high-precision mass measurements allowing for the investigation of neutron-separation energies. Unfortunately, approaching $^{130}$Cd with the mass-spectrometry techniques mentioned in this article is very complicated due to the low production rates and the large amount of isobaric contaminations. So far only the masses of the Cd isotopes up to $A=128$ could be addressed directly \cite{Brei2010}.

Going up in $Z$ to even heavier systems up to Rn, more exciting results were obtained, especially at the Penning-trap facilities ISOLTRAP for Sn \cite{Dwor2008}, Xe \cite{Neid2009b}, and Rn \cite{Neid2009}, at JYFLTRAP for Sr, Mo, Zr \cite{Hage2006}, Tc, Ru, Rh, Pd \cite{Hage2007}, and Sn, Sb, Te \cite{Haka2012}, at CPT where a large amount of new data has been taken within the CARIBU project on neutron-rich nuclides from $Z=51$ to 64 \cite{VanS2012}, and at the storage ring ESR on Sb, Te and I isotopes \cite{Sun2008}. The latter show significant deviations of the experimental mass values to the predictions of the ETFSI-Q mass model \cite{Pear1996}, in which the $N=82$ shell quenching is introduced explicitly. More data is expected to come in the next years either from CPT, JYFLTRAP or TRIGATRAP, the latter is the only facility installed at a research reactor (TRIGA-Mainz, Germany) \cite{Ketelaer2008}.

\section{Fundamental Studies}
Fundamental studies including tests of the Standard Model have the highest demands on accuracy for masses of short- and long-lived exotic nuclides. For example, precise studies of superallowed $\beta$-emitters \cite{Town2010} and the investigation of isotopes relevant for neutrino physics \cite{Blau2010} require mass measurements with an accuracy of 1~keV or less. Measurements on parent and daughter nuclides of $\beta$-decay transitions give stringent $Q$-values and complement nuclear spectroscopy measurements.

\subsection{Test of the Unitarity of the Quark Mixing Matrix}
\label{sec:CKMUnitarity}
\begin{figure}[h!]
\begin{center}
\includegraphics[width=\textwidth]{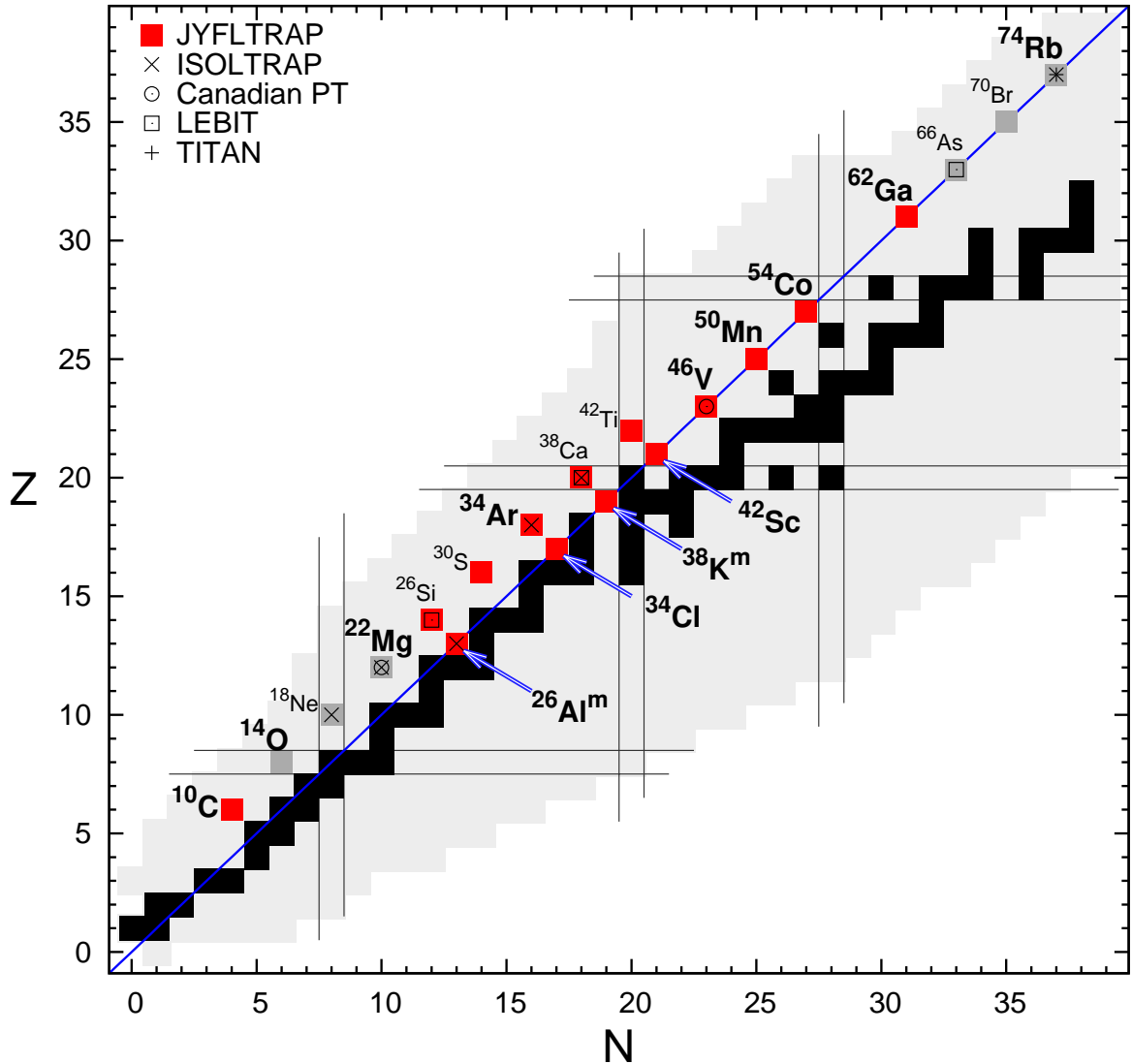}
\end{center}
\caption{(Colour online) Shown are all superallowed-beta decaying nuclides that have so-far being studied by Penning-trap mass spectrometry. The 13 nuclides named in bold are those contributing to the to-date best average $Ft$-value in order to test the conserved vector current hypothesis.}
\label{superallowed}
\end{figure}

The test of the unitarity of the Cabibbo-Kobayashi-Maskawa (CKM) quark mixing matrix, which relates the quark weak-interaction eigenstates to the quark mass eigenstates with the assumption of three quark generations, and thus of the Standard Model (SM) in general, was one of the most active research fields in high-precision Penning-trap mass spectrometry on short-lived nuclides in the past years. Many groups have contributed in measuring $Q$-values of superallowed $\beta$-emitters, but it is obvious from figure~\ref{superallowed} that JYFLTRAP \cite{Joki2006} played the leading role \cite{Eron2011}.

The first CKM maxtrix element $V_{ud}$ is the leading element in the top-row unitarity test which should fulfill
\begin{equation}
\sum_j |V_{uj}|^2 = |V_{ud}|^2+|V_{us}|^2+|V_{ub}|^2 = 1 .
\end{equation}
Any deviation from 1 can be related to concepts beyond the SM such as the existence of an additional $Z$-boson or the existence of right-handed currents in the weak interaction. $V_{ud}$ contributes to this test about $95\%$ and its determination is thus of particular importance. In nuclear physics it can be extracted from the vector coupling constant $G_V$. This constant is derived from the mean $Ft$ value of $0^+ \rightarrow \, 0^+$ superallowed nuclear $\beta$-decays, connecting the analogue states of specific nuclei without change of spin, parity and isospin \cite{Hard2009}, in conjunction with the Fermi coupling constant from the muon decay $G_{\mu}$
\begin{equation}
 |V_{ud}|^2=\frac{G^2_V}{G^2_\mu}.
\end{equation}
The $Ft$-value is the product of a corrected value $F$ of the phase space integral $f$, containing the lepton kinematics, and the partial half-life $t$ for the superallowed $\beta$-transition under study. According to the conserved vector current (CVC) hypothesis \cite{Feyn1958}, the vector part of the weak interaction is not influenced by the strong interaction. The comparative half-life $ft$ of a superallowed $\beta$-transition should therefore be only a function of the nuclear matrix element and the vector coupling constant \cite{Orma1989}. However, corrections have to be applied to these two quantities, including nuclear-dependent and nuclear-independent radiative corrections as well as isospin-symmetry-breaking corrections. Only the so-called corrected-$ft$-value, denoted by the symbol $Ft$, is expected to be constant. Fortunately, all corrections are small -- of the order of $1\%$ -- but they must be calculated with an accuracy of 10\% of their central value. This is presently the limiting factor in the CVC test.

It should be mentioned that the isospin-symmetry breaking term depends on the size of the nuclear charge distribution. This connects the CVC tests also with laser spectroscopic investigations and was the motivation for the determination of the hyperfine structure and isotope shift of $^{74}$Rb at TRIUMF. The measurements resulted in a 20\% reduction of the charge-radius dependent $\delta _{C}$ correction part \cite{Mane2011c}. To further improve the uncertainty of the $Ft$ value an improved measurement of the $\beta $-decay $Q$ value is required since the charge radius uncertainty is not the limiting factor anymore.

In order to determine $Ft$-values, not only the decay energy $Q$, i.e.\ the mass difference of the mother and daughter nuclide, of the superallowed $\beta$-transition needs to be measured, but also the half-life and the branching ratio, yielding the partial half-life $t$. Since $F$ depends on the fifth power of $Q$, stringent demands are put on the Penning-trap mass measurements calling for relative mass uncertainties of $10^{-8}$ and below, despite the fact that most of these nuclides have very short half-lives of less than 1\,s requiring the use of on-line systems.

In the most recent compilation of all superallowed $\beta$-transitions by Hardy and Towner in 2010 \cite{Town2010}, 13 transitions (see figure~\ref{superallowed}) contribute to the to-date best average $\overline{Ft}$-value of 3072.08(79)~s with a normalized $\chi^2$ of 0.29. Among the involved nuclides $^{74}$Rb with a half-life of only 65\,ms is the most exotic one that got addressed by Penning-trap mass spectrometry \cite{Ettenauer2011,Kellerbauer2004}. This $\overline{Ft}$-result provides a confirmation of CVC at the $2.5\cdot10^{-4}$ level. If it is taken to calculate $V_{ud}$ one obtains $|V_{ud}|=0.97425(22)$. Together with $V_{us}$ and $V_{ub}$ from particle physics $K$ and $B$ decays, respectively, this yields for the top-row unitarity test of the CKM matrix
\begin{equation}
|V_{ud}|^2+|V_{us}|^2+|V_{ub}|^2 = 0.999\,90(60) ,
\end{equation}
showing unitarity to be satisfied to a precision of $0.06\%$ \cite{Town2010}. Being fully consistent with unity it provides strong limits on different types of ``New Physics beyond the Standard Model". For more details and also on the status of tests using $T=1/2$ mirror transitions \cite{Navi2009} we refer to the contribution of N.~Severijns within this book of proceedings.

\subsection{Masses for Neutrino Physics}

Since the discovery of neutrino oscillations it is certain that they are massive particles. However, neither their absolute mass values nor the mass hierarchy are known yet. In addition, the question wether neutrinos and antineutrinos are identical or different, i.e.\ Majorana or Dirac type particles, is still open. Thus, the physical mechanism responsible for the non-zero neutrino mass is unknown and there are numerous activities in the neutrino-physics sector worldwide to answer these important questions. Penning-trap mass spectrometry provides important information for many of these neutrino-physics experiments, primarily via $Q$-value measurements of the involved nuclides in the decay processes.

\subsubsection{Neutrino Masses and Neutrino-Mass Hierarchy}
The most promising attempt to get the absolute mass value of the electron-antineutrino $\bar{\nu}_e$ is the investigation of the $\beta$-decay spectrum of tritium
\begin{equation}
^3\mbox{H} \rightarrow \, ^3\mbox{He}^+ + e^- + \bar{\nu}_e.
\end{equation}
The KArlsruhe TRItium Neutrino experiment (KATRIN) \cite{Wein2002} aims for a direct measurement of the antielectron neutrino rest mass or at least to improve its present upper limit of 2.0~eV (95\%~c.l.) \cite{Otte2006} to about 0.2~eV (90\%~c.l.) by investigating the $\beta$-decay of tritium gas molecules. The excess energy $Q$ is shared between the kinetic energy of the electron, the recoil of the nucleus and the total energy of the antineutrino. The shape of the energy spectrum for the decay electrons would be modified near the endpoint $Q_0$ in case of a finite electron antineutrino rest mass. $Q_0$ is determined by the mass difference of the mother and the daughter nucleus, i.e.\ $Q_0 = m(^3\mbox{H})- m(^3\mbox{He})$, minus the electronic excitation energy of the final state of $^3$He. Though electron retardation spectrometers, like KATRIN, can determine this quantity in addition to the neutrino mass from the analysis of the electron spectral shape, independent measurements help to rule out possible systematic errors in the neutrino mass measurement. Furthermore, as a future perspective, a significantly improved uncertainty of $Q_0$ compared to the currently best value from SMILETRAP (Stockholm, Sweden) of $Q_0=18\,589.8(1.2)$~eV \cite{Nagy2006} might allow for an absolute calibration and thus for an improved limit on the neutrino mass from KATRIN \cite{Wein2002}. An attempt to achieve $\delta Q_0 < 100$~meV is presently pushed forward at the Max-Planck-Institute f\"ur Kernphysik (MPIK Heidelberg, Germany) within the THe-TRAP project \cite{Dieh2011} as well as at the Florida State University by E.~Myers \cite{Myer2012}.

A comparable limit of 0.2~eV (90\%~c.l.) for the $\bar{\nu}_e$ mass is aimed for in the MARE project \cite{Monf2006}, which studies similarly to KATRIN the $\beta^-$-decay of $^{187}$Re. Here, the $Q_0$ value determined by the mass difference between $^{187}$Re and $^{187}$Os needs to be known with an unprecedented relative uncertainty of a few parts in $10^{12}$. The high-precision cryogenic multi-Penning-trap mass spectrometer PENTATRAP \cite{Repp2012,Roux2012}, also at MPIK Heidelberg, aims for such a precision employing highly charged ions.

\subsubsection{Neutrinoless Double-$\beta$ Decay and Double-Electron Capture Processes}
In the last few years tremendous efforts have been put into experiments to clarify the ambiguity whether the neutrino is a Dirac or a Majorana particle. The observation of a neutrinoless double $\beta^-$-decay ($0\nu\beta^-\beta^-$) or a neutrinoless double-electron capture decay ($0\nu$ECEC) would give an unambiguous answer since these decays can only exist if the neutrino is of Majorana type, i.e.\ its own antiparticle ($\nu_e \equiv \bar\nu_e$). This would violate the conservation of the total lepton number and manifest a clear sign for physics beyond the Standard Model.

There are presently several experiments in operation or under preparation aiming for the direct observation of the neutrinoless double-beta decay $(0\nu\beta\beta)$ \cite{Gome2012} for the clarification of the nature of the neutrino-antineutrino duality. Investigations concerning neutrinoless double-electron capture $(0\nu$ECEC) are still on the theory level \cite{Verg2011,Kriv2011,Elis2012} since they are even more challenging as will be discussed below.

\begin{figure}[h!]
\begin{center}
\includegraphics[width=\textwidth]{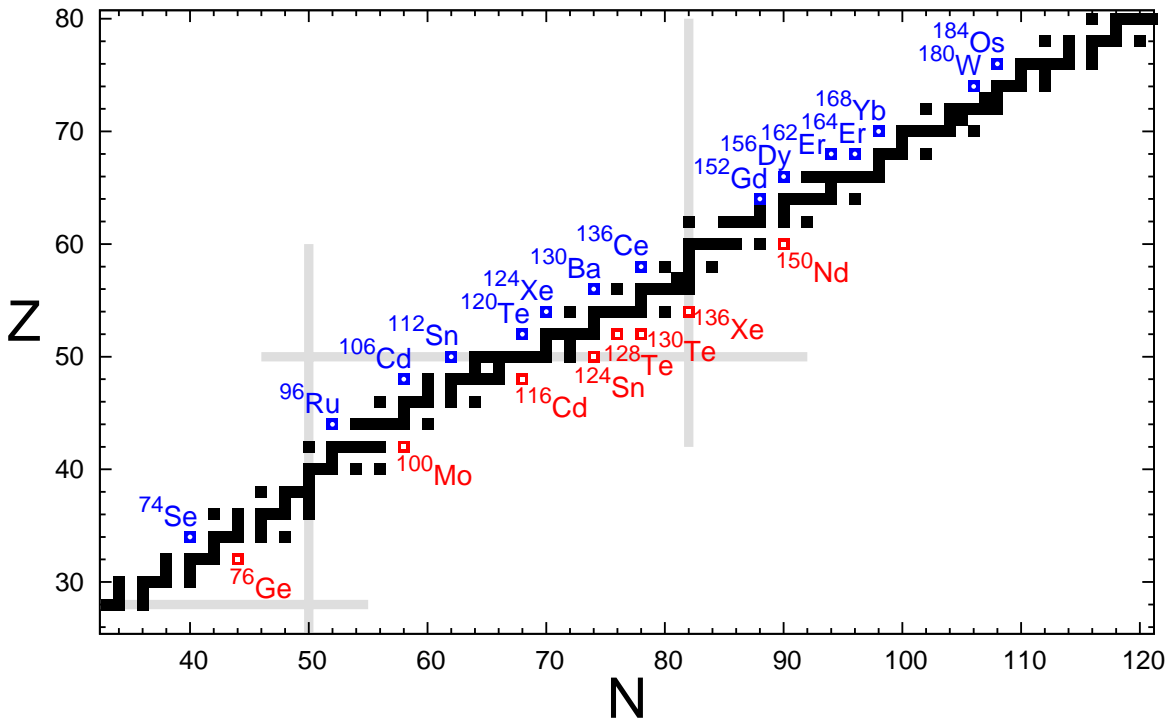}
\end{center}
\caption{(Colour online, updated figure from \cite{Blau2012}) Double-beta decay (red) and double-electron capture (blue) nuclei that have been addressed by high-precision Penning-trap mass spectrometry.}
\label{neutrinophysics}
\end{figure}

A survey of all masses of known nuclides in the nuclear chart reveals that only 35 of them can undergo a $0\nu\beta^-\beta^-$ decay due to energetic reasons. But only eleven of them are of practical use for the search of such a decay in large scale detectors due to their sufficiently large $Q$-values \cite{Zube2010}. The experimental signature would be the observation of an increased event-rate at the $Q_0$-value of the decay in a sum-energy spectrum of the two emitted electrons. Due to the huge background from the orders-of-magnitude stronger decay under the emission of two neutrinos ($2\nu\beta^-\beta^-$), the knowledge of the $Q_0$-value with an uncertainty of less than 1~keV is of utmost importance in order to search for $0\nu\beta^-\beta^-$ events at the right energy. To date, as shown in figure~\ref{neutrinophysics}, the $Q_0$-values of eight out of the eleven most interesting double $\beta^-$-decay transitions have been determined by high-precision Penning-trap mass spectrometry with uncertainties as low as a few 100~eV. Among them the four transitions under investigation in large-scale experiments, i.e.\ the double $\beta^-$-decays of $^{76}$Ge \cite{Douy2001,Moun2010}, $^{100}$Mo \cite{Raha2008}, $^{130}$Te \cite{Reds2009,Scie2009}, and $^{136}$Xe \cite{Reds2007,Kolh2011}.

An experiment for the observation of the $0\nu$ECEC process is not yet planned, although it would allow for the same physics conclusions as the neutrinoless double $\beta^-$-decay experiments. The reason lies in the expected 4-6 orders of magnitude longer half-lives and thus in the amount of material needed to search for such a virtually unobservable process. But as pointed out already in 1983 by Bernabeu \emph{et al.\ } \cite{Bern1983}, the process can be resonantly enhanced by a favorable nuclear and atomic structure: An energy degeneracy between the initial and the final state of the transition, i.e.\ when the $Q$-value is close to the energy of an excited state in the daughter atom, the decay probability can be substantially increased. The determination of such resonance conditions for specific transitions \cite{Kriv2011} -- in order to find suitable candidates for $0\nu$ECEC searches -- called for high-precision Penning-trap mass measurements.
In total 15 $Q$-value measurements of double-electron capture transitions, as indicated in figure~\ref{neutrinophysics}, have been performed since the first measurement by JYFLTRAP on $^{112}$Sn in 2009 \cite{Raha2009}, all with uncertainties well below 1~keV. However, so far only one system, namely $^{152}$Gd-$^{152}$Sm, meets the criteria of a full energy degeneracy, which results in a half-life estimate of $10^{26}$ years when normalizing to an effective neutrino mass of 1~eV \cite{Elis2011a}. For a detailed overview on all so-far investigated double-electron capture transitions by Penning-trap mass spectrometry we refer to the recent review by Eliseev \emph{et al.\ } \cite{Elis2012}. Among those one other system is worth mentioning due to its uniqueness: $^{156}$Dy-$^{156}$Gd. Here, a multi-resonance phenomenon has been discovered with four resonantly enhanced transitions to nuclear excited states in the daughter nuclide \cite{Elis2011b}. Still, the estimated half-life of $^{156}$Dy exceeds that of $^{152}$Gd by two orders of magnitude and thus excludes it as a suitable candidate for the search of $0\nu$ECEC-decays.

\section{Conclusion and Outlook}

Nuclear physics with radioactive beams opens up new venues to investigate nature on a deeper level then ever possible before. This includes tests of refined theoretical approaches where novel or amplified phenomena occur. Moreover, tests of symmetries become possible due to access to radioactive isotopes with specific and distinct features. The field of nuclear astrophysics is completely revolutionized due to beams of short-lived exotic species that can now be produced in laboratories and govern the nucleosynthesis of the chemical elements in the stellar objects. All of these sub-disciplines of nuclear physics benefited enormously from progress in adopting atomic physics techniques to the specific requirements posed by radioactive isotopes. The determination of ground state properties from mass spectrometry and laser spectroscopy is possible with unrivaled precision and sensitivity. These techniques have become key tools in detailed studies of physics with radioactive beams.

The success of these tools is reflected in their broad usage and large number of applications.
All new and upgrading facilities have an active or planned program for laser spectroscopy or at least laser ion sources. At JYFL a new cyclotron has been installed that is dedicated to low-energy studies of nuclear ground-state and isomer properties. At ISOLDE the COLLAPS and ISOLTRAP setups await the installation of HIE-ISOLDE, and at TRIUMF the collinear beamline is increasingly being used for laser spectroscopy studies. The TITAN facility at TRIUMF has just started its program and provided already the masses of halo nuclei with unprecedented accuracy, as well as first Penning-trap mass measurements with short-lived highly charged ions. With the upcoming ARIEL facility even more exotic isotopes will be in reach. New collinear laser spectroscopy and Penning trap installations are arising at TRIGA-Mainz \cite{Ketelaer2008}, which are the prototypes of the LASPEC \cite{Noertershaeuser2006,Rodriguez2006} and MATS experiments \cite{Rodriguez2006} at FAIR, and the BECOLA experiment \cite{Minamisono2009} at FRIB. Laser ion sources are planned also at ALTO and GANIL and a laser spectroscopy program (LUMIERE) as well as a Penning-trap program is foreseen at DESIR (SPIRAL2). Penning-trap mass measurements are already ongoing at the recently pre-commissioned CARIBU facility \cite{VanS2012} and a collinear laser spectroscopy setup is under consideration.
As mentioned earlier (compare table~\ref{tab:PT_world}) basically all current facilities and planned facilities use or will use Penning trap system for mass determinations, and more storage ring experimental set-ups are being planned or considered.

\section{Acknowledgments}
This work is partly supported by the Max-Planck Society, by the Helmholtz association through the Nuclear Astrophysics Virtual Institute (VH-VI-417/NAVI) and the HYIG VH-NG-148, the Federal Ministry of Education and Research (BMBF, 06MZ7169 and 06MZ7171I), through the ExtreMe Matter Institute (HA216/EMMI), and by the EU (ERC Grant No. 290870 - MEFUCO). Moreover, support was received from the Natural Sciences and Engineering Research Council of Canada (NSERC), Canadian Foundation of Innovation (CFI), and by the National Research Council of Canada (NRC). We appreciate the help of Dr.\ R.\ Sanchez in preparing figures and references.
\\

%TCIDATA{Version=5.50.0.2890}
%TCIDATA{LaTeXparent=0,0,BDN.tex}

\end{document}